\documentclass[aps,twocolumn]{revtex4}
\usepackage{graphicx}
\usepackage{amsmath}
\usepackage{amssymb}
\usepackage{mathtools}
\usepackage{hyperref}
\usepackage{xcolor}
\usepackage{bm}
\usepackage[T1]{fontenc}
\usepackage{upquote}

\DeclareMathOperator{\ch}{ch}
\DeclareMathOperator{\sh}{sh}
\DeclareMathOperator{\diag}{diag}
\newcommand{\rn}{{\rm n}}

\begin{document} 

\title{\bf Path-reversal, Doi-Peliti generating functionals, and
dualities between dynamics and inference for stochastic processes} 

\author{Eric Smith}

\affiliation{Earth-Life Science Institute, Tokyo Institute of
Technology, 2-12-1-IE-1 Ookayama, Meguro-ku, Tokyo 152-8550, Japan}

\affiliation{Department of Biology, Georgia Institute of
Technology, 310 Ferst Drive NW, Atlanta, GA 30332, USA}

\affiliation{Santa Fe Institute, 1399 Hyde Park Road, Santa Fe, NM
87501, USA}

\affiliation{Ronin Institute, 127 Haddon Place, Montclair, NJ 07043,
USA} 

\author{Supriya Krishnamurthy}

\affiliation{Department of Physics, Stockholm University, SE- 106 91,
Stockholm, Sweden} 

\date{\today}
\begin{abstract}

A variety of fluctuation theorems, concerning probability ratios of
rare versus common trajectories in stochastic processes, are defined
in terms of reversed paths and driving protocols.  A subset of these
apply the measure of the original stochastic process to reversed
paths, and have an interpretation in terms of time-reversed dynamics.
More generally, the adjoint measure of the fluctuation theorems,
constructed to reverse probability currents, may define a new
stochastic process, which is not inherently dynamical and may even be
incompatible with reverse-time dynamics.  Here we develop a general
interpretation of fluctuation theorems based on the adjoint process by
considering the duality of the Kolmogorov-forward and backward
equations that define the adjoint generator, as these equations apply
asymmetrically to distributions and to observables.
Kolmogorov-backward propagation of observables is related to problems
of statistical inference, so we characterize the adjoint construction
as a duality between dynamics and inference.

The adjoint process corresponds to the Kolmogorov backward equation in
a generating functional that erases memory from the dynamics of its
underlying distribution.  We derive this result for general
correlation functions by showing that fluctuation-theorem duality
exchanges the roles of advanced and retarded Green's functions.  We
work at the level of stochastic Chemical Reaction Networks, and show
that dualization acts on the \emph{finite} representation of the
generating event-set, in a manner similar to the usual action of
similarity transform on the (potentially infinite) set of state
transitions.  We use the Doi-Peliti functional integral representation
of generating functionals, within which duality transformation takes a
remarkably simple form as a change of integration variable.  Our
Green's function analysis is used to recover the Extended
Fluctuation-Dissipation Theorem of Seifert and Speck for
non-equilibrium steady states, and to show that the causal structure
responsible for it applies also to dualization about non-steady
states.

\end{abstract}

\maketitle

\section{Introduction: fluctuation theorems from different approaches
to path reversal}

Given any two probability distributions $P$ and $P^{\ast}$ defined on
a set of extended time trajectories for some system, it is a tautology
that the average of $P^{\ast} / P$ in the distribution $P$ must be
unity, because it is the same as the trace of the distribution
$P^{\ast}$.  In symbols, $E_P \left( P^{\ast} / P \right) \equiv 1$,
where $E_P$ denotes expectation in measure $P$.  At its simplest, the
relation could be so little as the change of measure on the space of
trajectories, and $P^{\ast} / P$ could be the Jacobean of the
corresponding change of variables.

Starting in the early 1990s for dynamical
systems~\cite{Evans:shear_SS:93,Gallavotti:dyn_ens_NESM:95,Gallavotti:dyn_ens_SS:95,Cohen:NESM_2_thms:99},
and a few years later for stochastic
processes~\cite{Jarzynski:eq_FE_diffs:97,Jarzynski:neq_FE_diffs:97,Kurchan:fluct_thms:98,Searles:fluct_thm:99,Crooks:NE_work_relns:99,Crooks:path_ens_aves:00,Kurchan:NEWRs:07,Jarzynski:fluctuations:08,Chetrite:fluct_diff:08,Esposito:fluct_theorems:10},
a carefully chosen subset of these tautologies have been exploited to
gain information about complex, potentially non-stationary
non-equilibrium distributions in terms of simpler stationary or
equilibrium distributions and functions along paths that can be
measured mechanically or calorimetrically.  What the family of
tautologies have in common that have been put to this use, is that the
probability $P^{\ast}$ is defined from $P$ in some way that involves
reversing the direction in which each trajectory is traversed,
generally along with the externally-imposed protocol of control
parameters under which the distributions evolve.

When constructing a probability $P^{\ast}$ to study based on reversal,
the most basic choice faced is whether to use the original measure $P$
while only reversing trajectories and protocols, or whether to
construct a new measure that will reverse probability
currents~\cite{Hatano:NESS_Langevin:01,Chernyak:PI_fluct_thms:06,Chetrite:fluct_diff:08}.
Relative to the process that generates $P$, the former choice results
in what is conventionally termed the ``backward'' process, while the
result of the latter choice is termed the ``adjoint''
process~\cite{Harris:fluct_thms:07}.  If the systems in question are
Markovian stochastic processes, the backward and adjoint processes are
the same precisely in the case that the transition probabilities
satisfy a condition of detailed balance, in which case their steady
states are true equilibrium distributions in the strict sense of
Boltzmann/Gibbs thermodynamics.

The general class of theorems of the form $E_P \left( P^{\ast} / P
\right) \equiv 1$ based on path reversal for stochastic processes are
called Integral Fluctuation Theorems.  Their refinements, in which the
ratios $P^{\ast} / P$ are given a functional form for individual
trajectories, are called Detailed Fluctuation
Theorems~\cite{Esposito:fluct_theorems:10}.  Integral and detailed
fluctuation theorems have been derived to describe rare-event
statistics in thermal
ensembles~\cite{Searles:fluct_thm:99,Crooks:NE_work_relns:99}, to
estimate equilibrium free energy values by sampling mechanical work
under non-equilibrium conditions, where they are known as
Non-Equilibrium Work Relations
(NEWRs)~\cite{Jarzynski:eq_FE_diffs:97,Jarzynski:neq_FE_diffs:97,Crooks:NE_work_relns:99,Crooks:path_ens_aves:00,Kurchan:NEWRs:07},
to separate out the contributions to heat flows that maintain a
non-equilibrium driven state from those that provide entropy to change
the state~\cite{Oono:NESS_thermo:98,Hatano:NESS_Langevin:01}, and to
derive generalized fluctuation-dissipation relations applicable to
non-equilibrium steady
states~\cite{Chetrite:fluct_diff:08,Chetrite:fluct_nonstat_bath:09,Seifert:FDT:10,Chetrite:refreshing_fluct:11,Verley:FDT_motors:11,Verley:FDT_ising:11,Verley:HS_FDT:12}.
The literature on fluctuation theorems is now large; for reviews
see~\cite{Evans:fluct_thm:02,Harris:fluct_thms:07,Chetrite:fluct_diff:08,Seifert:stoch_thermo_rev:12}.

Markov processes with detailed balance are a very restricted class,
about which a great deal is already known, and so much of the study of
probabilities under path reversal aims at understanding more general
cases.  For these, different information is gained, and different
limitations are faced, in the study of the backward versus the adjoint
processes~\cite{Harris:fluct_thms:07}.  The backward
process,\footnote{Note that the term ``backward process'', which we
adopt from~\cite{Harris:fluct_thms:07}, is not a reference to the
Kolmogorov-backward equation, which also figures in our development.}
requiring only a permutation of the probabilities defined in the
original process, works directly from the original specification of
dynamics, without reference to any particular solutions that they
generate.  The construction is limited, however, to cases in which
every trajectory with a measure in the original system has a reverse
which also has a measure in that system.  Even when that is the case,
the averages of $P^{\ast} / P$ produced for the backward process may
not carry useful information about quite ordinary properties such as
the Shannon entropy difference between non-equilibrium steady states
connected by a slow parameter change~\cite{Hatano:NESS_Langevin:01}.

Adjoint processes are more flexible than backward processes, but they
are also more arbitrary.  The adjoint process constructed from an
underlying Markov process is essentially a new process; the
trajectories to which it assigns probabilities are simply chosen to be
the reverses of those in the original process, whether or not they
themselves have probabilities in that process.  However, the property
of reversal of probability currents is not defined solely in terms of
the transition probabilities of the original process: it must make
reference to some state of the system from which those currents
originate.  Typically for non-stationary Markov chains, the reference
distribution at any time is taken to be the stationary distribution
that would result if the parameters in the Markov chain at that time
were fixed.\footnote{For stationary Markov chains, this is the actual
steady-state solution, and may be used directly as such.}  This choice
permits a definition of the adjoint that at least does not depend on
extended-time solutions for probability distributions, but even the
stationary distribution generally depends on probability flows among
many or all system states, and as such is non-local in the state
space~\cite{Harris:fluct_thms:07}.  In general it may also be
difficult to construct explicitly.  What is gained from the choice to
study averages of $P^{\ast} / P$ for the adjoint process may be the
ability to separate the maintenance-associated and change-associated
costs of altering a non-equilibrium
state~\cite{Oono:NESS_thermo:98,Hatano:NESS_Langevin:01}, as well as
new identities among correlation functions and response functions in
the non-equilibrium
background~\cite{Chetrite:fluct_diff:08,Chetrite:fluct_nonstat_bath:09,Seifert:FDT:10,Chetrite:refreshing_fluct:11,Verley:FDT_motors:11,Verley:FDT_ising:11,Verley:HS_FDT:12,Espigares:inf_2nd_law_ineq:12}.

The great majority of work done in the emerging discipline of
Stochastic Thermodynamics~\cite{Seifert:stoch_thermo_rev:12} involves
path trajectories $P^{\ast}$ that can be somehow constructed from the
backward process, seemingly for two reasons.  One is that the backward
process has an immediate interpretation in terms of events and their
time-reverses in the original process, and thus can be used to
understand the origin of thermodynamic arrows of
time~\cite{Evans:fluct_thm:02}, even as these are manifested in quite
complex
systems~\cite{Perunov:adaptation:15,Horowitz:fine_tuning_NESM:17}.
The second is that the major body of work in statistical physics since
the
1930s~\cite{Onsager:RRIP1:31,Onsager:RRIP2:31,Nicolis:fluct_NEQ:71,Glansdorff:structure:71,DeGroot:NET:84},
to formulate a theory of non-equilibrium thermodynamics, has been
based on the effort to derive constraints on non-equilibrium dynamics
from changes in the equilibrium entropy.  While not possible in
general, and not as general as the pursuit of a full non-equilibrium
large-deviation
theory~\cite{Touchette:large_dev:09,Touchette:LDP_NEQ_sys:13}, this
strategy is highly successful for systems with a separation of
timescales, in which a slow process with microscopically resolved
discrete events evolves in the presence of a fast-relaxing thermal
bath.  For such systems the trajectory probabilities in the original
and backward processes can be expressed in terms of changes in the
bath's thermodynamic entropy and the system's Shannon
entropy~\cite{Harris:fluct_thms:07,Seifert:FDT:10}, providing a
conservative but still very productive extension of equilibrium
thermodynamic methods to non-equilibrium systems.

Even cases in which $P^{\ast}$ was constructed as an adjoint process
for an underlying system without detailed balance, such as the
irreversible Langevin system studied by Hatano and
Sasa~\cite{Hatano:NESS_Langevin:01}, have been interpreted with
respect to heat flows and entropy-production (with the equilibrium
entropy state variable defining the entropy ``produced'' in the bath),
and thus implicitly with reference to time reversal at least for the
bath~\cite{Seifert:entP_traj:05}.  The very heavy dependence on time
reversal for interpretations of fluctuation theorems or NEWRs has
limited the study of adjoint processes.  This is reflected in the
caution given by Seifert~\cite{Seifert:stoch_thermo_rev:12} (p.~41)
that, given only the Langevin equation itself without a further
specification of context, the assignment of a quantity such as
dissipated work based on analogy with physically reversible cases, may
be ``a purely formal one without real physical meaning.''

\subsubsection*{Adjoint fluctuation theorems, beyond time reversal}

Often one wishes to study adjoint stochastic processes in which the
transition probabilities for dynamics are not interpreted in terms of
bath-entropy production, or they do not support detailed-balance
equilibria, or where the reverses of allowed trajectories are not even
allowed in the original process.  Models of this form naturally arise
in Chemical Reaction Network (CRN)
theory~\cite{Horn:mass_action:72,Feinberg:def_01:87} for networks that
produce steady states without detailed balance.  While, for
applications to chemistry, entropy-production interpretations may
still be valid, this class of models equally well represents
evolutionary population processes\footnote{This result is implicit in
that the representation we have used
elsewhere~\cite{Smith:evo_games:15} for stochastic population
processes is the same one produced by
CRNs~\cite{Krishnamurthy:CRN_moments:17,Smith:CRN_moments:17}, which
we will use below.} in which events of death and reproduction can
reflect inputs of a completely different origin than physical
thermodynamics.  Even within chemistry, the choice to omit a reverse
reaction from a CRN, which immediately precludes defining the backward
process, is a commitment that the dynamics of the system is controlled
by factors other than rare fluctuations against the likely directions
of flow; it should be possible to understand these in their own terms
as an effective theory~\cite{Weinberg:phenom_Lagr:79}.

For all cases it is desirable to assign a meaning to the adjoint
process that follows directly from its construction from the original
dynamics, without reference to time reversal and without dependence on
additional framing outside the dynamical equations themselves.  This
is true even if the adjoint process may \emph{also} have a
case-dependent interpretation in terms of time reversal.

Our aim in this paper is to explain a natural semantics for
fluctuation theorems based on the adjoint construction of $P^{\ast}$,
which follows from the transposed representations of the underlying
generator of the stochastic process.  Introduced by Kolmogorov as the
``forward'' and ``backward'' evolution
operators~\cite{Gardiner:stoch_meth:96}, these act respectively on
probability distributions or on observables averaged in those
distributions.

Transposing the action of a generator defines the sense of path- and
protocol-reversal beneath the construction of the adjoint process, but
by itself, this transposition is simply a regrouping of terms in a
sum, with no new information.  The nontrivial step in constructing the
adjoint process is recognizing that the ratio $P^{\ast} / P$
constitutes a \textit{tilt} that forms a generating
functional~\cite{Touchette:LDP_NEQ_sys:13} on the evolving probability
distribution of the underlying process.  The adjoint processes that
produce interesting fluctuation theorems are identified by going
beyond the identity $E_P \left( P^{\ast} / P
\right) \equiv 1$ to study the behavior of correlation functions in
the presence of the tilt\footnote{An example is the standard
formulation of the Transient Fluctuation Theorems in terms of a
generating functional for the path work; see~\cite{Kurchan:NEWRs:07}.}.
The particular adjoint process produced by dualizing about the
instantaneous steady states under the Markov generator is one that
erases memory of the causal history of the distribution from
correlation functions.  Mathematically, this criterion is equivalent
to interchanging the forms of the Kolmogorov forward and backward
generators, in a sense we will make precise below.  Other results
known for the adjoint process, such as the extended
Fluctuation-Dissipation Theorems
(FDTs)~\cite{Chetrite:fluct_diff:08,Chetrite:fluct_nonstat_bath:09,Seifert:FDT:10,Chetrite:refreshing_fluct:11,Verley:HS_FDT:12},
follow from this memory-erasing property, not only for stationary
nonequilibrium states, but also about non-stationary backgrounds.

If the natural interpretation of the backward process is physical, in
terms of time-inverted sampling of dynamical paths, the natural
interpretation of the adjoint process, suggested by the relation of
Kolmogorov forward and backward equations, should be informational.
The inverted time order by which operators depend on states under the
Kolmogorov-backward equation has no relation to reversal of dynamics;
it arises because in a causal process the expectations of operators
depend on the conditions imprinted on states at earlier times.  A
generating functional alters the weight with which trajectories are
sampled to produce average values, in a manner closely related to the
procedure of importance sampling from statistical inference.
Therefore we characterize the duality of fluctuation theorems using
the adjoint process as being dualities between dynamics and inference
for a stochastic process.

\subsubsection*{Green's functions from generating functionals, and the
Doi-Peliti 2-field representation}

The tools to efficiently study the propagation of disturbances either
forward or backward in time are the Green's functions of the
stochastic process.  Familiar as the Green-Kubo relations for linear
response to small perturbations, these are typically discussed in the
context of perturbations near equilibrium.  Within a generating
functional, however, it is possible to compute fluctuation-response
relations for perturbations about any background that the tilt of the
generating functional makes into a saddle point.  In cases where a
perturbative expansion in higher-order moments of the fluctuation
converges, the Green's functions form a basis for higher orders of
nonlinear response, while continuing to express the causal structure
in the underlying process.

Our preferred system for working with Green's functions is the
Doi-Peliti (DP) functional integral
framework~\cite{Doi:SecQuant:76,Doi:RDQFT:76,Peliti:PIBD:85,Peliti:AAZero:86}
for the representation of arbitrary generating functionals.  The DP
formalism is one of a larger family of 2-field functional integral
(2FFI) methods~\cite{Kamenev:DP:02} that include the
Martin-Siggia-Rose formalism for stochastic differential
equations~\cite{Martin:MSR:73} and the Schwinger-Keldysh
time-loop~\cite{Schwinger:MBQO:61,Keldysh:noneq_diag:65} for quantum
systems with decoherence.  The Doi operator representation abstracts
the linear algebra of generating functions, and the Peliti
construction then evaluates the time integrals of these on continuous
basis sets derived from Poisson distributions\footnote{In this
respect the Peliti functional integral may be understood as an
extension of the Poisson representation of Gardiner and
Chaturvedi~\cite{Gardiner:Poisson_rep:77,Chaturvedi:Poisson_rep:78}.
For an application of their method to Large Deviations and fluctuation
theorems, see~\cite{Petrosyan:NEQ_Lyap:14}.}, which become the field
variables of functional integrals.  This additional change of
representation leads to very direct derivations of Green's functions
and fluctuation-dissipation relations~\cite{Kamenev:DP:02}, as well as
very elegant routes to ray-theoretic or other semiclassical
approximation
methods~\cite{Freidlin:RPDS:98,Smith:LDP_SEA:11,Smith:evo_games:15}.

The most important feature of the DP construction for us will be that
the two fields of this 2FFI formalism directly represent the action of
both the forward and backward generators.  Time-dependent correlations
can be computed either by propagating distributions forward in time
with \textit{retarded} Green's functions which are causal, or by
propagating observables backward in time with \textit{advanced}
Green's function which are anti-causal.  The relations between these
and the equal-time correlation function is the basis for
non-equilibrium fluctuation-dissipation
relations~\cite{Kamenev:DP:02,Smith:LDP_SEA:11}.  The anti-causality
of the advanced Green's function, in the context of saddle-point
approximations~\cite{Goutis:saddlepts:95} provides our connection
between the backward equation and methods of importance
sampling~\cite{Owen:mcbook:13}.

The elements we believe are new in this paper include the DP
construction of fluctuation theorems for general discrete-state
stochastic processes, which is a straightforward extension from the
study of Langevin systems (equivalent to Gaussian-order fluctuations)
in~\cite{Chernyak:PI_fluct_thms:06}.  The elegant form of the DP
generator for CRNs (Eq.~(\ref{eq:L_psi_from_A}) below) and the
significance of its modular structure, although introduced
previously~\cite{Krishnamurthy:CRN_moments:17,Smith:CRN_moments:17},
will play a new role in the context of fluctuations theorems.  It was
recognized earlier by Baish~\cite{Baish:DP_duality:15} that the
construction of the adjoint process corresponds in the DP functional
integral to a simple change of the dummy variable of integration.  We
will show that this change of variable produces a similarity
transformation of the adjacency matrix of the CRN, exactly reflecting
on a finite generator set the similarity transform that defines the
adjoint process on the matrix of transition probabilities which has
the dimension of the state space.  Although the extended FDT that we
derive was chosen to recover the result of~\cite{Seifert:FDT:10} for
non-equilibrium steady states, our Green's function derivation makes
contact between two different notions of FDT, and also shows that the
results extend to non-stationary backgrounds and non-linear
fluctuations.

The presentation is organized as follows.  Sec.~\ref{sec:gen_bkgnd}
provides basic notation for discrete-state stochastic processes, and
reviews the standard construction for fluctuation theorems using the
adjoint process.  We then review Kolmogorov forward/backward duality,
and introduce the concepts and notations for the DP generating
functional construction, emphasizing the particularly simple form it
takes for stochastic CRNs.

Sec.~\ref{sec:duality_SPs} develops a very simple class of examples
that can be solved exactly, to illustrate the meaning and roles of the
conjugate fields in 2-field functional integrals, their connection to
concepts from importance sampling, the way duality appears as a change
of variables in the integral representation, and how tilting to form
generating functionals interacts with dynamics, particularly for
NEWRs.

Sec.~\ref{sec:caus_anticaus} extends the results of
Sec.~\ref{sec:duality_SPs} from the stationary-point approximation to
the Green's functions for response and fluctuation, and to general
CRNs for which a second-order expansion of the rate equations is
defined.  Here we introduce the versions of FDTs that emerge in
functional integrals as a consequence of internal symmetries, and
relate them to the more familiar forms used
in~\cite{Seifert:FDT:10,Polettini:trans_fluct:14}.  We provide
explicit solution forms for the simple models of
Sec.~\ref{sec:duality_SPs} in order to show the subtle manner in which
path weights, by exchanging the roles of distributions and
observables, harness causal and anticausal response functions that
were already present as a reflection of the Kolmogorov transpose
relation.

Sec.~\ref{sec:duality_gen_CRN} derives the appropriate generalization
of NEWRs in the CRN framework, emphasizing the connection between the
required form of the path-weight function to produce anti-causal
dynamics, and the original duality between the forward and backward
equations with which we began the discussion.

\section{Background: discrete-state stochastic processes, dualization,
and generating functionals}
\label{sec:gen_bkgnd}

We begin by defining the class of stochastic processes to be
considered and introducing notation for the master equation and
generator. 

\subsection{Discrete-state stochastic processes and generator}

Let a state be given by a vector $\rn \equiv \left[ {\rn}_p
\right]$ with non-negative integer-valued components, for $p \in 1,
\ldots , P$.  We will refer to $p$ as an index of the \textit{species}
in the system, and ${\rn}_p$ as the count of species $p$.  Let
${\rho}_{\rn}$ be a probability density indexed on $\rn$.  Dynamics
for $\rho$ are governed by a continuous-time
\textit{master equation} of the form
\begin{align}
  \frac{d \rho}{d \tau} 
& = 
  {\rm T} \rho 
\nonumber \\ 
\mbox{shorthand for } \quad
  \frac{d {\rho}_{\rn}}{d \tau} 
& = 
  \sum_{{\rn}^{\prime}}
  {\rm T}_{\rn {\rn}^{\prime}} 
  {\rho}_{{\rn}^{\prime}} .
\label{eq:ME_genform}
\end{align}
${\rm T} \equiv \left[ {\rm T}_{\rn {\rn}^{\prime}}
\right]$, called the \textit{transition rate matrix} of the stochastic
process, is a representation of its generator.  In general we regard
${\rm T}$ as having a fixed form -- for instance reflecting a fixed
set of reactions in a CRN -- but adjustable parameters such as rate
constants representing controls or interactions with an environment.
Sometimes it is convenient to re-express the control parameters in
terms of quantities such as mean values of the steady state they would
produce; when we do so we denote those values with underbars, to
distinguish them from mean values of dynamical states actually
produced.  Following~\cite{Hatano:NESS_Langevin:01}, denote these
parameters $\underline{\alpha}$; in general they may vary with time as
the process runs.

\subsection{Fluctuation theorems based on the adjoint process} 

The construction of the adjoint process to some underlying system
follows from a relation between a path-weight that can be used to
modify the measure on paths, and a dual generator constructed from the
underlying Kolmogorov backward operator by absorption of the weighting
terms.  The path weight is derived from the stationary distribution at
the instantaneous parameter values $\underline{\alpha}$.  For systems
with detailed balance evolving under a Hamiltonian, the logarithm is
proportional to the energy.  To review the construction as given in
Equations~(5--9) of~\cite{Hatano:NESS_Langevin:01}, with minimal
notation, we reduce the stochastic process to a discrete-time process
by evolving for finite time intervals, and treating the parameters in
the generator as fixed within any short interval.

Let $\delta \tau$ be a small enough time interval that the parameters
in ${\rm T}$ are effectively constant over that interval, and
introduce the matrix of one-step transition probabilities 
\begin{equation}
  {\rm W} \equiv e^{\delta \tau {\rm T}}
\label{eq:W_def_from_T}
\end{equation}
Let $\underline{\rho}$ denote the steady-state distribution
annihilated by ${\rm T}$ at parameters $\underline{\alpha}$.  The
one-step transition matrix satisfies
\begin{align}
  \sum_{\rn} {\rm W}_{{\rn}^{\prime} \rn}
  {\underline{\rho}}_{\rn} 
& = 
  {\underline{\rho}}_{{\rn}^{\prime}} 
\nonumber \\ 
  \sum_{{\rn}^{\prime}} {\rm W}_{{\rn}^{\prime} \rn} 
& = 
  1 \; ; \forall \rn , 
\label{eq:W_properties}
\end{align}
where the second line states that ${\rm W}$ is a stochastic matrix. 

% Dualization concerns averages over ensembles of trajectories.  
Denote a discrete-time trajectory in the lattice of species counts by
$\left[ \rn \right]$, which is a sequence of states ${\rn}_0, {\rn}_1,
\ldots {\rn}_K$ over a real-time interval of length $K \delta \tau
\equiv T$.  Let $\left[ \underline{\alpha} \right]$ likewise denote a
protocol, a sequence of values of the control parameters
${\underline{\alpha}}_0, {\underline{\alpha}}_1, \ldots
{\underline{\alpha}}_K$.  Write ${\rm W}^{\left( k \right)}_{{\rn}_{k+1}
{\rn}_k} \equiv {\rm W}_{{\rn}_{k+1} {\rn}_k} \!
\left( {\underline{\alpha}}_k \right)$ for the matrix element in the
one-step matrix at step $k$, between the $k$th and $\left( k+1
\right)$th state in trajectory $\left[ \rn \right]$, and likewise
denote by ${\underline{\rho}}^{\left( k \right)}_{\rn}$ the density at
any $\rn$ in the steady state under ${\rm W}^{\left( k
\right)}$.

The expectation of any function $g_{\left[ \rn \right]}$ on
trajectories, starting from the steady-state density at step $k = 0$,
is defined in the measure
\begin{equation}
  \left< g \right> \equiv 
  \sum_{\left[ \rn \right]}
  g_{\left[ \rn \right]}
  \left( 
    \prod_{k = 0}^{K-1}
    {\rm W}^{\left( k \right)}_{{\rn}_{k+1} {\rn}_k}
  \right) 
  {\underline{\rho}}^{\left( 0 \right)}_{{\rn}_0} . 
\label{eq:DT_traj_measure}
\end{equation}
By Eq.~(\ref{eq:W_properties}), at each step $k$ 
\begin{equation}
  \sum_{{\rn}_k} 
  {\rm W}^{\left( k \right)}_{{\rn}_{k+1} {\rn}_k}
  {\underline{\rho}}^{\left( k \right)}_{{\rn}_k} = 
  {\underline{\rho}}^{\left( k \right)}_{{\rn}_{k+1}} , 
\label{eq:per_step_preservation}
\end{equation}
from which it follows that 
\begin{align}
  \left<
    \prod_{k = 0}^{K-1}
    \frac{
      {\underline{\rho}}^{\left( k+1 \right)}_{{\rn}_{k+1}}
    }{
      {\underline{\rho}}^{\left( k \right)}_{{\rn}_{k+1}}
    }    
  \right> = 
  \left<
    \exp 
    \left( 
      \sum_{k = 0}^{K-1}
      \log {\underline{\rho}}^{\left( k+1 \right)}_{{\rn}_{k+1}} - 
      \log {\underline{\rho}}^{\left( k \right)}_{{\rn}_{k+1}}
    \right) 
  \right> = 
  1 . 
\label{eq:tilted_identity}
\end{align}
$- \log {\underline{\rho}}^{\left( k \right)}$ plays the role of a
non-equilibrium generalization of the free energy
in~\cite{Hatano:NESS_Langevin:01}.  Although in general we may not be
able to express ${\underline{\rho}}^{\left( k \right)}_{{\rn}_k}$ as a
function of ${\rn}_k$ and ${\underline{\alpha}}_k$, if we take the
difference $\log {\underline{\rho}}^{\left( k+1 \right)}_{{\rn}_{k+1}}
- \log {\underline{\rho}}^{\left( k \right)}_{{\rn}_{k+1}}$ to define
$d\tau \left( d{\underline{\alpha}} / d\tau \right) \partial
\underline{\rho} \! \left( \rn ; \underline{\alpha} \right) / \partial
\alpha$ in the limit $\delta \tau \rightarrow 0$, we may write
Eq.~(\ref{eq:tilted_identity}) as
\begin{align}
  \left<
    \exp 
    \int_0^T d\tau 
    \frac{d \underline{\alpha}}{d\tau}
    \frac{
      \partial
      \log \underline{\rho} \! \left( \rn ; \underline{\alpha} \right)
    }{
      \partial \underline{\alpha}
    }
  \right> = 
  1 . 
\label{eq:tilted_identity_ctm}
\end{align}
holding $T$ fixed as $\delta \tau \rightarrow 0$. 

Re-grouping terms in the products in Eq.~(\ref{eq:tilted_identity}),
the same path sum may be written
\begin{equation}
  \left<
    \prod_{k = 0}^{K-1}
    \frac{
      {\underline{\rho}}^{\left( k+1 \right)}_{{\rn}_{k+1}}
    }{
      {\underline{\rho}}^{\left( k \right)}_{{\rn}_{k+1}}
    }    
  \right> = 
  \sum_{\left[ \rn \right]}
  {\underline{\rho}}^{\left( K \right)}_{{\rn}_K} 
  \left( 
    \prod_{k = 0}^{K-1}
    \frac{
      1 
    }{
      {\underline{\rho}}^{\left( k+1 \right)}_{{\rn}_{k+1}}
    }
    {\rm W}^{\left( k \right)}_{{\rn}_{k+1} {\rn}_k}
    {\underline{\rho}}^{\left( k \right)}_{{\rn}_k}
  \right) . 
\label{eq:re_arranged}
\end{equation}
The chain of sums produces a sequence of normalized distributions when
evaluated from left to right in Eq.~(\ref{eq:re_arranged}), whereas in
Eq.~(\ref{eq:DT_traj_measure}) normalized distributions result from
summing from right to left.  Hence we identify the transpose of a new
one-step operator ${\hat{\rm W}}^T$ as
\begin{align}
  {\left( {\hat{\rm W}}^T \right)}_{{\rn}^{\prime} \rn} 
& \equiv 
    \frac{
      1 
    }{
      {\underline{\rho}}_{{\rn}^{\prime}}
    }
    {\rm W}_{{\rn}^{\prime} \rn}
    {\underline{\rho}}_{\rn}
\nonumber \\ 
\mbox{or} \quad 
  {\hat{\rm W}}_{\rn {\rn}^{\prime}} 
& \equiv 
    {\underline{\rho}}_{\rn}
    {\left( {\rm W}^T \right)}_{\rn {\rn}^{\prime}}
    \frac{
      1 
    }{
      {\underline{\rho}}_{{\rn}^{\prime}}
    } . 
\label{eq:adjoint_update}
\end{align}
From Eq.~(\ref{eq:W_properties}) for ${\rm W}$, it follows that 
\begin{align}
  \sum_{{\rn}^{\prime}} {\hat{\rm W}}_{\rn {\rn}^{\prime}}
  {\underline{\rho}}_{{\rn}^{\prime}} 
& = 
  {\underline{\rho}}_{\rn} 
\nonumber \\ 
  \sum_{\rn} {\hat{\rm W}}_{\rn {\rn}^{\prime}} 
& = 
  1 \; ; \forall {\rn}^{\prime}
\label{eq:hatW_properties}
\end{align}
In the limit $\delta \tau \rightarrow 0$,
Eq.~(\ref{eq:adjoint_update}) is the one-step matrix for a
transformed rate-matrix or generator
\begin{align}
  {\hat{\rm T}}_{\rn {\rn}^{\prime}} 
& \equiv 
    {\underline{\rho}}_{\rn}
    {\left( {\rm T}^T \right)}_{\rn {\rn}^{\prime}}
    \frac{
      1 
    }{
      {\underline{\rho}}_{{\rn}^{\prime}}
    } . 
\label{eq:adjoint_rate_T}
\end{align}
with the properties that 
\begin{align}
  \sum_{{\rn}^{\prime}} {\hat{\rm T}}_{\rn {\rn}^{\prime}}
  {\underline{\rho}}_{{\rn}^{\prime}} 
& = 
  0 
\nonumber \\ 
  \sum_{\rn} {\hat{\rm T}}_{\rn {\rn}^{\prime}} 
& = 
  0 \; ; \forall {\rn}^{\prime}
\label{eq:hatT_properties}
\end{align}
The \textit{adjoint} stochastic process, generated by ${\hat{\rm T}}$,
has the same steady states as the process defined by ${\rm T}$, and it
has nonzero probabilities on each trajectory that is the time-reverse
of some trajectory for which ${\rm T}$ produces non-zero
probabilities.  

Combining terms across the two lines in Eq.~(\ref{eq:adjoint_update}),
the adjoint process could alternatively have been defined as the one
that reverses probability flow between any two states in the
stationary distribution~\cite{Crooks:path_ens_aves:00},
\begin{align}
  {\hat{\rm W}}_{\rn {\rn}^{\prime}} 
  {\underline{\rho}}_{{\rn}^{\prime}} = 
  {\rm W}_{{\rn}^{\prime} \rn}
  {\underline{\rho}}_{\rn} . 
\label{eq:prob_reversal}
\end{align}
It is immediate that the condition of detailed balance is the
statement that ${\hat{\rm W}} = {\rm W}$, in which case
${\underline{\rho}}$ is an equilibrium distribution at parameters
$\underline{\alpha}$.  Clearly there is no implication of reversed
dynamics in this construction; the index $k$ on $\rn$ and
$\underline{\alpha}$ is simply read in the reverse order from the
original process to define the ``forward'' direction for ${\hat{\rm
W}}$.

\subsection{Interpreting the adjoint process in terms of Kolmogorov
forward-backward duality and memory erasure} 

To understand what the adjoint process means, accepting that part of
its definition was our arbitrary choice of the stationary
$\underline{\rho}$ as the reference for reversal of currents, we
consider the use of a similar transpose to relate the
Kolmogorov-forward and backward equations for any stochastic process.
In particular, we make explicit not only the role of distributions,
but also that of observables which may be averaged to produce moments
or correlation functions. 

Consider the expectation of an arbitrary observable $\mathcal{O}$ at a
single time in a distribution 
$\rho$, given by
\begin{equation}
  \left< \mathcal{O} \right> \equiv 
  \sum_{\rn}
  {\mathcal{O}}_{\rn} {\rho}_{\rn} . 
\label{eq:EA_min_form}
\end{equation}
The ${\mathcal{O}}_{\rn}$ are chosen not to be explicit functions of
time, so all time dependence in $\left< \mathcal{O} \right>$ results
from the evolution of $\rho$.  This evolution can be expressed in two
ways, as
\begin{align}
  \frac{d}{d \tau} 
  \left< \mathcal{O} \right> 
& = 
  \sum_{\rn}
  {\mathcal{O}}_{\rn} 
  {
    \left( 
      {\rm T} \rho
    \right) 
  }_{\rn} 
\nonumber \\
& = 
  \sum_{\rn}
  {
    \left( 
      {\rm T}^{\ast} \mathcal{O}
    \right) 
  }_{\rn} 
  {\rho}_{\rn} 
\label{eq:EA_evol_fwd_bck}
\end{align}
The first line, in which ${\rm T}$ acts on $\rho$, is called the
\textit{forward equation}, and the second line where ${\rm T}^{\ast}$
acts on $\mathcal{O}$ -- also formally an adjoint when written this
way, but we will think of it in terms of a matrix transpose to avoid
confusion with the fluctuation-theorem adjoint -- is called the
\textit{backward equation}\footnote{The relation
between the forward and backward equations was developed by Kolmogorov
for stochastic differential equations, where the forward equation is
also known as the \textit{Fokker-Planck}
equation.~\cite{Gardiner:stoch_meth:96}.}.  If these equations are
integrated over a finite interval separating an initial condition
${\rho}^{\left( 0 \right)}$ from ${\mathcal{O}}$ evaluated at a later
time $\tau$, the forward equation evolves the distribution up to
$\tau$, whereas the backward equation evolves the dependence of
${\mathcal{O}}$ (metaphorically, a ``shadow cast by ${\mathcal{O}}$''
on earlier times) down to the initial distribution.

The Kolmogorov-backward generator ${\rm T}^{\ast}$ is not yet the
generator of the adjoint process; it is a stochastic matrix on the
``wrong'' index.  The rescaling of Eq.~(\ref{eq:adjoint_update}) has
the effect of exchanging the behavior of the indices that contract
with distributions and with observables. To show how the the path
weight~(\ref{eq:tilted_identity}) responsible for the rescaling is
constructed as a tilt of a dynamically evolving distribution $\rho$
that is \emph{not} generally equal to $\underline{\rho}$, we begin by
reviewing the use of generating functions and functionals to study the
time evolution of distributions and correlation functions, and show
how it is instantiated in the DP functional integral formalism.

\subsection{Generating functions, and the Doi-Peliti framework}
\label{sec:text_DP_intro}

\subsubsection*{The Doi Hilbert space}

It is often more convenient than working directly with the density
${\rho}_{\rn}$, to work with the
\textit{moment-generating function}.  Historically, the generating
function was introduced as a Laplace transform of the density
\begin{equation}
  \Phi \! 
  \left( z \right) \equiv 
  \sum_{\rn}
  \left( 
    \prod_p 
    z_p^{{\rn}_p}
  \right) 
  {\rho}_{\rn} , 
\label{eq:gen_fn_multi_arg}
\end{equation}
in which $z \equiv \left[ z_p \right]$ is a vector of complex-valued
arguments to $\Phi$\footnote{The properties of analytic functions can
lead in some cases to powerful and elegant methods to derive the
asymptotic behavior of coefficients in the series of generating
functions~\cite{Flajolet:anal_comb:09}.  Many results do not require
the full machinery of analyticity, however, and employ only the formal
power series~\cite{Wilf:gen_fun:06}.  We introduce the analytic
representation here because it clarifies some interpretations that are
often implicit in the abstract linear-algebra formalism, in which the
operator $a$ (introduced below) is treated as a ``continuous-valued''
argument, an interpretation that is manifest for its antecedent $z$
and otherwise obscure.},\footnote{An unfortunate collision of terms
causes the adjective {\tt com\textquotesingle plex} of complex
variables and the noun {\tt \textquotesingle com,plex} for CRN
reactants or products to take the same English orthography.  We will
attempt to use language that avoids ambiguity.}.  In this
representation, The moment-generating function evolves in time under a
\textit{Liouville equation} induced by the master
equation~(\ref{eq:ME_genform}), of the form
\begin{align}
  \frac{\partial}{\partial \tau} 
  \Phi \! \left( z \right)
& = 
  - \mathcal{L} \! 
  \left( z , \frac{\partial}{\partial z} \right)
  \Phi \! \left( z \right) , 
\label{eq:Liouville_eq_multi_arg}
\end{align}
in which $\mathcal{L} \! \left( z , \partial / \partial z
\right)$ is termed the \textit{Liouville operator}.  

For many uses it is not necessary to evaluate $\Phi$ as an analytic
function of argument $z$; only the linear algebra induced by its
formal power series is required.  Here we adopt a representation due
to Doi~\cite{Doi:SecQuant:76,Doi:RDQFT:76} that replaces complex
(analytic) variables and their derivatives by abstract
\textit{raising} and
\textit{lowering operators}, under the mapping 
\begin{align}
  z_p 
& \rightarrow 
  a^{\dagger}_p ; & 
  \frac{\partial}{\partial z_p} 
& \rightarrow 
  a_p .
\label{eq:a_adag_defs}
\end{align}
The algebra of these operators, induced by the action of partial
derivatives on analytic functions, is given by
\begin{align}
  \left[ 
    a_p , a^{\dagger}_q 
  \right] = 
  {\delta}_{pq} . 
\label{eq:comm_relns}
\end{align}

The Doi construction of a Hilbert space of generating functions, with
inner product corresponding to the trace of the underlying density, is
a standard exercise~\cite{Smith:LDP_SEA:11}, for which we provide a
brief summary  in App.~\ref{sec:Doi_algebra_rev}.  The result is
that the analytic generating function $\Phi \! \left( z \right)$ is
replaced by a state vector $\left| \Phi \right)$, on which time
evolution under the Liouville
operator~(\ref{eq:Liouville_eq_multi_arg}) becomes
\begin{align}
  \frac{\partial}{\partial \tau} 
  \left| \Phi \right) 
& = 
  - \mathcal{L} \! 
  \left( a^{\dagger} , a \right)
  \left| \Phi \right) . 
\label{eq:Liouville_eq_aadag}
\end{align}

The evaluation of $\Phi \! \left( z \right)$ at $z = 1$, equivalent to
tracing over the underlying density $\rho$, is accomplished with a
projection operator that defines the inner product, given in
Eq.~(\ref{eq:Glauber_is_trace}).  Crucially, the expectations of
observables~(\ref{eq:EA_min_form}) are represented by operator
insertions in the inner product, establishing a correspondence between
distributions and observables on the index $\rn$, with states and
operators in the Doi Hilbert space.  One can invert the
mapping~(\ref{eq:a_adag_defs}) to restore the analytic structure of
the generating function using its complex-variable argument in a
modified version of the inner product, as
\begin{align}
  \left( 0 \right|
  e^{\sum_p z_p a_p}
  \left| \Phi \right) = 
  \Phi \! \left( z \right) . 
\label{eq:Glauber_to_FG}
\end{align}
This transformation will be useful in defining the functional-integral
representation of generating functions next.

\subsubsection*{The Peliti 2-field functional integral}

A functional-integral representation for the Hilbert space of
time-dependent generating functions and functionals was introduced by
Peliti~\cite{Peliti:PIBD:85,Peliti:AAZero:86}, building on the Doi
algebra.  It facilitates a variety of stationary-point approximations
(related to the ray methods of Freidlin and Wentzel for diffusion
equations~\cite{Freidlin:RPDS:98}), and is one of a larger class of
\textit{2-field functional-integral} (2FFI) methods including the
Schwinger-Keldysh time-loop
formalism~\cite{Schwinger:MBQO:61,Keldysh:noneq_diag:65,Kamenev:DP:02}
for quantum systems, and the Martin-Siggia-Rose formalism for
dissipative dynamical systems~\cite{Martin:MSR:73}.  

The Peliti method makes use of \textit{coherent states} as basis
elements for the expansion of arbitrary generating functions.  For a
vector $\phi \equiv \left[ {\phi}_p \right]$ of complex-valued
coefficients, the coherent state
\begin{align}
  \left| \phi \right) 
& \equiv 
  e^{
    \left( a^{\dagger} - 1 \right) \phi
  } 
  \left| 0 \right) 
\label{eq:coh_state_def}
\end{align}
is the generating function for a Poisson distribution with mean
$\phi$.  

In the Doi algebra, dual to each right-hand coherent state is a
\textit{projection operator}, which is a function of the
Hermitian conjugate vector ${\phi}^{\dagger}$.  Constructed from the
left-hand ground state $\left( 0 \right|$ defined in
Eq.~(\ref{eq:null_states}), it is given by
\begin{align}
  \left( \phi \right|
& \equiv 
  e^{
    \left( 1 - {\phi}^{\dagger} \right) 
    \phi 
  }
  \left( 0 \right|
  e^{
    {\phi}^{\dagger} a 
  } . 
\label{eq:dual_phi_states}
\end{align}

A representation of unity is obtained from the integral over coherent
states and their conjugate projection operators, as 
\begin{equation}
  \int 
  \frac{
    d^p \! {\phi}^{\dagger}
    d^p \! \phi
  }{
    {\pi}^p
  }
  \left| \phi \right) 
  \left( \phi \right| = 
  I . 
\label{eq:field_int_ident}
\end{equation}
When the representation~(\ref{eq:field_int_ident}) is inserted into an
expression to provide an expansion in the basis $\left(
  {\phi}^{\dagger} , \phi \right)$, these vectors become field
variables of integration.  The field $\phi$ is termed the
\textit{observable field}, because it corresponds to the mean of a
Poisson distribution, while ${\phi}^{\dagger}$ is termed the
\textit{response field}, for reasons that will become clear when we
study Greens functions in Sec.~\ref{sec:caus_anticaus}.

Evolution under the Liouville equation~(\ref{eq:Liouville_eq_aadag})
can formally be reduced to quadrature and converted into a functional
integral through repeated insertion of copies of the
representation~(\ref{eq:field_int_ident}) of unity at small increments
of time; more detail on this construction is provided in
App.~\ref{sec:Peliti_integral_rev}.  If the inner
product~(\ref{eq:Glauber_to_FG}) is used at late time to re-establish
a connection between the complex field variables, and the complex
surface arguments $z$ that can be used to probe $\Phi$, the integral
representation of the generating function for a distribution evolved
from time $\tau = 0$ to $\tau = T$ can be written 
\begin{equation}
  {\Phi}_T \! \left( z \right) = 
  \int_0^T 
  \mathcal{D} {\phi}^{\dagger}
  \mathcal{D} {\phi}
  e^{
    \left( z - {{\phi}^{\dagger}}_T \right) {\phi}_T 
  }
  e^{- S}
  {\Phi}_0 \! 
  \left( 
    {\phi}_0^{\dagger}
  \right) . 
\label{eq:PI_genform}
\end{equation}
In Eq.~(\ref{eq:PI_genform}) a new functional $S$ with the form of a
Lagrange-Hamilton \textit{action functional} appears, which is defined
in terms of the Liouville operator from
Eq.~(\ref{eq:Liouville_eq_aadag}) as
\begin{align}
  S 
& = 
  \int_0^T d\tau 
  \left\{ 
    - \left( d_{\tau} {\phi}^{\dagger} \right)
    \phi + 
    \mathcal{L} \! \left( {\phi}^{\dagger} , \phi \right)
  \right\} .
\label{eq:CRN_L_genform}
\end{align}

There are many reasons to adopt an extended-time, functional-integral
representation for a function such as ${\Phi}_T \! \left( z \right)$
evaluated at a single time, beyond simply extracting the moments of an
underlying evolved density ${\rho}_T$.  The integral offers a way to
insert operators at times $\tau < T$ to study the dependence of
late-time observables on regions in the distribution at earlier times,
giving a representation of the backward equation.  The density can
also be given incremental weights in continuous time, making
Eq.~(\ref{eq:PI_genform}) an extended-time generating
function\textit{al}.  We will use all these below to understand the
origin of reverse-time evolution and anti-causality in NEWRs and their
generalizations.

\subsection{Nonlinear rate laws and concurrency: forms of the
  generators for Chemical Reaction Networks}
\label{sec:2FFI_basics}

A very general class of discrete-state stochastic processes are those
for which each elementary event can remove a set of members from one
or more species, and then introduce another set.  The removal of all
members in the removed set happens concurrently, and if a state has
too few members of some species to populate the removed set, the event
cannot occur.  Processes in this class are of mathematical interest
because the condition of concurrency makes the graphical
representation of the process model a \emph{directed
multi-hypergraph}~\cite{Andersen:comp_rules:13,Andersen:generic_strat:14},
on which problems of constraint satisfaction are often computationally
complex~\cite{Andersen:NP_autocat:12}, and because the rate laws are
generally nonlinear.  They are of practical interest because they
include models of \textit{Chemical Reaction Networks}
(CRNs)~\cite{Horn:mass_action:72,Feinberg:notes:79,Feinberg:def_01:87},
though the class is rich enough to include a wide variety of other
population processes as well.  We will develop them here because
removal and addition, performed respectively by lowering and raising
operators, in the Doi-Peliti formalism stand in the relation of
projection operators and states.

We will adopt a set of concepts and terms from the CRN literature,
which are reviewed in App.~\ref{app:CRN_bkgnd}.  Elementary events are
termed \textit{reactions}, and the input and output sets in each
reaction are termed \textit{complexes}.  The representation of
reactions in the transition rate matrix is decomposed into three
components:
\begin{trivlist}

\item A \textbf{stoichiometric matrix} denoted $Y$ that gives the
  numbers of each species in any input or output complex;

\item an \textbf{adjacency/rate matrix} denoted ${\mathbb{A}}_k$ ($k$
  are the rate constants), which acts as the equivalent of a graph
  Laplacian between complexes; and

\item an \textbf{activity function} in two forms, denoted ${\Psi}_Y$
  and ${\psi}_Y$.  ${\Psi}_Y$, typically representing a sampling
  process, expresses the probability to form a complex in terms of the
  numbers of species $\rn$, and ${\psi}_Y$ is a corresponding function
  defined on shift operators or the arguments of the generating
  function.

\end{trivlist}

In terms of these quantities, for a simple CRN where complexes are
formed by sampling without replacement from the pools of species (the
rule underpinning the usual \textit{mass-action rate law}), the matrix
${\rm T}$ that generates the forward equation can be written
\begin{align}
  {\rm T} 
& = 
  {\psi}_Y^T \! \left( e^{-\partial / \partial \rn} \right) 
  {\mathbb{A}}_k
  \left[ 
    {\psi}_Y \! \left( e^{\partial / \partial \rn} \right) \cdot 
    {\Psi}_Y \! \left( \rn \right) 
  \right]
\nonumber \\
& = 
  \sum_{\left( i , j \right)}
  \left[
    {\psi}_Y^j \! \left( e^{-\partial / \partial \rn} \right) - 
    {\psi}_Y^i \! \left( e^{-\partial / \partial \rn} \right) 
  \right]
  k_{ji}
  {{\psi}_Y}_i \! \left( e^{\partial / \partial \rn} \right) 
  {{\Psi}_Y}_i \! \left( \rn \right) .
\label{eq:T_psi_from_A}
\end{align}
Here sub/superscripts $i$ and $j$ index complexes, the ordered pair
$\left( i , j \right)$ indexes a reaction from complex $i$ to complex
$j$ with associated rate constant $k_{ji}$, and ${\Psi}_{Yi}$ is the
count of ways in which the set in $i$ can be sampled from a state
$\rn$.  We indicate shift operators on functions indexed by $\rn$ with
$e^{\partial / \partial \rn}$; the activity functions ${\psi}_{Yi}$,
defined in Eq.~(\ref{eq:psi_of_shift_def}), are simply those that,
with argument $e^{\partial / \partial \rn}$, shift the indices of all
functions to their right upward in ${\rn}_p$ by the stoichiometric
coefficient $y^i_p$ of species $p$ in complex $i$.

The Liouville operator in the Doi representation, corresponding to the
generating matrix~(\ref{eq:T_psi_from_A}), takes the strikingly simple
form
\begin{align}
  - \mathcal{L} 
  \! \left( a^{\dagger} , a \right)
& = 
  {\psi}^T_Y 
  \! \left( a^{\dagger} \right)
  {\mathbb{A}}_k \, 
  {\psi}_Y
  \! \left( a \right) . 
\label{eq:L_psi_from_A}
\end{align}
We have developed several consequences of the symmetric
form~(\ref{eq:L_psi_from_A}) for the solution of moment hierarchies
in~\cite{Krishnamurthy:CRN_moments:17,Smith:CRN_moments:17}.  The fact
that such a formal symmetry exists between raising and lowering
operators -- or between states and projectors -- will give the adjoint
construction of the generator a comparably simple form whether or not
the dynamics comes from a Hamiltonian.

\subsubsection*{The ``form'' of transition matrices and their adjoints}

Eq.~(\ref{eq:T_psi_from_A}) shows the typical form of generators of
the forward equation: shift operators act on the distribution and also
on functions such as combinatorial factors that determine reaction
rates (here ${\Psi}_Y \! \left( \rn \right)$).  The
backward generator ${\rm T}^{\ast}$ may be formed by
shifting indices in the sum~(\ref{eq:ME_genform}); for this class of
processes, it amounts to changing $\partial / \partial \rn \rightarrow
- \partial / \partial \rn$, and having shift operators act to the
left, where they no longer transform ${\Psi}_Y \! \left( \rn \right)$.

The surprising result of the NEWRs and their generalizations is that
by adding a generating-functional weight to an evolving distribution,
the characteristic forms of ${\rm T}$ and ${\rm T}^{\ast}$ can be
interchanged.  Although the inner product~(\ref{eq:Glauber_inn_prod})
for stochastic processes is inherently asymmetric between states and
projection operators, the formal asymmetry in the Doi-Peliti
functional integral is (up to a choice of boundary conditions) all
contained within the kernel matrix ${\mathbb{A}}_k$ in the Liouville
operator~(\ref{eq:L_psi_from_A}).  A suitable generalization of the
transpose of this kernel can thus exchange the behavior of states and
operators.

We show below in Eq.~(\ref{eq:tildeA_constr_genform}) that dualization
of ${\mathbb{A}}_k$ is carried out by a rescaling similarity
transformation of the same form as the adjoint relation in
Eq.~(\ref{eq:adjoint_update}), but instead of using probabilities
${\rho}_{\rn}$ indexed on states, it uses complex-activities
${\psi}_Y$ related to average particle number.  The important feature
of the 3-factor symmetric product form~(\ref{eq:L_psi_from_A}) is to
encapsulate the network topology (a simple graph) within the factor
${\mathbb{A}}_k$, allowing the activities ${\psi}_Y$ and ${\psi}_Y^T$
to act symmetrically on observable and response fields.

\section{Duality between states and operators in the Doi-Peliti
  formalism}
\label{sec:duality_SPs}

We have shown how in the DP formalism distributions take the form of
states, and observables correspond to operators.  We show now how
states and operators are represented respectively by the observable
and response fields of the DP representation.  In the context of a
simple $2$-state model, for which many expectations can be obtained
exactly by stationary-point methods, we then demonstrate how the
similarity transform of Eq.~(\ref{eq:adjoint_update}) may be effected
by a simple change of variables, which has then the effect of
exchanging the response and observable fields.  

The change of variables also results in an extra term, which is a
functional weight added to ${\mathcal{L}}$.  This term corresponds to
what has been called the \textit{excess heat} in Langevin
approximation~\cite{Oono:NESS_thermo:98,Hatano:NESS_Langevin:01}
(though here no Gaussian limit of fluctuations is required), and
reduces to the \textit{total entropy production} in the integral
fluctuation theorem for systems with detailed
balance~\cite{Esposito:fluct_theorems:10,Seifert:stoch_thermo_rev:12}.
Though forward and reverse dynamics are always present in a generating
function, simply as a consequence of the duality between the forward
and backward actions of the generator, the adjoint construction of the
fluctuation theorems exchanges the forms of the forward and reverse
generators as mentioned earlier.  We show in the next section that the
forward and reverse dynamics derived here for stationary paths extends
to causality and anticausality of more general Green's functions.  We
also point out connections of this procedure to the \textbf{tilting}
transformation performed in \textit{Importance Sampling}, which
exchanges a weight function between a distribution and an observable
while preserving expectation values.

\subsection{States, operators, and duality: a single-time generating
  function in the DP representation}
\label{sec:single_time_GF}

We begin not with dynamics, but simply with the construction of a
generating function for a static distribution, in the DP
representation.  This will introduce the idea of a \textit{nominal
distribution}, for which the mean is reported by the expectations of
bilinear forms ${\phi}^{\dagger} \phi$ in the Peliti field integral,
and will show how it differs from the distribution reported by the
stationary point $\bar{\phi}$ of the observable field alone.  The
nominal distribution will correspond to the tilted density in the
generating function, while the state associated with $\bar{\phi}$ and
the operator associated with the stationary point
${\bar{\phi}}^{\dagger}$ of the response field can vary depending on
how the tilting is performed.  Ideas and terminology associated with
Importance Sampling are reviewed in App.~\ref{sec:IS_SP}.

Our example will be a class of binomial distributions for a 2-state
system, which can be written 
\begin{equation}
  \rho \! 
  \left( {\rm n}_a , {\rm n}_b \right) \equiv 
  {\nu}_a^{{\rm n}_a}
  {\nu}_b^{{\rm n}_b}
  \frac{
    N!
  }{
    {\rm n}_a! 
    {\rm n}_b! 
  } .
\label{eq:binom_genform}
\end{equation}
${\rn}_a$ and ${\rn}_b$ range over non-negative values with $N \equiv
{\rn}_a + {\rn}_b$ fixed, and we take ${\nu}_a + {\nu}_b = 1$.
${\nu}_a$ and ${\nu}_b$ are respectively the mean values of ${\rn}_a
/ N$ and ${\rn}_b / N$ under $\rho$.

The one degree of freedom in such distributions is a variable we
denote $x$, with 
\begin{align}
  {\nu}_a 
& = 
  \frac{1}{2}
  \left( 1 - x \right) ; 
&
  {\nu}_b 
& = 
  \frac{1}{2}
  \left( 1 + x \right) . 
\label{eq:bar_x_to_nus}
\end{align}
Two other quantities $\mu$ and $\xi$ that, in the case where $\rho$ is
an equilibrium Gibbs distribution, have the interpretations
respectively of a chemical potential and a free energy, are related to
$x$ as
\begin{align}
  \frac{
    {\nu}_a 
  }{
    {\nu}_b 
  } 
& \equiv 
  e^{\beta \mu} 
& 
  x 
& =   
  - {\rm th} 
  \frac{\beta \mu}{2} 
& 
  \beta \xi 
& \equiv 
  \log {\rm ch}
  \frac{\beta \mu}{2} , 
\label{eq:bar_mu_conversions}
\end{align}
where $\beta$ is inverse temperature. 

We consider a starting distribution ${\rho}_0$ with parameters
${{\nu}_a}_0$ and ${{\nu}_b}_0$, and its generating function,
constructed as in Eq.~(\ref{eq:gen_fn_multi_arg}),
\begin{align}
  {\Phi}_0 \! 
  \left( z_a , z_b \right) 
& \equiv 
  \sum_{{\rm n}_a = 0}^N
  z_a^{{\rm n}_a}
  z_b^{{\rm n}_b}
  {\rho}_0 \! 
  \left( {\rm n}_a , {\rm n}_b \right) 
\nonumber \\ 
& = 
  {
    \left[ 
      {{\nu}_a}_0 z_a + 
      {{\nu}_b}_0 z_b 
    \right]
  }^N . 
\label{eq:psi_twoarg_eval}
\end{align} 
We will look only at contours for $z_a$ and $z_b$ which leave
${\Phi}_0$ normalized; these can be written 
\begin{align}
  z_a 
& = 
  \frac{
    1 - x_T
  }{
    1 - x_0
  } \equiv 
  \frac{{{\nu}_a}_T}{{{\nu}_a}_0} ; & 
  z_b 
& = 
  \frac{
    1 + x_T
  }{
    1 + x_0
  } \equiv 
  \frac{{{\nu}_b}_T}{{{\nu}_b}_0} , 
\label{eq:z_shift_vals_assn}
\end{align}
for some parameter $x_T$.  The density $ z_a^{{\rm n}_a} z_b^{{\rm
n}_b} {\rho}_0 \!  \left( {\rm n}_a , {\rm n}_b
\right)$ determines the mean values of ${\rm n}_a$ and ${\rm n}_b$
under ${\Phi}_0$, and will serve as the nominal distribution for the
rest of the discussion of this static generating function.

\subsubsection*{The connection of incremental tilting with concepts
  from Importance Sampling}

The generating function can be formed from ${\rho}_0$ by one discrete
tilt as in Eq.~(\ref{eq:psi_twoarg_eval}), or the tilt can be
accumulated incrementally along a contour.  Introduce an interval
$\left[ 0 , T \right]$, an increment $\delta \tau$, and a sequence of
values $\tau = k \delta \tau$ for $k \in 0 , \ldots , T / \delta
\tau$.  Then introduce a sequence of values ${\underline{x}}_{\tau}$
with ${\underline{x}}_0 = x_0$ and ${\underline{x}}_T = x_T$, which we
will take to converge to a smooth function (except possibly in the
final step) as $\delta \tau \rightarrow 0$.  Then $z_a$ and $z_b$ can
be factored as
\begin{align}
  z_a 
& = 
  \prod_{\tau = d\tau}^T
  \left( 
    \frac{
      {{\underline{\nu}}_a}_{\tau}
    }{
      {{\underline{\nu}}_a}_{\tau - d\tau}
    } 
  \right) ; & 
  z_b 
& = 
  \prod_{\tau = d\tau}^T
  \left( 
    \frac{
      {{\underline{\nu}}_b}_{\tau}
    }{
      {{\underline{\nu}}_b}_{\tau - d\tau}
    } 
  \right) .
\label{eq:z_factorization}
\end{align}

Denote the last term in the product ~Eq.(\ref{eq:z_factorization}) 
\begin{align}
  z_{aT} 
& = 
  \left( 
    \frac{
      {{\underline{\nu}}_a}_T
    }{
      {{\underline{\nu}}_a}_{T - \delta \tau}
    } 
  \right) ; & 
  z_{bT} 
& = 
  \left( 
    \frac{
      {{\underline{\nu}}_b}_T
    }{
      {{\underline{\nu}}_b}_{T - \delta \tau}
    } 
  \right) .
\label{eq:z_surface_terms}
\end{align}
We will use these as boundary terms in the
conversion~(\ref{eq:Glauber_to_FG}) from the Doi algebra back to
analytic functions.  We will consider the two cases where $z_{aT}$,
$z_{bT} = 1 + \mathcal{O} \! \left( \delta \tau \right)$ so that
${{\underline{\nu}}_a}_{T - \delta \tau} \rightarrow
{{\underline{\nu}}_a}_T$ and ${{\underline{\nu}}_b}_{T - \delta \tau}
\rightarrow {{\underline{\nu}}_b}_T$, or where $z_{aT}$, $z_{bT} \sim
\mathcal{O} \! \left( {\delta \tau}^0 \right)$ to impose a finite
shift at the final value $\tau = T$.

The generating function~(\ref{eq:psi_twoarg_eval}), if tilted
incrementally using the factorization~(\ref{eq:z_factorization}), can
be decomposed at any intermediate value $\tau$ into factors 
\begin{align}
  {\Phi}_0 \! 
  \left( 
    z_a , z_b 
  \right) 
& = 
  \sum_{{\rm n}_a = 0}^N
  \prod_{\tau = d\tau}^T
    {
      \left( 
        \frac{
          {{\underline{\nu}}_a}_{\tau}
        }{
          {{\underline{\nu}}_a}_{\tau - d\tau}
        } 
      \right) 
    }^{{\rm n}_a}
    {
      \left( 
        \frac{
          {{\underline{\nu}}_b}_{\tau}
        }{
          {{\underline{\nu}}_b}_{\tau - d\tau}
        } 
      \right) 
    }^{{\rm n}_b}
  {\rho}_0 \! 
  \left( {\rm n}_a , {\rm n}_b \right) 
% \label{eq:gen_fn_prod_expand} \\
\nonumber \\ 
& =   \sum_{{\rm n}_a = 0}^N
  \prod_{{\tau}^{\prime} = \tau + d\tau}^T 
    {
      \left( 
        \frac{
          {{\underline{\nu}}_a}_{{\tau}^{\prime}}
        }{
          {{\underline{\nu}}_a}_{{\tau}^{\prime} - d\tau}
        } 
      \right) 
    }^{{\rm n}_a}
    {
      \left( 
        \frac{
          {{\underline{\nu}}_b}_{{\tau}^{\prime}}
        }{
          {{\underline{\nu}}_b}_{{\tau}^{\prime} - d\tau}
        } 
      \right) 
    }^{{\rm n}_b}
  {\rho}_{\tau} \! 
  \left( {\rm n}_a , {\rm n}_b \right) , 
\label{eq:psi_prod_part_evolve}
\end{align}
where we have defined 
\begin{equation}
  {\rho}_{\tau} \! 
  \left( {\rm n}_a , {\rm n}_b \right) \equiv 
  \prod_{{\tau}^{\prime} = d\tau}^{\tau}
    {
      \left( 
        \frac{
          {{\underline{\nu}}_a}_{{\tau}^{\prime}}
        }{
          {{\underline{\nu}}_a}_{{\tau}^{\prime} - d\tau}
        } 
      \right) 
    }^{{\rm n}_a}
    {
      \left( 
        \frac{
          {{\underline{\nu}}_b}_{{\tau}^{\prime}}
        }{
          {{\underline{\nu}}_b}_{{\tau}^{\prime} - d\tau}
        } 
      \right) 
    }^{{\rm n}_b}
  {\rho}_0 \! 
  \left( {\rm n}_a , {\rm n}_b \right) .
\label{eq:gen_fn_intermed_rho}
\end{equation}
Referring to the review of Importance Sampling in
App.~\ref{sec:IS_SP}, each of the distributions ${\rho}_{\tau}$
behaves as an importance distribution, and the residual factor in the
second line of Eq.~(\ref{eq:psi_prod_part_evolve}) behaves as its
conjugate likelihood ratio, with respect to the nominal distribution
in Eq.~(\ref{eq:psi_twoarg_eval}).  We will show next how these
factors are carried by observable and response fields in the
functional integral.  

In the limit $\delta \tau \rightarrow 0$, the intermediate density
${\rho}_{\tau}$ evolves along the parameter $\tau$ under the equation
(making the state-index $\rn$ explicit and suppressing $\tau$ from the
notation)
\begin{align}
  \frac{d {\rho}_{\rn}}{d \tau} 
& = 
  \frac{d\underline{x}}{d\tau} 
  \frac{
    {\rm n}_b 
    \left( 1 - \underline{x} \right) - 
    {\rm n}_a 
    \left( 1 + \underline{x} \right) 
  }{
    \left( 1 - {\underline{x}}^2 \right)
  }
  {\rho}_{\rn} 
\nonumber \\
& = 
  - \frac{
    \beta \dot{\underline{\mu}}
  }{
    2 
  }
  \left[
    \left( {\rn}_b - {\rn}_a \right) - N \underline{x} 
  \right]
  {\rho}_{\rn} ,
\label{eq:part_rho_tau_form}
\end{align}
in which $\underline{x}$ is the function with values
${\underline{x}}_{\tau}$ introduced at the beginning of the section to
define Eq.~(\ref{eq:z_factorization}).  (Here and below we use overdot
$\dot{\mbox{ }}$ as a shorthand for $d/d\tau$.)  We have introduced
the ``physical'' variables~(\ref{eq:bar_mu_conversions}) to make
contact with the Hamiltonian construction of
Crooks~\cite{Crooks:NE_work_relns:99}.  $\left[ \mu \left( {\rn}_b -
{\rn}_a \right) / 2 - N \xi \right]$ is the Hamiltonian that will
produce the binomial distribution~(\ref{eq:binom_genform}) at
parameter $x$ as a Gibbs equilibrium, net of the instantaneous Gibbs
free energy, and its $\tau$-derivative -- a path ``work'' -- acts as
the generator of $\tau$-translation for $\rho$.

\subsection{A functional integral representation for static generating
  functions}
\label{sec:static_GF_FI}

The parameter $\tau$ provides a coordinate along which a Doi-Peliti
functional integral representation for ${\Phi}_0$ can be built.
Depending on the contour ${\underline{x}}_{\tau}$ assumed, the
functional integral can represent either the one-shot tilt of
Eq.~(\ref{eq:psi_twoarg_eval}), or a smooth incremental accumulation
as in Eq.~(\ref{eq:psi_prod_part_evolve}).  Following the steps
outlined in Sec.~\ref{sec:text_DP_intro} and
App.~\ref{app:DP_details}, the 2FFI representation is given by
\begin{widetext}
\begin{equation}
  {\Phi}_0 \! \left( z_a , z_b \right) = 
  \int 
  \mathcal{D} {\phi}_a^{\dagger}
  \mathcal{D} {\phi}_a
  \mathcal{D} {\phi}_b^{\dagger}
  \mathcal{D} {\phi}_b
  e^{
    \left( z_{aT} - {{\phi}_a^{\dagger}}_T \right) {{\phi}_a}_T + 
    \left( z_{bT} - {{\phi}_b^{\dagger}}_T \right) {{\phi}_b}_T
  }
  e^{- S}
  {\Phi}_0 \! 
  \left( 
    {{\phi}_a^{\dagger}}_0 , 
    {{\phi}_b^{\dagger}}_0
  \right) . 
\label{eq:PI_twospecform}
\end{equation}
\end{widetext}
The action $S$ depends functionally on $\underline{x}$ over the range
$0 \le \tau \le T - \delta \tau$, with the last factors $z_{aT}$,
$z_{bT}$ from Eq.~(\ref{eq:z_surface_terms}) appearing in the boundary
terms.  We now compare two cases.

\subsubsection{Identity map, followed by a discrete tilt}

First consider the case $z_{aT} = z_a$, $z_{bT} = z_b$,
${\underline{x}}_{\tau} = {\underline{x}}_0$ which we set equal to
$x_0$ of the starting distribution.  Because $\dot{\underline{\mu}}
\equiv 0$, the action in Eq.~(\ref{eq:PI_twospecform}) is given by
\begin{equation}
  S_{\rm null} = 
  \int d\tau 
  \left\{
    - \left( d_{\tau} {\phi}_a^{\dagger} \right)
    {\phi}_a - 
    \left( d_{\tau} {\phi}_b^{\dagger} \right)
    {\phi}_b 
  \right\} .
\label{eq:S_null_form}
\end{equation}
A functional integral with $S_{\rm null}$ propagates ${\rho}_0$
through a sequence of identity maps~(\ref{eq:field_int_ident}), and
the boundary terms apply the discrete tilt of
Eq.~(\ref{eq:psi_twoarg_eval}) at $\tau = T$.

App.~\ref{app:null_SP} derives the stationary-point solutions for the
observable and response fields.  These, and the number field given in
the stationary-point approximation by ${\bar{n}}_a \approx
{\bar{\phi}}^{\dagger}_a {\bar{\phi}}_a$, ${\bar{n}}_b \approx
{\bar{\phi}}^{\dagger}_b {\bar{\phi}}_b$, take values 
\begin{align}
  {\bar{\phi}}_a
& = 
  \frac{N}{2}
  \left( 1 - {x}_0 \right) & 
  {\bar{\phi}}_b
& = 
  \frac{N}{2}
  \left( 1 + {x}_0 \right) , 
\nonumber \\
  {\bar{\phi}}^{\dagger}_a
& = 
  z_a & 
  {\bar{\phi}}^{\dagger}_b
& = 
  z_b , 
\nonumber \\
  {\bar{n}}_a
& = 
  \frac{N}{2}
  z_a 
  \left( 1 - {x}_0 \right) & 
  {\bar{n}}_b
& = 
  \frac{N}{2}
  z_b 
  \left( 1 + {x}_0 \right) 
\nonumber \\ 
& = 
  \frac{N}{2}
  \left( 1 - {x}_T \right) & 
& = 
  \frac{N}{2}
  \left( 1 + {x}_T \right) . 
\label{eq:SPs_null_evals}
\end{align}
The stationary coherent-state distribution at $\bar{\phi}$ remains the
input distribution ${\rho}_0$ at all $\tau < T$; the coherent-state
projection operator at parameter ${\bar{\phi}}^{\dagger}$ multiplies
each state $\left| \rn \right)$ by the weight $z_a^{{\rn}_a}
z_b^{{\rn}_b}$ of the generating function~(\ref{eq:psi_twoarg_eval}).
The numbers $\bar{n}$ reflect the expectations in the nominal
distribution $z_a^{{\rn}_a} z_b^{{\rn}_b} {\rho}_0 \! \left( {\rn}_a,
  {\rn}_b \right)$.

\subsubsection{Static generating function, accumulated continuously} 

Next consider the complementary case where 
$z_{aT}$, 
$z_{bT} = 1 + \mathcal{O} \! \left( \delta \tau \right)$, 
and 
${\underline{x}}_{\tau}$ interpolates smoothly between 
${\underline{x}}_0 = {x}_0$ of the initial distribution, and
${\underline{x}}_{\tau} \rightarrow {x}_T$ set by $z_a$,
$z_b$ as $\tau \rightarrow T$. 
The action in Eq.~(\ref{eq:PI_twospecform}) for this case is given by 
\begin{widetext}
\begin{equation}
  S_{\rm tilt} = 
  \int d\tau 
  \left\{
    - \left( d_{\tau} {\phi}_a^{\dagger} \right)
    {\phi}_a - 
    \left( d_{\tau} {\phi}_b^{\dagger} \right)
    {\phi}_b + 
    \frac{
      \beta \dot{\underline{\mu}}
    }{
      2 
    }
    \left[
      \left( 1 - \underline{x} \right)
      {\phi}_b^{\dagger} {\phi}_b - 
      \left( 1 + \underline{x} \right) 
      {\phi}_a^{\dagger} {\phi}_a 
    \right]
  \right\} , 
\label{eq:S_triv_form}
\end{equation}
\end{widetext}
where the term in $\dot{\underline{\mu}}$ implements the
$\tau$-evolution of Eq.~(\ref{eq:part_rho_tau_form}).  The existence
of $\tau$-derivatives in the ``kinetic'' term of $S_{\rm tilt}$, which
merely reflects the overlap $\left( {\phi}_{\tau - \delta \tau} \mid
  {\phi}_{\tau} \right)$ between adjacent representations of
unity~(\ref{eq:field_int_ident}), suggests a way to remove the
time-derivative term by a change of variable, which is the 2FFI
expression of the duality transform of the NEWRs and their
generalizations.

The change of variables that expresses duality in Doi-Peliti
functional integrals is one already recognized by
Baish~\cite{Baish:DP_duality:15}.  From the original field variables
$\left( {\phi}^{\dagger}, \phi \right)$, introduce two new variables
$\left( {\varphi}^{\dagger}, \varphi \right)$ defined by
\begin{align}
  {\phi}_a^{\dagger} 
& \equiv 
  \frac{
    {\varphi}_a^{\dagger} 
  }{
    1 - \underline{x} 
  }
& 
  {\phi}_a 
& \equiv 
  \left( 1 - \underline{x} \right)
  {\varphi}_a
\nonumber \\
  {\phi}_b^{\dagger} 
& \equiv 
  \frac{
    {\varphi}_b^{\dagger} 
  }{
    1 + \underline{x} 
  }
& 
  {\phi}_b 
& \equiv 
  \left( 1 + \underline{x} \right)
  {\varphi}_b . 
\label{eq:varphis_ulines}
\end{align}
In the dual variables the action~(\ref{eq:S_triv_form}) becomes 
\begin{equation}
  S = 
  \int d\tau 
  \left\{
    - \left( d_{\tau} {\varphi}_a^{\dagger} \right)
    {\varphi}_a - 
    \left( d_{\tau} {\varphi}_b^{\dagger} \right)
    {\varphi}_b 
  \right\} . 
\label{eq:S_triv_var}
\end{equation}
We recover the form~(\ref{eq:S_null_form}) of the null action, in a
basis which is tilted to absorb the generating-functional weight in
each interval $\delta \tau$.  

Stationary-point solutions in the original and dual variables are
derived in App.~\ref{app:tilt_only_SP}.  ${\bar{\varphi}}^{\dagger}$
and $\bar{\varphi}$ are constant as in the last example, though at
different values because the change of variables alters their boundary
conditions.  The stationary-point solutions in the original fields are
now non-trivial functions of $\tau$:
\begin{align}
  {
    \left( 
      {\bar{\phi}}_a
    \right) 
  }_{\tau}
& = 
  \frac{N}{2}
  \left( 1 - {\underline{x}}_{\tau} \right) & 
  {
    \left( 
      {\bar{\phi}}_b
    \right) 
  }_{\tau}
& = 
  \frac{N}{2}
  \left( 1 + {\underline{x}}_{\tau} \right) , 
\nonumber \\
  {
    \left( 
      {\bar{\phi}}^{\dagger}_a
    \right) 
  }_{\tau}
& = 
  \frac{
    1 - {x}_T
  }{
    1 - {\underline{x}}_{\tau}
  } & 
  {
    \left( 
      {\bar{\phi}}^{\dagger}_b
    \right) 
  }_{\tau}
& = 
  \frac{
    1 + {x}_T
  }{
    1 + {\underline{x}}_{\tau}
  } , 
\nonumber \\
  {\bar{n}}_a
& \equiv  
  \frac{N}{2}
  \left( 1 - {x}_T \right) & 
  {\bar{n}}_b
& \equiv  
  \frac{N}{2}
  \left( 1 + {x}_T \right) . 
\label{eq:SPs_tilt_evals}
\end{align}
Note that $\bar{n}$ continues to report the mean in the nominal
distribution $z_a^{{\rn}_a} z_b^{{\rn}_b} {\rho}_0 \! \left( {\rn}_a,
  {\rn}_b \right)$, while now $\bar{\phi}$ is the mean in
${\rho}_{\tau}$ from Eq.~(\ref{eq:gen_fn_intermed_rho}), and
${\bar{\phi}}^{\dagger}$ produces the conjugate $\tau$-dependent
likelihood ratio in Eq.~(\ref{eq:psi_prod_part_evolve}).

\subsection{Generating functions and functionals of dynamically
  evolving distributions}
\label{sec:GF_dyn_2cases}

The main result from Sec.~\ref{sec:single_time_GF} and
Sec.~\ref{sec:static_GF_FI} is that, for the generating function of a
static distribution, the nominal distribution is determined, though it
can be factored into importance distributions and likelihood ratios in
a continuum of ways\footnote{See~\cite{Espigares:inf_2nd_law_ineq:12}
for an application of this freedom to replace pathological cases of
Hatano-Sasa duality, where for instance the steady-state distribution
may be everywhere non-smooth, with reference distributions that have
the same support in the state space as the dynamical distribution.}.
The expression of that freedom in the functional integral is important
for understanding the meaning and roles of the observable and response
fields in the Doi-Peliti construction.  In this section we will show
that for the generating function or functional of an evolving state,
the nominal distribution becomes dynamical.  The new feature is that
the nominal distribution itself, as well as the way it is factored
into importance distributions and likelihood ratios, now depends on
the way weights are accumulated along paths.

We will show that, in the static generating function of an evolved
distribution, the nominal distribution is a simple sum of the sample
distribution propagated with retarded dynamics, and the
generating-function weight propagated (in a suitable measure) with
advanced dynamics.  However, if an incremental tilting protocol is
matched to the dynamics, the evolution of the sample distribution can
be made memoryless, while the nominal distribution evolves entirely
with advanced dynamics.  The required matching protocol defines the
construction of the adjoint process, and it has the effect of
transposing the forms of forward and backward generators.

To keep the example as simple as possible, we introduce the minimal
non-trivial dynamics for the 2-state system, which preserves the
binomial form~(\ref{eq:binom_genform}) of distributions under
arbitrary time-dependent rate parameters.  A minimal transition rate
matrix~(\ref{eq:T_psi_from_A}) is given by
\begin{equation}
  {\rm T} = 
    {\underline{\nu}}_b 
    \left(
      e^{
        \partial / \partial {\rm n}_a - 
        \partial / \partial {\rm n}_b
      } - 1  
    \right)
    {\rm n}_a + 
    {\underline{\nu}}_a 
    \left(
      e^{
        \partial / \partial {\rm n}_b - 
        \partial / \partial {\rm n}_a
      } - 1  
    \right)
    {\rm n}_b , 
\label{eq:T_two_state_genform}
\end{equation}
describing single-particle hops with per-particle rates
${\underline{\nu}}_b$ (for $A \rightharpoonup B$) and
${\underline{\nu}}_a$ and (for $B \rightharpoonup A$).  We will take
${\underline{\nu}}_a$ and ${\underline{\nu}}_b$ to define the contour
${\underline{x}}_{\tau}$, where $\tau$ is now a time coordinate and
not simply an arbitrary parameter.  The 2-state system of course
possesses detailed balance, making this a standard
NEWR~\cite{Crooks:NE_work_relns:99}.  However, the tilting protocols,
2FFI variable changes, and causality arguments of this and the next
section also go through more generally, in the same form except that a
non-equilibrium steady state must be computed.

\subsubsection{Generating function of an evolved distribution, formed
discretely at the end of evolution}

We first consider a single-time generating function, like
Eq.~(\ref{eq:psi_twoarg_eval}), but applied to a distribution evolved
to time $\tau = T$ under the master
equation~(\ref{eq:T_two_state_genform}).  This illustrates a case
where the observable field $\bar{\phi}$ corresponds to the expected,
time-dependent coherent state, but the nominal distribution differs
from this state because it also reflects the existence of the
late-time tilt in the generating function.  Write this generating
function from Eq.~(\ref{eq:PI_twospecform}) now as ${\Phi}_T \! \left(
  z_{aT} , z_{bT} \right)$.  The arguments $\left( z_{aT} , z_{bT}
\right)$ will in general be different from $\left( z_a , z_b \right)$
of Eq.~(\ref{eq:z_shift_vals_assn}), because they will shift the
distribution from a time-dependent contour $\bar{x}$ different from
the initial value $x_0$. 

The action in the integral~(\ref{eq:PI_genform}) for
this case is
\begin{widetext}
\begin{equation}
  S_{\rm dyn} = 
  \int d\tau 
  \left\{
    - \left( d_{\tau} {\phi}_a^{\dagger} \right)
    {\phi}_a - 
    \left( d_{\tau} {\phi}_b^{\dagger} \right)
    {\phi}_b + 
    \frac{1}{2}
    \left( 
      {\phi}_b^{\dagger} - 
      {\phi}_a^{\dagger}
    \right) 
    \left[
      \left( 1 - \underline{x} \right) {\phi}_b - 
      \left( 1 + \underline{x} \right) {\phi}_a
    \right] 
  \right\} . 
\label{eq:S_plain_form}
\end{equation}
\end{widetext}
The terms including a factor $\underline{x}$ come from the Liouville
operator corresponding to ${\rm T}$ in
Eq.~(\ref{eq:T_two_state_genform}). 

Stationary-point solutions under the action $S_{\rm dyn}$ are given in
App.~\ref{app:untilted_SP}.  We introduce a new function
${\bar{x}}_{\tau}$ with initial value $x_0$, and satisfying
\begin{equation}
  \frac{d \bar{x}}{d\tau} = 
  - \left( \bar{x} - \underline{x} \right) .
\label{eq:binoms_MFT_evol}
\end{equation}
In terms of its solution, 
\begin{align}
  {
    \left( 
      {\bar{\phi}}_a
    \right)
  }_{\tau}
& = 
  \frac{N}{2}
  \left( 1 - {\bar{x}}_{\tau} \right) & 
  {
    \left( 
      {\bar{\phi}}_b
    \right)
  }_{\tau}
& = 
  \frac{N}{2}
  \left( 1 + {\bar{x}}_{\tau} \right) .
\label{eq:SPs_dyn_evals}
\end{align}
Comparing this to the first line of Eq.~(\ref{eq:SPs_tilt_evals}),
$\underline{x}$ for $S_{\rm tilt}$ has been replaced with the retarded
solution $\bar{x}$ for $S_{\rm dyn}$.

The stationary-path solutions for the response fields are more
complicated, and are given in Eq.~(\ref{eq:plain_phi_dag_sol}).  The
number fields, however, can be shown to satisfy
\begin{equation}
  {
    \left. 
      \frac{
        \left( 
          {\bar{n}}_b - {\bar{n}}_a
        \right) - 
        N \bar{x} 
      }{
        1 - {\bar{x}}^2
      }
    \right|
  }_{\tau} = 
  e^{- \left( T - \tau \right)}
  {
    \left. 
      \frac{
        \left( 
          {\bar{n}}_b - {\bar{n}}_a
        \right) - 
        N \bar{x} 
      }{
        1 - {\bar{x}}^2
      }
    \right|
  }_T . 
\label{eq:plain_genfun_num_sol}
\end{equation}
If $z_{aT} = z_{bT} = 1$ in the generating function ${\Phi}_T$,
$\bar{n}$ coincides with the coherent-state mean $\bar{\phi}$ from
Eq.~(\ref{eq:SPs_dyn_evals}), as in the usual
expositions~\cite{Mattis:RDQFT:98,Cardy:FTNEqSM:99}.  If $z_{bT} -
z_{aT} \neq 0$ in the generating function, the left-hand side of
Eq.~(\ref{eq:plain_genfun_num_sol}) is nonzero, and the deviation of
$\left( {\bar{n}}_b - {\bar{n}}_a \right)$ from $N \bar{x}$ decays
exponentially \emph{backward} in time, with a measure $\left( 1 -
  {\bar{x}}^2 \right)$ which gives the instantaneous variance in the
coherent state at $\bar{\phi}$.

\subsubsection{Cumulative generating functional matched to the time
evolution} 
\label{sec:dual_GF_SPsol}

Finally we consider the interaction between dynamics and incremental
tilting, so that we form a time-dependent generating functional and
not simply a single-time generating function.  Return to the argument
decomposition $z_{aT} = 1 + \mathcal{O} \! \left( \delta \tau
\right)$, $z_{bT} = 1 + \mathcal{O} \! \left( \delta \tau \right)$,
and replace the generating function from
Eq.~(\ref{eq:psi_twoarg_eval}) with one written
${\Phi}_{\underline{\hat{x}}} \! \left( z_{aT} , z_{bT} \right)$, to
indicate that the tilting protocol will depend functionally on a
contour $\underline{\hat{x}}$ that we have yet to specify.

To identify the appropriate contour for incrementally tilting an
evolving distribution, note that for a binomial density advancing
along a parameter $\tau$ under Eq.~(\ref{eq:part_rho_tau_form}), one
can freely add a factor of the instantaneous generator ${\rm T}$ with
the rate constants depending on ${\underline{x}}_{\tau}$ without
altering the result, because the tilted distribution is
instantaneously annihilated by ${\rm T}$:
\begin{align}
  \frac{d \rho}{d \tau} 
& = 
  \left\{ 
    {\rm T} - 
    \beta \dot{\underline{\mu}}
    \left(
      {\rm n} - \frac{N}{2} \underline{x} 
    \right)
  \right\} 
  \rho . 
\label{eq:part_rho_tau_form_wL}
\end{align}
The action in Eq.~(\ref{eq:PI_twospecform}), with
$\underline{\hat{x}}$ set equal to $\underline{x}$ as suggested by
Eq.~(\ref{eq:part_rho_tau_form_wL}), becomes 
\begin{widetext}
\begin{equation}
  S_{\rm Crooks} = 
  \int d\tau 
  \left\{
    - \left( d_{\tau} {\phi}_a^{\dagger} \right)
    {\phi}_a - 
    \left( d_{\tau} {\phi}_b^{\dagger} \right)
    {\phi}_b + 
    \frac{1}{2}
    \left( 
      {\phi}_b^{\dagger} - 
      {\phi}_a^{\dagger}
    \right) 
    \left[
      \left( 1 - \underline{x} \right) {\phi}_b - 
      \left( 1 + \underline{x} \right) {\phi}_a
    \right] + 
    \frac{
      \beta \dot{\underline{\mu}}
    }{
      2 
    }
    \left[
      \left( 1 - \underline{x} \right)
      {\phi}_b^{\dagger} {\phi}_b - 
      \left( 1 + \underline{x} \right) 
      {\phi}_a^{\dagger} {\phi}_a 
    \right]
  \right\} .
\label{eq:S_Crooks_form}
\end{equation}
Although the matrix ${\rm T}$ is invisible in
Eq.~(\ref{eq:part_rho_tau_form_wL}) when acting on the instantaneous
equilibrium distribution, the Liouville term remains in the action,
and governs the evolution of more general distributions and
correlation functions.  We label this action $S_{\rm Crooks}$ because,
as we noted following Eq.~(\ref{eq:part_rho_tau_form}),
$\dot{\underline{\mu}} \left[ \left( 1 - \underline{x} \right)
{\phi}_b^{\dagger} {\phi}_b - \left( 1 + \underline{x} \right)
{\phi}_a^{\dagger} {\phi}_a \right] / 2$ is the time derivative of the
difference between a Hamiltonian and local free energy, which is the
normalization convention of~\cite{Crooks:NE_work_relns:99}.

To solve this case, we observe that the
transformation~(\ref{eq:varphis_ulines}) removes the explicit
time-derivative term from $S_{\rm Crooks}$ as it does for $S_{\rm
  tilt}$ (it is the same term), giving the form
\begin{equation}
  S_{\rm Crooks} = 
  \int d\tau 
  \left\{
    - \left( d_{\tau} {\varphi}_a^{\dagger} \right)
    {\varphi}_a - 
    \left( d_{\tau} {\varphi}_b^{\dagger} \right)
    {\varphi}_b + 
    \frac{1}{2}
    \left[
      \left( 1 - \underline{x} \right) 
      {\varphi}_b^{\dagger} - 
      \left( 1 + \underline{x} \right) 
      {\varphi}_a^{\dagger}
    \right] 
    \left(
      {\varphi}_b - 
      {\varphi}_a
    \right) 
  \right\} .
\label{eq:S_Crooks_var}
\end{equation}
\end{widetext}
The new observation is what the duality transform does to the term
originating from the Liouville operator: it interchanges the roles of
$\phi$ and ${\phi}^{\dagger}$ with those of ${\varphi}^{\dagger}$ and
$\varphi$, respectively.  If the Liouville term is written as a
bilinear form (first introduced by
Keldysh~\cite{Keldysh:noneq_diag:65,Kamenev:DP:02}), duality has the
effect of transposing the matrix kernel of this form.  We show in
Sec.~\ref{sec:duality_gen_CRN} how this result generalizes for
non-linear and non-Hamiltonian systems.

Stationary-point solutions for $S_{\rm Crooks}$ in the original and
dual variables are derived in App.~\ref{app:Crooks_tilted_SP}.  As for
the static generating functional with continuous tilting, and as
suggested by Eq.~(\ref{eq:part_rho_tau_form_wL}), fields $\bar{\phi}$
again take the values in the first line of
Eq.~(\ref{eq:SPs_tilt_evals}).

To describe the $\bar{n}$ fields and the behavior of the nominal
distribution, we introduce a new mean field contour $\bar{x}$ with
${\bar{x}}_T = x_T$, evolving under the \emph{advanced} dynamics
\begin{equation}
  \frac{d \bar{x}}{d\tau} = 
  \left( \bar{x} - \underline{x} \right) ; 
\label{eq:binoms_MFT_adv_evol}
\end{equation}
this is contrasted with retarded dynamics under the unweighted
stochastic process shown in Eq.~(\ref{eq:binoms_MFT_evol}).  In terms
of the solution $\bar{x}$ to Eq.~(\ref{eq:binoms_MFT_adv_evol}),
${\bar{n}}_a + {\bar{n}}_b = N$ at all times, and
\begin{equation}
  {
    \left( 
      {\bar{n}}_b - {\bar{n}}_a 
    \right)
  }_{\tau} = 
  N {\bar{x}}_{\tau} . 
\label{eq:SP_n_crooks}
\end{equation}
Thus the nominal distribution evolves under time-reversed dynamics,
but the exact form of the solution for $\tau < T$ depends on the fact
that the generating functional has been continued to a time $\tau = T$
in the future.  The shift from retarded to advanced dynamics reflects
the reverse-time evolution that the backward equation produces for any
operator, as applied to the particular case where the operator is a
path weight chosen to erase memory from the instantaneous distribution
given by the coherent state at $\bar{\phi}$.

The surprising result is that this path-weighting protocol does not
only time-reverse the dynamics of binomial distributions, but
transposes the whole kernel in the Liouville operator.  In the usual
DP theory for unweighted distributions, the stationary values of
response fields are fixed points of the backward equation.  Under the
transform~(\ref{eq:varphis_ulines}), the stationary value for the dual
\emph{observable} field becomes a fixed point of the (dual) forward
equation.  The consequence is that the generator of the backward
equation takes on the equivalent form to the generator of a forward
equation, rather than the adjoint form.  It implies, as we show next, a
transposition of causal and anti-causal propagation in general
correlation functions.

\subsection{Dynamics, tilting, and duality working directly in number
fields} 
\label{sec:AA_duality}

Before studying fluctuations, however, we make a brief digression to
introduce a change of variables from the coherent-state field
variables $\left( {\phi}^{\dagger} , \phi \right)$, to a set $\left( n
  , \eta \right)$ that stand in the relation of \textit{action-angle}
variables~\cite{Goldstein:ClassMech:01} to the original fields.  As
explained elsewhere~\cite{Smith:LDP_SEA:11}, this is a
\textit{canonical transformation} with respect to the functional
integral \footnote{This means that the na{\"{\i}}ve measure in the new
variables produces the correct fluctuation statistics, and is free of
anomalies.  The representation of unity in these fields is derived
directly in Li \textit{et al.}~\cite{Li:CRN_path_integrals:16}.} and is
thus usable for the study of Green's functions and FDTs.  It allows us
to work directly in the number field as an elementary (rather than
composite) variable, with a conjugate field that has the
interpretation of a chemical potential \footnote{Note that this means
the Hamiltonian-conjugate fields with respect to the conserved volume
element of the DP construction correspond to the Legendre dual fields
with respect to the equilibrium entropy.}.

The transformation from coherent-state to number fields for the
two-state system is given by
\begin{align}
  {\phi}_a^{\dagger}
& \equiv 
  e^{{\eta}_a}
& 
  {\phi}_a
& \equiv 
  e^{-{\eta}_a}
  n_a 
\nonumber \\
  {\phi}_b^{\dagger}
& \equiv 
  e^{{\eta}_b}
& 
  {\phi}_b
& \equiv 
  e^{-{\eta}_b}
  n_b .
\label{eq:AA_std_defs}
\end{align}
The combination ${\eta}_b + {\eta}_a$ does not appear in
$\mathcal{L}$, because the stochastic
process~(\ref{eq:T_two_state_genform}) conserves total particle
number, so we perform a second shift to diagonal variables
\begin{align}
  h 
& \equiv 
  \frac{1}{2}
  \left( {\eta}_b + {\eta}_a \right)
& 
  N 
& \equiv 
  \left( n_b + n_a \right)
\nonumber \\
  \eta
& \equiv 
  \left( {\eta}_b - {\eta}_a \right)
& 
  n 
& \equiv 
  \frac{1}{2}
  \left( n_b - n_a \right) .
\label{eq:AA_diag_shift}
\end{align}
The un-tilted dynamical action~(\ref{eq:S_plain_form}) becomes, in
these variables, 
\begin{widetext}
\begin{equation}
  S_{\rm dyn} = 
  \int d\tau
  \left\{
    - N \, 
    d_{\tau} h - 
    n \, 
    d_{\tau} \eta + 
    \frac{1}{2}
    \left[
      \left( 1 + \underline{x} \right)
      \left( 1 - e^{\eta} \right)
      \left( \frac{N}{2} - n \right) + 
      \left( 1 - \underline{x} \right)
      \left( 1 - e^{-\eta} \right)
      \left( \frac{N}{2} + n \right)
    \right]
  \right\} . 
\label{eq:S_plain_AA_form}
\end{equation}
The tilted action~(\ref{eq:S_Crooks_form}) is then 
\begin{equation}
  S_{\rm Crooks} = 
  \int d\tau
  \left\{
    - N \, 
    d_{\tau} \! 
    \left(
      h - \beta \underline{\xi}
    \right) - 
    n \, 
    d_{\tau} \! 
    \left(
      \eta - \beta \underline{\mu}
    \right) + 
    \frac{1}{2}
    \left[
      \left( 1 + \underline{x} \right)
      \left( 1 - e^{\eta} \right)
      \left( \frac{N}{2} - n \right) + 
      \left( 1 - \underline{x} \right)
      \left( 1 - e^{-\eta} \right)
      \left( \frac{N}{2} + n \right)
    \right]
  \right\} , 
\label{eq:S_Crooks_AA_form}
\end{equation}
where for the first time $\xi$ from Eq.~(\ref{eq:bar_mu_conversions})
is written explicitly, showing its role~\cite{Crooks:NE_work_relns:99}
as a normalizing factor.

The duality transformation in action-angle variables is simpler than
in coherent-state variables, and is suggested immediately by the
removal of explicit time-dependence from the kinetic term in
Eq.~(\ref{eq:S_Crooks_AA_form}); we simply shift the fields conjugate
to the particle numbers:
\begin{align}
  \tilde{h}
& \equiv 
  h - \beta \underline{\xi} , 
& 
% \nonumber \\
  \tilde{\eta}
& \equiv 
  \eta - 
  \beta \underline{\mu} . 
\label{eq:Crooks_AA_shifted}
\end{align}
Using the variable relations~(\ref{eq:bar_mu_conversions}) to simplify
terms, the dual representation of the action becomes
\begin{equation}
  S_{\rm Crooks} = 
  \int d\tau
  \left\{
    - \left(
      d_{\tau}
      \tilde{h}
    \right) 
    N - 
    \left(
      d_{\tau} 
      \tilde{\eta}
    \right) 
    n + 
    \frac{1}{2}
    \left[
      \left[ 
        \left( 1 + \underline{x} \right) - 
        \left( 1 - \underline{x} \right) 
        e^{\tilde{\eta}}
      \right]
      \left( \frac{N}{2} - n \right) + 
      \left[
        \left( 1 - \underline{x} \right) - 
        \left( 1 + \underline{x} \right)
        e^{-\tilde{\eta}}
      \right]
      \left( \frac{N}{2} + n \right)
    \right]
  \right\} 
\label{eq:S_Crooks_AA_primes}
\end{equation}
\end{widetext}
$\tilde{h}$ and $\tilde{\eta}$ are the action-angle variables that
would have been obtained from Eq.~(\ref{eq:S_Crooks_var}) by
transforming directly in the dual ${\varphi}^{\dagger}$, $\varphi$
variables.  Note that $n$ and $N$ are not affected by the duality
transformation at all, though their stationary-path equations of
motion and Greens functions will be different when tilting leads to
action~(\ref{eq:S_Crooks_AA_primes}), relative to the un-tilted
action~(\ref{eq:S_plain_AA_form}).

\subsection{Connection of the field-integral approach to the
similarity-transform derivation of Crooks}
\label{sec:Crooks_sim_trans}

To close the section, we show how the transformation of the 2FFI
action above is recovered directly from the similarity-transform
construction as given by Crooks~\cite{Crooks:NE_work_relns:99}.

In that construction, final-time parameters ${{\nu}_a}_T$,
${{\nu}_b}_T$ are imposed to create the desired reference state, and
representations of unity are then inserted in the form of
multiplication and division by Gibbs factors at intermediate times.
The similarity transform over each segment of Liouville evolution
converts the generator matrix ${\rm T}$ into its dual, while the
incomplete cancellation of the Gibbs factors at different times
generates an explicit time-dependence that can be canceled if an
appropriate weight is multiplied onto the evolving density at each
time-step.

For these binomial distributions, following
Eq.~(\ref{eq:part_rho_tau_form}), the Hamiltonian is 
\begin{equation}
  H = 
  \underline{\mu} 
  \left( 
    {\rm n}_b - 
    {\rm n}_a 
  \right) . 
\label{eq:H_is}
\end{equation}
Initial distributions, which are assumed to be of Gibbs form with
Hamiltonian $H_0$ at $\tau = 0$ are evolved with the quadrature of the
master equation~(\ref{eq:ME_genform}), which we write as the
time-ordered product of one-step operators from
Eq.~(\ref{eq:W_def_from_T}), giving $\prod_{\tau = 0}^{T - \delta
\tau} e^{ \delta \tau {\rm T}_\tau }$.

The relevant similarity transform, as in Eq.~(\ref{eq:re_arranged}),
is carried out with a matrix product that is the same for either plain
or tilted Liouville evolution:
\begin{align}
  \prod_{\tau = 0}^{T - \delta \tau}
  e^{
    \delta \tau
    {\rm T}_\tau
  }
  e^{- \beta H_0} 
& = 
  e^{- \beta H_T}
  \prod_{\tau = 0}^{T - \delta \tau}
  \left(
    e^{\beta H_{\tau + \delta \tau}}    
    e^{
      \delta \tau 
      {\rm T}_{\tau}
    }
    e^{- \beta H_{\tau}}
  \right) . 
\label{eq:time_rev_slices}
\end{align}
We suppose that $\delta \tau$ is sufficiently small to write 
\begin{align}
  e^{\beta H_{\tau + \delta \tau}}    
  e^{
    \delta \tau 
    {\rm T}_{\tau}
  }
  e^{- \beta H_{\tau}}
& \approx 
  1 + 
  \delta \tau 
  \left[
    e^{\beta H_{\tau}}    
    {\rm T}_{\tau}
    e^{- \beta H_{\tau}} + 
    d_{\tau}
    \left( 
      \beta H_{\tau}
    \right)
  \right]
\nonumber \\ 
& \equiv 
  1 + 
  \delta \tau 
  \left[
    {\tilde{\rm T}}_{\tau} + 
    d_{\tau}
    \left( 
      \beta H_{\tau}
    \right)
  \right]
\nonumber \\
& \approx 
  e^{
    \delta \tau 
    \left[
      {\tilde{\rm T}}_{\tau} + 
      d_{\tau}
      \left( 
        \beta H_{\tau}
      \right)
    \right]
  } . 
\label{eq:convert_T}
\end{align}
Here the notation $\tilde{\rm T}$ designates the generator matrix
acted on through similarity transform with $e^{\beta H}$.

Applied to the matrix product in Eq.~(\ref{eq:time_rev_slices}), the
similarity transform~(\ref{eq:convert_T}) produces
\begin{align}
  \prod_{\tau = 0}^{T - \delta \tau}
  e^{
    \delta \tau 
    {\rm T}_{\tau} 
  }
  e^{- \beta H_0} 
& = 
  e^{- \beta H_T}
  \prod_{\tau = 0}^{T - \delta \tau}
  e^{
    \delta \tau 
    \left[
      {\tilde{\rm T}}_{\tau} + 
      d_{\tau}
      \left( 
        \beta H_{\tau}
      \right)
    \right]
  } .
\label{eq:time_rev_slices_plain}
\end{align}
The similarity transform is equivalent to the change of integration
variables~(\ref{eq:varphis_ulines}) in the functional integral, and by
itself has no affect on any computed quantities.  If, however, the
tilting factor $- d_{\tau} \left( \beta H_{\tau} \right)$
from Eq.~(\ref{eq:part_rho_tau_form}) had been introduced in the
original evolution equation, that would then cancel the corresponding
term from the similarity transform, leaving only $\tilde{\rm T}$, to
give
\begin{align}
  \prod_{\tau = 0}^{T - \delta \tau}
  e^{
    \delta \tau 
    \left[
      {\rm T}_{\tau} - 
      d_{\tau}
      \left( 
        \beta H_{\tau}
      \right)
    \right]
  }
  e^{- \beta H_0} 
& = 
  e^{- \beta H_T}
  \prod_{\tau = 0}^{T - \delta \tau}
  e^{
    \delta \tau 
    {\tilde{\rm T}}_{\tau}
  } 
\label{eq:time_rev_slices_insert}
\end{align}

For the two-state model $e^{\beta H}$ can be written
\begin{equation}
  e^{\beta H} = 
  {
    \left( 
      \frac{{\underline{\nu}}_a}{{\underline{\nu}}_b} 
    \right)
  }^{\rn} = 
  {
    \left( 
      \frac{1 - \underline{x}}{1 + \underline{x}} 
    \right)
  }^{\rn} , 
\label{eq:exp_as_power}
\end{equation}
in which case the similarity-transformed generator becomes
\begin{align}
  e^{\beta H} 
  {\rm T}
  e^{- \beta H} 
& = 
  \left(
    e^{- \partial / \partial {\rm n}}
    {\underline{\nu}}_a - 
    {\underline{\nu}}_b
  \right)
  {\rm n}_a + 
  \left(
    e^{\partial / \partial {\rm n}}
    {\underline{\nu}}_b - 
    {\underline{\nu}}_a
  \right)
  {\rm n}_b
\nonumber \\
& \equiv
  \tilde{\rm T} . 
\label{eq:T_dag_eval}
\end{align}
If we denote by $\tilde{\mathcal{L}}$ the Liouville operator obtained
by the usual construction using $\tilde{\rm T}$ rather than T, that
evaluation gives 
\begin{align}
  \tilde{\mathcal{L}} 
& = 
  \left(
    {\underline{\nu}}_b - 
    {\underline{\nu}}_a 
    e^{\eta}
  \right)
  n_a + 
  \left(
    {\underline{\nu}}_a - 
    {\underline{\nu}}_b 
    e^{- \eta}
  \right)
  n_b
\nonumber \\
& = 
  \frac{1}{2}
  \left\{
    \left[
      \left( 1 + \underline{x} \right) - 
      \left( 1 - \underline{x} \right)
      e^{\eta}
    \right]
    n_a + 
    \left[
      \left( 1 - \underline{x} \right) - 
      \left( 1 + \underline{x} \right)
      e^{- \eta}
    \right]
    n_b
  \right\} . 
\label{eq:tilde_L_eval}
\end{align}
Thus we correctly recover the Liouvillian term in the transformed
$S_{\rm Crooks}$ of Eq.~(\ref{eq:S_Crooks_AA_primes}).

\section{Beyond stationary points: tilting from causality to
anti-causality in the full distribution} 
\label{sec:caus_anticaus}

Sec.~\ref{sec:duality_SPs} used several simplifications from a 2-state
example and a stationary-point analysis to show where the sense of
time reversal in NEWRs originates in the duality between distributions
and observables.  In this section we show for general correlation
functions and general stochastic CRNs (as long as the effects of
nonlinearity can be expanded in perturbation series for fluctuations),
that the time-reversal illustrated above for stationary points extends
to a full transposition in the roles of observable and response
fields, and of causal with anti-causal response functions.  We
demonstrate this by studying the Green's function expansion obtained
from the second-order approximation to the action functional, which is
also the source of \textit{fluctuation-dissipation relations} both for
unweighted evolving distributions and for their generating
functionals.  We show how these are derived from internal symmetries
of 2-field integrals, and relate them to the Extended FDTs of Seifert
and Speck~\cite{Seifert:FDT:10}.  The key step in all such
constructions is the replacement of correlation functions derived from
the dynamical action with others derived from the Hatano-Sasa tilt
term, concisely reviewed and placed in context of related approaches
to extended FDTs in~\cite{Verley:HS_FDT:12}.

\subsection{Green's function expansions, Ward identities, and
Fluctuation-Dissipation Theorems}
\label{sec:GF_expand_WI}

In a field theory, there may be relations among the expectations of
different operator products in a time-dependent distribution implied
by internal symmetries of the theory.  These are known as the
\textit{Ward identities} of the field theory~\cite{Weinberg:QTF_I:95}.
In 2FFI representations, some of these identities are produced by
shifts of the dummy variables of integration.  One important group are
the Green's functions, which describe the propagation of disturbances
in response to idealized point-like perturbing events.  Here we
consider the \textit{free Green's functions}, which describe the
propagation of disturbances at leading (Gaussian) order; more general
response functions can often be constructed from these by common
perturbative methods~\cite{Weinberg:QTF_I:95}.

In the study of free Green's functions, a condensed notation greatly
simplifies the presentation.  Regard $\phi$ as a column vector and
${\phi}^{\dagger}$ as its conjugate row vector, and likewise for dual
fields $\varphi$ and ${\varphi}^{\dagger}$.  Then for any field
action~(\ref{eq:CRN_L_genform}) the quadratic-order expansion in
fields which controls Gaussian fluctuations may be cast in the
form~\cite{Kamenev:DP:02}
\begin{equation}
  S = 
  \int d\tau 
  \left\{
    - \left( d_{\tau} {\phi}^{\dagger} \right)
    \phi + 
    {\phi}^{\dagger} D_{\tau} \phi - 
    {\phi}^{\dagger} 
      {\Delta}_{\tau}
    {{\phi}^{\dagger}}^T / 2 
  \right\} .
\label{eq:action_kin_cond_not}
\end{equation}
Here $D_{\tau}$ is the \textit{drift} matrix for the hopping rates in
the stochastic process -- generally time-dependent through a varying
parameter such as ${\underline{x}}_{\tau}$ -- and ${\Delta}_{\tau}$ is
a possible source for stochastic fluctuations.  For the free theories
produced by the two-state example, the quadratic-order expansion is
the whole action, and $\Delta \equiv 0$ in the coherent-state
variables.

\subsubsection{The free Green's function arrived at as a variational
identity, and the most basic FDT} 
\label{sec:GF_from_WI}

Let $\left< \mbox{ } \right>$, when bracketing field variables, denote
expectation in the functional integral~(\ref{eq:PI_genform}).
Starting from the expectation
\begin{displaymath}
  \left<
    {
      \left[
        \begin{array}{cc}
          {\phi}^{\dagger} & {\phi}^T
        \end{array}
      \right]
    }_{{\tau}^{\prime}}
  \right> 
\end{displaymath}
at a time ${\tau}^{\prime}$, consider the variation produced by the
pair of shifts of dummy variable of integration at some (generally
different) time $\tau$:
\begin{align}
  0 
& = 
  {
    \left[
      \begin{array}{c}
        \delta / \delta {\phi}^{\dagger} \\
        \delta / \delta {\phi}^T
      \end{array}
    \right]
  }_{\tau}
  \left<
    \begin{array}{c}
      {
        \left[
          \begin{array}{cc}
            {\phi}^{\dagger} & {\phi}^T
          \end{array}
        \right]
      }_{{\tau}^{\prime}} \\
      \phantom{X}
    \end{array}
  \right> 
\nonumber \\
& = 
  I 
  {\delta}_{\tau {\tau}^{\prime}} - 
  \left[
    \begin{array}{cc}
      0 & 
      d_{\tau} + D_{\tau} \\
      - d_{\tau} + D_{\tau}^T & 0 
    \end{array}
  \right]
  \left<
    {
      \left[
        \begin{array}{l}
          {{\phi}^{\dagger}}^T \\
          \phi
        \end{array}
      \right]
    }_{\tau}
    \begin{array}{c}
      {
        \left[
          \begin{array}{cc}
            {\phi}^{\dagger} & {\phi}^T
          \end{array}
        \right]
      }_{{\tau}^{\prime}} \\
      \phantom{X}
    \end{array}
  \right> .
\label{eq:Greens_var_phi}
\end{align}
Here $I$ stands for the $2P \times 2P$ identity matrix and
${\delta}_{\tau {\tau}^{\prime}}$ is the Dirac $\delta$-function that
results from the Kronecker $\delta$ scaled by $1 / \delta \tau$ as
$\delta \tau \rightarrow 0$ in the skeletonized
measure~(\ref{eq:ctm_measure}).

In the first line of Eq.~(\ref{eq:Greens_var_phi}), the variation is
zero because a shift of a dummy variable of integration produces no
change in the value of an integral.  In the second line, the
$\delta$-function variation comes from the direct action of the shift
on the argument of the expectation value at ${\tau}^{\prime}$, while
the second term comes from functional variation of $S$ in the
exponential of Eq.~(\ref{eq:PI_genform}).  

The causal structure of a stochastic process in forward time
ensures~\cite{Kamenev:DP:02} that the expectation has a form first
propounded by Keldysh~\cite{Keldysh:noneq_diag:65} for dissipative
quantum field theories,
\begin{equation}
  \left<
    {
      \left[
        \begin{array}{l}
          {{\phi}^{\dagger}}^T \\
          \phi
        \end{array}
      \right]
    }_{\tau}
    \begin{array}{c}
      {
        \left[
          \begin{array}{cc}
            {\phi}^{\dagger} & {\phi}^T
          \end{array}
        \right]
      }_{{\tau}^{\prime}} \\
      \phantom{X}
    \end{array}
  \right> = 
  {
    \left[ 
      \begin{array}{cc}
        0   & G^A \\
        G^R & G^K
      \end{array}
    \right]
  }_{\tau {\tau}^{\prime}} , 
\label{eq:tri_diag_causal}
\end{equation}
in which $G^R$, $G^A$, and $G^K$ are the \textit{retarded},
\textit{advanced} and \textit{Keldysh} Green's functions of the
theory.  The retarded and advanced Green's functions are solutions to
the inhomogeneous differential equations
\begin{align}
  \left( 
    d_{\tau} + D_{\tau} 
  \right) 
  G^R_{\tau {\tau}^{\prime}} = 
  I {\delta}_{\tau {\tau}^{\prime}} , 
\nonumber \\ 
  \left( 
    - d_{\tau} + D_{\tau}^T 
  \right) 
  G^A_{\tau {\tau}^{\prime}} = 
  I {\delta}_{\tau {\tau}^{\prime}}
\label{eq:G_RA_solve}
\end{align}
from the diagonal blocks of Eq.~(\ref{eq:Greens_var_phi}), (where $I$
now stands for the $P \times P$ identity matrix), while $G^K$ is the
solution to the homogeneous differential equation
\begin{align}
  \left( 
    d_{\tau} + D_{\tau} 
  \right) 
  G^K_{\tau {\tau}^{\prime}} = 
  0 
\label{eq:G_K_EOM}
\end{align}
from the upper-right off-diagonal block.  The general form that is
admitted~\cite{Kamenev:DP:02} for such a homogeneous solution is
\begin{equation}
  G^K_{\tau {\tau}^{\prime}} = 
  G^R_{\tau {\tau}^{\prime}}
  M_{{\tau}^{\prime}} + 
  M_{\tau}
  G^A_{\tau {\tau}^{\prime}} , 
\label{eq:G_K_genform}
\end{equation}
where $M_{\tau}$ satisfies the differential equation
\begin{equation}
  d_{\tau} M_{\tau} + 
  D_{\tau} M_{\tau} + 
  M_{\tau} D_{\tau}^T = 
  0 . 
\label{eq:M_diff_eq}
\end{equation}

Equations~(\ref{eq:G_K_genform}, \ref{eq:M_diff_eq}) are the essential
relations defining the \textit{Fluctuation-Dissipation Theorem (FDT)}.
The retarded and advanced Green's functions, which govern the response
of fields to external perturbations, also determine the rate of decay
of endogenous noise and thus the level of self-maintained
fluctuations.  For equilibrium states, these reduce to the familiar
FDT.  For non-equilibrium distributions, whether steady or
time-dependent, they capture the essential relation between
dissipative relaxation under the retarded and advanced Green's
functions, and the kernel $M_{\tau}$ which serves as the source of
fluctuations.  

For the action~(\ref{eq:S_plain_form}), the matrix $D_{\tau}$ is 
\begin{equation}
  D_{\tau} = 
  \left[ 
    \begin{array}{cc}
      0 & 0 \\
      - {\underline{x}}_{\tau} & 1 
    \end{array}
  \right] 
\label{eq:plain_D_form}
\end{equation}
Because the stationary-path backgrounds $\bar{\phi}$ and
${\bar{\phi}}^{\dagger}$ are homogeneous solutions to the equations of
motion, we may subtract them out, and the
equations~(\ref{eq:G_RA_solve}--\ref{eq:M_diff_eq}) are satisfied for
residual fluctuations ${\phi}^{\prime} \equiv \phi - \bar{\phi}$ and
${{\phi}^{\dagger}}^{\prime} \equiv {\phi}^{\dagger} -
{\bar{\phi}}^{\dagger}$.  The explicit forms for $G^R$ and $G^A$ in
the convenient basis of even and odd symmetry are then
\begin{widetext}
\begin{align}
  \frac{1}{2}
  \left<
    {
      \left[
        \begin{array}{l}
          {\phi}_b^{\prime} + {\phi}_a^{\prime}  \\
          {\phi}_b^{\prime} - {\phi}_a^{\prime}  \\
        \end{array}
      \right]
    }_{\tau}
    \begin{array}{c}
      {
        \left[
          \begin{array}{cc}
            \left( 
              {{\phi}^{\dagger}_b}^{\prime} + {{\phi}^{\dagger}_a}^{\prime}
            \right) &
            \left( 
              {{\phi}^{\dagger}_b}^{\prime} - {{\phi}^{\dagger}_a}^{\prime} 
            \right) 
          \end{array}
        \right]
      }_{{\tau}^{\prime}} \\
      \phantom{X}
    \end{array}
  \right> 
& \equiv 
  G^R_{\tau {\tau}^{\prime}} = 
  {\Theta}_{\tau > {\tau}^{\prime}}
  \left[
    \begin{array}{ll}
      1 & 0 \\
      \int_{{\tau}^{\prime}}^{\tau}
      d {\tau}^{\prime \prime}
      e^{ - 
        \left( \tau - {\tau}^{\prime \prime} \right)
      }
      {\underline{x}}_{{\tau}^{\prime \prime}} & 
      e^{ - 
        \left( \tau - {\tau}^{\prime} \right)
      }
    \end{array}
  \right] , 
\nonumber \\
  \frac{1}{2}
  \left<
    {
      \left[
        \begin{array}{l}
          {{\phi}^{\dagger}_b}^{\prime} + {{\phi}^{\dagger}_a}^{\prime} \\
          {{\phi}^{\dagger}_b}^{\prime} - {{\phi}^{\dagger}_a}^{\prime} 
        \end{array}
      \right]
    }_{\tau}
    \begin{array}{c}
      {
        \left[
          \begin{array}{cc}
            \left( {\phi}_b^{\prime} + {\phi}_a^{\prime} \right) & 
            \left( {\phi}_b^{\prime} - {\phi}_a^{\prime} \right) 
          \end{array}
        \right]
      }_{{\tau}^{\prime}} \\
      \phantom{X}
    \end{array}
  \right> 
& \equiv 
  G^A_{\tau {\tau}^{\prime}} = 
  {\Theta}_{{\tau}^{\prime} > \tau}
  \left[
    \begin{array}{ll}
      1 & 
      \int_{\tau}^{{\tau}^{\prime}}
      d {\tau}^{\prime \prime}
      e^{ - 
        \left( {\tau}^{\prime \prime} - {\tau}^{\prime} \right)
      }
      {\underline{x}}_{{\tau}^{\prime \prime}} \\
      0 & 
      e^{ - 
        \left( \tau - {\tau}^{\prime} \right)
      }
    \end{array}
  \right] .
\label{eq:plain_G_RA_eval}
\end{align}
The Keldysh Green's function at equal time, which coincides with the
value of $M_{\tau}$, is given by
\begin{align}
  \frac{1}{2}
  \left<
    {
      \left[
        \begin{array}{l}
          {\phi}_b^{\prime} + {\phi}_a^{\prime}  \\
          {\phi}_b^{\prime} - {\phi}_a^{\prime}  \\
        \end{array}
      \right]
    }_{\tau}
    \begin{array}{c}
      {
        \left[
          \begin{array}{cc}
            \left( {\phi}_b^{\prime} + {\phi}_a^{\prime} \right) & 
            \left( {\phi}_b^{\prime} - {\phi}_a^{\prime} \right) 
          \end{array}
        \right]
      }_{\tau} \\
      \phantom{X}
    \end{array}
  \right> 
& \equiv 
  G^K_{\tau \tau} = 
  - \frac{N}{2}
  \left[
    \begin{array}{l}
      1 \\ 
      {\underline{x}}_{\tau}
    \end{array}
  \right]
  \begin{array}{c}
    \left[
      \begin{array}{cc}
        1 & {\underline{x}}_{\tau}
      \end{array}
    \right] \\
    \phantom{X}
  \end{array} . 
\label{eq:plain_G_K_eval}
\end{align}
\end{widetext}
For a free theory, the expectations of general operators are obtained
by combinatorial contractions with the above Green's functions, by the
usual Wick expansion~\cite{Weinberg:QTF_I:95}.  Here we will evaluate
one such expectation using a shift of the variables of integration to
illustrate its relation to the simple FDT from the Green's function
alone.

\subsubsection{The Green's function expansion and Ward identities for
general observables}
\label{sec:GF_gen_expand}

Let ${\mathcal{O}}_{{\tau}^{\prime}}$ be some observable which is a
function of the fields in the functional integral at a time
${\tau}^{\prime}$.  Consider the same symmetries as above, under
shifts of the integration variables, but now for the quantity
\begin{displaymath}
  \left<
    {\mathcal{O}}_{{\tau}^{\prime}}
    {
      \left[
        \begin{array}{cc}
          {\phi}^{\dagger} & {\phi}^T
        \end{array}
      \right]
    }_{{\tau}^{\prime \prime}}
  \right> . 
\end{displaymath}
The Ward identity generalizing Eq.~(\ref{eq:Greens_var_phi}) for this
composite observable is
\begin{widetext}
\begin{align}
  0 
& = 
  {
    \left[
      \begin{array}{c}
        \delta / \delta {\phi}^{\dagger} \\
        \delta / \delta {\phi}^T
      \end{array}
    \right]
  }_{\tau}
  \left<
    {\mathcal{O}}_{{\tau}^{\prime}}
    \begin{array}{c}
      {
        \left[
          \begin{array}{cc}
            {\phi}^{\dagger} & {\phi}^T
          \end{array}
        \right]
      }_{{\tau}^{\prime \prime}} \\
      \phantom{X}
    \end{array}
  \right> 
\nonumber \\
& = 
  \left<
    {\mathcal{O}}_{{\tau}^{\prime}}
  \right> 
  I 
  {\delta}_{\tau {\tau}^{\prime \prime}} + 
  \left<
    {
      \left[
        \begin{array}{l}
          \partial \mathcal{O} / 
          \partial {\phi}^{\dagger} \\
          \partial \mathcal{O} / 
          \partial {\phi}^T
        \end{array}
      \right]
    }_{{\tau}^{\prime}}
    \begin{array}{c}
      {
        \left[
          \begin{array}{cc}
            {\phi}^{\dagger} & {\phi}^T
          \end{array}
        \right]
      }_{{\tau}^{\prime \prime}} \\
      \phantom{X}
    \end{array}
  \right> 
  {\delta}_{\tau {\tau}^{\prime}} - 
  \left[
    \begin{array}{cc}
      0 & 
      d_{\tau} + D_{\tau} \\
      - d_{\tau} + D_{\tau}^T & 0 
    \end{array}
  \right]
  \left<
    {\mathcal{O}}_{{\tau}^{\prime}}
    {
      \left[
        \begin{array}{l}
          {{\phi}^{\dagger}}^T \\
          \phi
        \end{array}
      \right]
    }_{\tau}
    \begin{array}{c}
      {
        \left[
          \begin{array}{cc}
            {\phi}^{\dagger} & {\phi}^T
          \end{array}
        \right]
      }_{{\tau}^{\prime \prime}} \\
      \phantom{X}
    \end{array}
  \right> .
\label{eq:Greens_phi}
\end{align}
It follows, from the definition of the Green's function as the inverse
of the kernel for the equations of motion, that the expectation must
have the form
\begin{align}
  \left<
    {\mathcal{O}}_{{\tau}^{\prime}}
    {
      \left[
        \begin{array}{l}
          {{\phi}^{\dagger}}^T \\
          \phi
        \end{array}
      \right]
    }_{\tau}
    \begin{array}{c}
      {
        \left[
          \begin{array}{cc}
            {\phi}^{\dagger} & {\phi}^T
          \end{array}
        \right]
      }_{{\tau}^{\prime \prime}} \\
      \phantom{X}
    \end{array}
  \right> = 
  {
    \left[ 
      \begin{array}{cc}
          0 & G^A \\
        G^R & G^K 
      \end{array}
    \right]
  }_{\tau {\tau}^{\prime}} 
  \left<
    {
      \left[
        \begin{array}{l}
          \partial \mathcal{O} / 
          \partial {\phi}^{\dagger} \\
          \partial \mathcal{O} / 
          \partial {\phi}^T
        \end{array}
      \right]
    }_{{\tau}^{\prime}}
    \begin{array}{c}
      {
        \left[
          \begin{array}{cc}
            {\phi}^{\dagger} & {\phi}^T
          \end{array}
        \right]
      }_{{\tau}^{\prime \prime}} \\
      \phantom{X}
    \end{array}
  \right> + 
  \left<
    {\mathcal{O}}_{{\tau}^{\prime}}
  \right>
  {
    \left[ 
      \begin{array}{cc}
          0 & G^A \\
        G^R & G^K 
      \end{array}
    \right]
  }_{\tau {\tau}^{\prime \prime}} 
\label{eq:Greens_phi_solve}
\end{align}
\end{widetext}
This is the implementation of the Wick expansion of arbitrary
observables in terms of contractions of their derivatives with the
free Green's function.

\subsection{Green's functions and Ward identities in the dual process}
\label{sec:dual_GF_WI}

Because the duality transform~(\ref{eq:varphis_ulines}) can be used to
absorb the explicit time derivative from
Eq.~(\ref{eq:part_rho_tau_form_wL}) or its generalizations, the
generating functional possesses an equivalent field theory to the one
for the underlying process.  The observation made following
Eq.~(\ref{eq:S_Crooks_var}), that duality transposes the structure of
the dynamical equations from fields to their conjugates has the
consequence that, for Green's functions, the block-diagonal
form~(\ref{eq:tri_diag_causal}) is likewise transposed, resulting in a
transposition from causal to anti-causal response in the Green's
functions.

\subsubsection{Anti-Keldysh form and anti-causality}
\label{sec:anti_Keldysh_GF}

In the tilted theory, after transformation to dual field variables,
the second-order action becomes
\begin{equation}
  S = 
  \int d\tau 
  \left\{
    - \left( d_{\tau} {\varphi}^{\dagger} \right)
    \varphi + 
    {\varphi}^{\dagger} {\tilde{D}}_{\tau} \varphi - 
    {\varphi}^{\dagger} 
      {\tilde{\Delta}}_{\tau}
    {{\varphi}^{\dagger}}^T / 2 
  \right\} .
\label{eq:action_kin_cond_dual}
\end{equation}
For the two-state example, ${\tilde{D}}_{\tau} = D_{\tau}^T$ though
this need not be the case more generally.  The variational condition
has the same form as Eq.~(\ref{eq:Greens_var_phi}) with $D$ replaced
by $\tilde{D}$ and $\left( {\phi}^{\dagger} , \phi \right)$ replaced
by $\left( {\varphi}^{\dagger} , \varphi \right)$.

In order for the stationary-path backgrounds to vanish under the dual
equations of motion, the expectation of the outer product of fields
must take the transposed block-diagonal, or ``anti-Keldysh'' form 
\begin{equation}
  \left<
    {
      \left[
        \begin{array}{l}
          {{\varphi}^{\dagger}}^T \\
          \varphi
        \end{array}
      \right]
    }_{\tau}
    \begin{array}{c}
      {
        \left[
          \begin{array}{cc}
            {\varphi}^{\dagger} & {\varphi}^T
          \end{array}
        \right]
      }_{{\tau}^{\prime}} \\
      \phantom{X}
    \end{array}
  \right> = 
  {
    \left[ 
      \begin{array}{cc}
        {\tilde{G}}^K & {\tilde{G}}^A \\
        {\tilde{G}}^R & 0 
      \end{array}
    \right]
  }_{\tau {\tau}^{\prime}} .
\label{eq:tri_diag_anticausal}
\end{equation}

The dual retarded and advanced Green's functions are the transposes of
those in the un-tilted theory, reflecting the fact that here
${\tilde{D}}_{\tau} = D_{\tau}^T$.
\begin{widetext}
\begin{align}
  \frac{1}{2}
  \left<
    {
      \left[
        \begin{array}{l}
          {\varphi}_b^{\prime} + {\varphi}_a^{\prime}  \\
          {\varphi}_b^{\prime} - {\varphi}_a^{\prime}  \\
        \end{array}
      \right]
    }_{\tau}
    \begin{array}{c}
      {
        \left[
          \begin{array}{cc}
            \left( 
              {{\varphi}^{\dagger}_b}^{\prime} + 
              {{\varphi}^{\dagger}_a}^{\prime}
            \right) &
            \left( 
              {{\varphi}^{\dagger}_b}^{\prime} - 
              {{\varphi}^{\dagger}_a}^{\prime} 
            \right) 
          \end{array}
        \right]
      }_{{\tau}^{\prime}} \\
      \phantom{X}
    \end{array}
  \right> 
& \equiv 
  {\tilde{G}}^R_{\tau {\tau}^{\prime}} = 
  {\Theta}_{\tau > {\tau}^{\prime}}
  \left[
    \begin{array}{ll}
      1 & 
      \int_{{\tau}^{\prime}}^{\tau}
      d {\tau}^{\prime \prime}
      e^{ - 
        \left( \tau - {\tau}^{\prime \prime} \right)
      }
      {\underline{x}}_{{\tau}^{\prime \prime}} \\
      0 & 
      e^{ - 
        \left( \tau - {\tau}^{\prime} \right)
      }
    \end{array}
  \right] , 
\nonumber \\
  \frac{1}{2}
  \left<
    {
      \left[
        \begin{array}{l}
          {{\varphi}^{\dagger}_b}^{\prime} + 
          {{\varphi}^{\dagger}_a}^{\prime} \\
          {{\varphi}^{\dagger}_b}^{\prime} - 
          {{\varphi}^{\dagger}_a}^{\prime} 
        \end{array}
      \right]
    }_{\tau}
    \begin{array}{c}
      {
        \left[
          \begin{array}{cc}
            \left( {\varphi}_b^{\prime} + {\varphi}_a^{\prime} \right) & 
            \left( {\varphi}_b^{\prime} - {\varphi}_a^{\prime} \right) 
          \end{array}
        \right]
      }_{{\tau}^{\prime}} \\
      \phantom{X}
    \end{array}
  \right> 
& \equiv 
  {\tilde{G}}^A_{\tau {\tau}^{\prime}} = 
  {\Theta}_{{\tau}^{\prime} > \tau}
  \left[
    \begin{array}{ll}
      1 & 0 \\
      \int_{\tau}^{{\tau}^{\prime}}
      d {\tau}^{\prime \prime}
      e^{ - 
        \left( {\tau}^{\prime \prime} - {\tau}^{\prime} \right)
      }
      {\underline{x}}_{{\tau}^{\prime \prime}} & 
      e^{ - 
        \left( \tau - {\tau}^{\prime} \right)
      }
    \end{array}
  \right] . 
\label{eq:dual_G_RA_eval}
\end{align}
These Green's functions still have the same causal structure with
respect to ordering of the time-indices as their
counterparts~(\ref{eq:plain_G_RA_eval}), but as noted above, the
dependence on $\underline{x}$ is shifted from the $\phi$ fields to the
counterpart ${\varphi}^{\dagger}$ fields.  As a result, the causal
structure couples very differently to the expectations of operators,
in the same way as the mean value ${\bar{x}}_{\tau}$ switches from
retarded dynamics~(\ref{eq:binoms_MFT_evol}) in the underlying
stochastic process to advanced dynamics~(\ref{eq:binoms_MFT_adv_evol})
in the tilted generating functional.

A Ward identity equivalent to Eq.~(\ref{eq:Greens_phi}) for an
arbitrary arbitrary operator $ \left< {\mathcal{O}}_{{\tau}^{\prime}}
\right>$ in the dual variables is
\begin{align}
  0 
& = 
  {
    \left[
      \begin{array}{c}
        \delta / \delta {\varphi}^{\dagger} \\
        \delta / \delta {\varphi}^T
      \end{array}
    \right]
  }_{\tau}
  \left<
    {\mathcal{O}}_{{\tau}^{\prime}}
    \begin{array}{c}
      {
        \left[
          \begin{array}{cc}
            {\varphi}^{\dagger} & {\varphi}^T
          \end{array}
        \right]
      }_{{\tau}^{\prime \prime}} \\
      \phantom{X}
    \end{array}
  \right> 
\nonumber \\
& = 
  \left<
    {\mathcal{O}}_{{\tau}^{\prime}}
  \right> 
  I 
  {\delta}_{\tau {\tau}^{\prime \prime}} + 
  \left<
    {
      \left[
        \begin{array}{l}
          \partial \mathcal{O} / 
          \partial {\varphi}^{\dagger} \\
          \partial \mathcal{O} / 
          \partial {\varphi}^T
        \end{array}
      \right]
    }_{{\tau}^{\prime}}
    \begin{array}{c}
      {
        \left[
          \begin{array}{cc}
            {\varphi}^{\dagger} & {\varphi}^T
          \end{array}
        \right]
      }_{{\tau}^{\prime \prime}} \\
      \phantom{X}
    \end{array}
  \right> 
  {\delta}_{\tau {\tau}^{\prime}} - 
  \left[
    \begin{array}{cc}
      0 & 
      d_{\tau} + {\tilde{D}}_{\tau} \\
      - d_{\tau} + {\tilde{D}}_{\tau}^T & 0 
    \end{array}
  \right]
  \left<
    {\mathcal{O}}_{{\tau}^{\prime}}
    {
      \left[
        \begin{array}{l}
          {{\varphi}^{\dagger}}^T \\
          \varphi
        \end{array}
      \right]
    }_{\tau}
    \begin{array}{c}
      {
        \left[
          \begin{array}{cc}
            {\varphi}^{\dagger} & {\varphi}^T
          \end{array}
        \right]
      }_{{\tau}^{\prime \prime}} \\
      \phantom{X}
    \end{array}
  \right> , 
\label{eq:Greens_varphi}
\end{align}
and its solution is 
\begin{align}
  \left<
    {\mathcal{O}}_{{\tau}^{\prime}}
    {
      \left[
        \begin{array}{l}
          {{\varphi}^{\dagger}}^T \\
          \varphi
        \end{array}
      \right]
    }_{\tau}
    \begin{array}{c}
      {
        \left[
          \begin{array}{cc}
            {\varphi}^{\dagger} & {\varphi}^T
          \end{array}
        \right]
      }_{{\tau}^{\prime \prime}} \\
      \phantom{X}
    \end{array}
  \right> = 
  {
    \left[ 
      \begin{array}{cc}
        {\tilde{G}}^K & {\tilde{G}}^A \\
        {\tilde{G}}^R & 0 
      \end{array}
    \right]
  }_{\tau {\tau}^{\prime}} 
  \left<
    {
      \left[
        \begin{array}{l}
          \partial \mathcal{O} / 
          \partial {\phi}^{\dagger} \\
          \partial \mathcal{O} / 
          \partial {\phi}^T
        \end{array}
      \right]
    }_{{\tau}^{\prime}}
    \begin{array}{c}
      {
        \left[
          \begin{array}{cc}
            {\phi}^{\dagger} & {\phi}^T
          \end{array}
        \right]
      }_{{\tau}^{\prime \prime}} \\
      \phantom{X}
    \end{array}
  \right> + 
  \left<
    {\mathcal{O}}_{{\tau}^{\prime}}
  \right>
  {
    \left[ 
      \begin{array}{cc}
        {\tilde{G}}^K & {\tilde{G}}^A \\
        {\tilde{G}}^R & 0 
      \end{array}
    \right]
  }_{\tau {\tau}^{\prime \prime}} . 
\label{eq:Greens_varphi_solve}
\end{align}
\end{widetext}

These Ward identities subsume all consequences of FDTs for 2-field
functional integrals, and the
solutions~(\ref{eq:plain_G_RA_eval},\ref{eq:dual_G_RA_eval})
demonstrating the shift from causality in the underlying process to
anti-causality in the dual generating functional extends the results
of Sec.~\ref{sec:duality_SPs} from stationary points to general
correlations.  However, formulations of FDTs may take many forms, and
the ``extended FDT'' for non-equilibrium steady states derived
in~\cite{Seifert:FDT:10} appears quite different from the above
formulae.  We will return in Sec.~\ref{sec:extend_FDTs} to show how
the Green's function expansion can be used to evaluate the response to
\emph{physical} perturbation of the boundary conditions, and how the
combination of the anticausality in the generating functional, with
the Ward identities resulting from internal symmetry, can be used to
produce another collection of operator equivalences which are the
extended FDTs.

\subsection{Causal and anti-causal Green's functions in number fields}
\label{sec:AA_anticaus_GF}

A derivation and comparison of the Green's functions in the
action-angle variables of Sec.~\ref{sec:AA_duality}, for the
unweighted distribution and the dual generating functional, gives
further insight into the coexistence of forward and reverse-time
propagation in all these functional integrals, and the way it is
harnessed under duality.

The number field corresponds to a bilinear operator in
${\phi}^{\dagger}$ and $\phi$, so it inherits both causal and
anti-causal responses from their respective dynamics.  The
anticausality of the nominal distribution shown in
equations~(\ref{eq:binoms_MFT_adv_evol},\ref{eq:SP_n_crooks}) results
from changing the way forward and backward responses couple to changes
in the physical boundary conditions, within a block-diagonal matrix
that retains overall Keldysh form.

Although the interpretation of the number fields is more direct than
that of coherent-state fields, the nonlinearity of the action-angle
variable transformation~(\ref{eq:AA_std_defs}) converts the bilinear
coherent-state action for a free field theory into a more complicated
form involving transcendental functions in the number fields, which is
more difficult to work with algebraically.  Therefore we will compute
here only the set of terms directly responsible for fluctuations in
the number field.

The shift of integration variable that defines Green's functions for
the number field is made in $\eta$ (or respectively, $\tilde{\eta}$)
in the plain or dual variables.  The two Ward identities corresponding
to the upper row in Eq.~(\ref{eq:Greens_var_phi}), from variation of
the actions~(\ref{eq:S_Crooks_AA_form})
and~(\ref{eq:S_Crooks_AA_primes}) respectively, are given by
\begin{widetext}
\begin{align}
  0 = 
  \frac{\delta}{\delta {\eta}_{\tau}}
  \left<
    {
      \left[
        \begin{array}{cc}
          \eta & n 
        \end{array}
      \right]
    }_{{\tau}^{\prime}}
  \right> 
& = 
  \left[
    \begin{array}{cc}
      1 & 0  
    \end{array}
  \right]
  {\delta}_{\tau {\tau}^{\prime}} - 
  \left<
    {
      \left[
        \left(
          d_{\tau} + 
          \ch \eta + \underline{x} \sh \eta 
        \right)
        n - 
        \left( 
          \sh \eta + \underline{x} \ch \eta 
        \right) 
        \frac{N}{2}
      \right]
    }_{\tau}
    {
      \left[
        \begin{array}{cc}
          \eta & n 
        \end{array}
      \right]
    }_{{\tau}^{\prime}}
  \right>
\nonumber \\
& = 
  \left[
    \begin{array}{cc}
      1 & 0  
    \end{array}
  \right]
  {\delta}_{\tau {\tau}^{\prime}} - 
  \left<
    {
      \left[
        - \frac{1}{2}
        \left( 
          N - 2 \bar{n} \underline{x}
        \right)
        {\eta}^{\prime} + 
        \left(
          d_{\tau} + 
          \ch \bar{\eta} + \underline{x} \sh \bar{\eta} 
        \right)
        n^{\prime} 
      \right]
    }_{\tau}
    {
      \left[
        \begin{array}{cc}
          {\eta}^{\prime} & n^{\prime} 
        \end{array}
      \right]
    }_{{\tau}^{\prime}}
  \right> , 
\nonumber \\
  0 = 
  \frac{\delta}{\delta {\tilde{\eta}}_{\tau}}
  \left<
    {
      \left[
        \begin{array}{cc}
          \tilde{\eta} & n 
        \end{array}
      \right]
    }_{{\tau}^{\prime}}
  \right> 
& = 
  \left[
    \begin{array}{cc}
      1 & 0  
    \end{array}
  \right]
  {\delta}_{\tau {\tau}^{\prime}} - 
  \left<
    {
      \left[
        \left(
          d_{\tau} + 
          \ch \tilde{\eta} - \underline{x} \sh \tilde{\eta} 
        \right)
        n - 
        \left( 
          \sh \tilde{\eta} - \underline{x} \ch \tilde{\eta} 
        \right) 
        \frac{N}{2}
      \right]
    }_{\tau}
    {
      \left[
        \begin{array}{cc}
          \tilde{\eta} & n 
        \end{array}
      \right]
    }_{{\tau}^{\prime}}
  \right>
\nonumber \\
& = 
  \left[
    \begin{array}{cc}
      1 & 0  
    \end{array}
  \right]
  {\delta}_{\tau {\tau}^{\prime}} - 
  \left<
    {
      \left[
        - \frac{1}{2}
        \left( 
          N - 2 \bar{n} \underline{x}
        \right)
        {\tilde{\eta}}^{\prime} + 
        \left(
          d_{\tau} + 
          \ch \bar{\tilde{\eta}} - \underline{x} \sh \bar{\tilde{\eta}} 
        \right)
        n^{\prime} 
      \right]
    }_{\tau}
    {
      \left[
        \begin{array}{cc}
          {\tilde{\eta}}^{\prime} & n^{\prime} 
        \end{array}
      \right]
    }_{{\tau}^{\prime}}
  \right> . 
\label{eq:etas_vary_AA_twoforms}
\end{align}
\end{widetext}
In the second line of each expression several simplifications are
made.  Terms involving fluctuations of $N$ are dropped, because $N$ is
constant.  Stationary-path values, which are homogeneous solutions to
the equations of motion, are subtracted out to leave expressions for
remainders ${\eta}^{\prime} \equiv \eta - \bar{\eta}$, $n^{\prime}
\equiv n - \bar{n}$, and ${\tilde{\eta}}^{\prime} \equiv \tilde{\eta} -
\bar{\tilde{\eta}}$, and terms linear in fluctuations about the 
stationary paths are also removed because by construction these have
zero mean.  Finally, it can be checked to follow from the
stationary-path equation of motion, that the fluctuation source term
for ${\left( {\eta}^{\prime} \right)}^2$ or ${\left(
{\tilde{\eta}}^{\prime} \right)}^2$ -- this is the term from
${\Delta}_{\tau}$ in the second-order expansion corresponding to
Eq.~(\ref{eq:action_kin_cond_not}), which would be zero in
coherent-state fields~\cite{Kamenev:DP:02} but is non-zero in
action-angle variables~\cite{Smith:LDP_SEA:11} -- takes the same form
$- \left( N - 2 \bar{n} \underline{x} \right) / 2$ in both the forward
and reverse-time solutions, though the functional form $\bar{n}$
appearing in this expression differs for the two solutions.

The formulae for the retarded and advanced Green's functions in the
$\left( \eta , n \right)$ sector are as for the coherent-state system.
The Keldysh Green's functions for fluctuations in $n$ in the two cases
have the standard forms $G^K_{\tau {\tau}^{\prime}} = G^R_{\tau
{\tau}^{\prime}} M_{{\tau}^{\prime}} + M_{\tau} G^A_{\tau
{\tau}^{\prime}}$, or ${\tilde{G}}^K_{\tau {\tau}^{\prime}} =
{\tilde{G}}^R_{\tau {\tau}^{\prime}} {\tilde{M}}_{{\tau}^{\prime}} +
{\tilde{M}}_{\tau} {\tilde{G}}^A_{\tau {\tau}^{\prime}}$, only now the
kernel functions $M$ or $\tilde{M}$ satisfy the inhomogeneous
equations
\begin{align}
  \left[ 
    d_{\tau} + 
    2 
    \left( 
      \ch \bar{\eta} + \underline{x} \sh \bar{\eta} 
    \right) 
  \right] M 
& = 
  \frac{1}{2}
  \left( 
    N - 2 \bar{n} \underline{x}
  \right) , 
\nonumber \\ 
  \left[ 
    d_{\tau} + 
    2 
    \left( 
      \ch \bar{\tilde{\eta}} - \underline{x} \sh \bar{\tilde{\eta}} 
    \right) 
  \right] \tilde{M} 
& = 
  \frac{1}{2}
  \left( 
    N - 2 \bar{n} \underline{x}
  \right) 
\label{eq:AA_GK_M_plain_dual}
\end{align}
(compare to Eq.~(\ref{eq:M_diff_eq}) for coherent state fields).

Further solution details are worked out in App.~\ref{app:AA_Gs_comps}.
The important result for understanding the nature of anti-causality in
the dual generating functional is the form of the Green's functions in
the dual theory,
\begin{align}
  {
    \left[
      \begin{array}{cc}
        0 & {\tilde{G}}^A \\ 
        {\tilde{G}}^R & {\tilde{G}}^K
      \end{array}
    \right]
  }_{\tau {\tau}^{\prime}} 
& = 
  {\Theta}_{\tau > {\tau}^{\prime}}
  e^{
    - \left( \tau - {\tau}^{\prime} \right) 
  }
  \frac{
    1 - {\bar{x}}^2_{\tau}
  }{
    1 - {\bar{x}}^2_{{\tau}^{\prime}}
  }
  \left[
    \begin{array}{cc}
      0 & 0 \\
      1 & {\tilde{M}}_{{\tau}^{\prime}}
    \end{array}
  \right]
\nonumber \\ 
& \phantom{=}
  \mbox{} + 
  {\Theta}_{{\tau}^{\prime} > \tau}
  e^{
    - \left( {\tau}^{\prime} - \tau \right)
  }
  \frac{
    1 - {\bar{x}}^2_{{\tau}^{\prime}}
  }{
    1 - {\bar{x}}^2_{\tau}
  }
  \left[
    \begin{array}{cc}
      0 & 1 \\
      0 & {\tilde{M}}_{\tau}
    \end{array}
  \right] , 
\label{eq:AA_dual_Gs_eval}
\end{align}
in which ${\tilde{M}}_{\tau} = N \left( 1 - {\bar{x}}_{\tau}^2 \right)
/ 4$ from Eq.~(\ref{eq:AA_M_both}), and ${\bar{x}}_{\tau}$ satisfies
Eq.~(\ref{eq:binoms_MFT_adv_evol}).  We return to use these forms in
Eq.~(\ref{eq:delt_x_insert_AA}) below.

\subsection{Extended FDTs from the anticausality and Ward identities
  of generating functionals}
\label{sec:extend_FDTs}

The term \textit{Fluctuation-Dissipation Theorem} can refer to
multiple concepts that are related but that differ in detailed form.
In the original work of Einstein, and in the Extended FDTs of Seifert
and Speck~\cite{Seifert:FDT:10}, the goal is to relate the response of
a system to perturbations in its \emph{physical} boundary conditions
or control parameters, to the magnitude of its fluctuations in the
absence of the perturbation, which define a measure of susceptibility.

In 2FFI representations, including the Schwinger-Keldysh time-loop for
quantum mechanics and the DP construction for stochastic processes,
the presence of the response fields places a layer of intermediate
variables between the operator that represents an external disturbance
and the measure of the system's response.  The advanced and retarded
Green's functions~(\ref{eq:tri_diag_causal}), written as correlation
functions between the observable and response fields, give the
response to any perturbation, and these are related to the fluctuation
spectrum through the Keldysh
relations~(\ref{eq:G_K_genform},\ref{eq:M_diff_eq}).  In 2FFI
theories, these Ward-identity relations are known as
FDTs~\cite{Kamenev:DP:02}.  In this section we relate them to the more
familiar extended~\cite{Seifert:FDT:10} and
generalized~\cite{Polettini:trans_fluct:14} FDTs usually derived from
state-space methods by working with the action $S_{\rm Crooks}$ of
Eq.~(\ref{eq:S_Crooks_form}).  Needless to say, the relations hold more
generally beyond any specific form of the action.

Starting from the action $S_{\rm Crooks}$ of
Eq.~(\ref{eq:S_Crooks_form}) in the original field variables $\left(
{\phi}^{\dagger} , \phi \right)$, we wish to compute the effect of a
perturbation in ${\underline{x}}_{\tau}$ on the expectation
${\mathcal{O}}_{{\tau}^{\prime}}$, as in Sec.~\ref{sec:GF_gen_expand}.
Although only the shift in the generating
matrix~(\ref{eq:T_two_state_genform}) is a ``physical'' perturbation
in the boundary conditions on the stochastic process, we vary the
generating functional along the contour where the path-weight factor
remains matched to the generator, which means varying
${\dot{\underline{\mu}}}_{\tau}$ and ${\underline{x}}_{\tau}$ in the
second term in $S_{\rm Crooks}$ as well.  The result is an expression
for the variation of ${\mathcal{O}}_{{\tau}^{\prime}}$ as a sum of two
operator products, one with the variational term from the Liouville
operator and the other from the path weight:
\begin{widetext}
\begin{equation}
  \frac{\delta}{\delta {\underline{x}}_{\tau}}
  \left<
    {\mathcal{O}}_{{\tau}^{\prime}}
  \right> = 
  - \frac{1}{2}
  \left<
    {\mathcal{O}}_{{\tau}^{\prime}}
    {
      \left(
        {\phi}^{\dagger}_b - 
        {\phi}^{\dagger}_a 
      \right)
    }_{\tau}
    {
      \left(
        {\phi}_b + 
        {\phi}_a 
      \right)
    }_{\tau}
  \right> + 
  \frac{d}{d\tau}
  \left\{ 
  \frac{1}{1 - {\underline{x}}_{\tau}^2}
  \left<
    {\mathcal{O}}_{{\tau}^{\prime}}
    {
    \left[
      {\phi}^{\dagger}_b 
      {\phi}_b 
      \left( 1 - \underline{x} \right) - 
      {\phi}^{\dagger}_a 
      {\phi}_a 
      \left( 1 + \underline{x} \right)
    \right]
    }_{\tau}
  \right> 
  \right\} . 
\label{eq:delt_x_insert_phi}
\end{equation}
\end{widetext}
In the second term we have used an integration by parts to transfer
the $d / d\tau$ from its original argument $\underline{\mu}$ in
Eq.~(\ref{eq:S_Crooks_form}) to the expectation value.

Variation of ${\underline{x}}_{\tau}$ is not inherently a symmetry as
the shift of a dummy integration variable is, so it is not apparent in
the original variables $\left( {\phi}^{\dagger} , \phi \right)$ that
when $\tau < {\tau}^{\prime}$ the variation on the left-hand side of
Eq.~(\ref{eq:delt_x_insert_phi}) is actually identically zero.  This
is true, however, as a consequence of anticausality of the Green's
functions in the generating functional, and this observation gives us
a way to compute the first operator product on the right-hand side of
Eq.~(\ref{eq:delt_x_insert_phi}) in terms of the second.

Although the variation leading to Eq.~(\ref{eq:delt_x_insert_phi}) was
performed within a class of matched generating functionals, in the
particular case where ${\underline{x}}_{\tau}$ is constant, the
background for $\dot{\underline{\mu}}$ in the
action~(\ref{eq:S_Crooks_form}) is zero, so both of the expectation
values on the right-hand side of Eq.~(\ref{eq:delt_x_insert_phi}) must
also be those of the underlying stochastic process.  Since we have
chosen the initial state ${\rho}_0$ to be annihilated by the
generator~(\ref{eq:T_two_state_genform}), these expectations are those
of a (non-equilibrium or equilibrium) steady state.  Relations of the
form~(\ref{eq:delt_x_insert_phi}) are the \textit{extended FDTs} for
non-equilibrium steady states introduced by Seifert and
Speck~\cite{Seifert:FDT:10} (see also Verley \textit{et
  al.}~\cite{Verley:FDT_motors:11,Verley:FDT_ising:11}).  They fit
within the class of ``generalized FDTs'' defined by Polettini and
Esposito~\cite{Polettini:trans_fluct:14}.

In the more general case, with ${\underline{x}}_{\tau}$ non-constant,
the relation~(\ref{eq:delt_x_insert_phi}) still holds, though now the
expectation values refer to those in the generating functional with
$\dot{\underline{\mu}}$ nonzero.  The total variation $\delta \left<
{\mathcal{O}}_{{\tau}^{\prime}} \right> / \delta
{\underline{x}}_{\tau}$ remains zero even in the dynamical case --
which is far from apparent in the original variables -- as a
consequence of the memory-erasing effect of the weighting term, and
the anti-causal correlation structure that results.

The duality transformation~(\ref{eq:varphis_ulines}), together with
the Green's function evaluations of Sec.~\ref{sec:anti_Keldysh_GF},
provide a way to show vanishing of the
variation~(\ref{eq:delt_x_insert_phi}) and thus the extended FDT for
steady states and its dynamical generalization.  Variation with
$\delta {\underline{x}}_{\tau}$ in the action~(\ref{eq:S_Crooks_var})
gives only one term:
\begin{equation}
  \frac{\delta}{\delta {\underline{x}}_{\tau}}
  \left<
    {\mathcal{O}}_{{\tau}^{\prime}}
  \right> = 
  - \frac{1}{2}
  \left<
    {\mathcal{O}}_{{\tau}^{\prime}}
    {
      \left(
        {\varphi}^{\dagger}_b + 
        {\varphi}^{\dagger}_a 
      \right)
    }_{\tau}
    {
      \left(
        {\varphi}_b - 
        {\varphi}_a 
      \right)
    }_{\tau}
  \right> , 
\label{eq:delt_x_insert_varphi}
\end{equation}
after which the Ward identity from Eq.~(\ref{eq:Greens_varphi_solve})
may be used to evaluate the expectation in terms of a Green's function
expansion,
\begin{widetext}
\begin{equation}
  \left< 
    {\mathcal{O}}_{{\tau}^{\prime}}
    {
      \left(
        {\varphi}^{\dagger}_b + 
        {\varphi}^{\dagger}_a 
      \right)
    }_{\tau}
    {
      \left(
        {\varphi}_b - 
        {\varphi}_a 
      \right)
    }_{\tau}
  \right> = 
  {\Theta}_{\tau > {\tau}^{\prime}}
  e^{
    - \left( \tau - {\tau}^{\prime} \right) 
  }
  \left< 
    {
      \left. 
        \frac{
          \partial \mathcal{O} 
        }{
          \partial 
          \left( {\varphi}^{\dagger}_b - {\varphi}^{\dagger}_a \right) 
        }
      \right|
    }_{{\tau}^{\prime}}
    {
      \left( {\varphi}^{\dagger}_b + {\varphi}^{\dagger}_a \right) 
    }_{\tau}
  \right> + 
  {\Theta}_{{\tau}^{\prime} > \tau}
  \left< 
    {
      \left. 
        \frac{
          \partial \mathcal{O} 
        }{
          \partial 
          \left( {\varphi}_b + {\varphi}_a \right) 
        }
      \right|
    }_{{\tau}^{\prime}}
    {
      \left( {\varphi}_b - {\varphi}_a \right) 
    }_{\tau}
  \right> . 
\label{eq:Ext_FDT_dual_solve}
\end{equation}
There is the potential for coupling with either ordering of $\tau$ and
${\tau}^{\prime}$, because both retarded and advanced Green's
functions make a contribution.  However, if the second term on the
right-hand side of Eq.~(\ref{eq:Ext_FDT_dual_solve}) is evaluated in a
Green's function expansion of $\partial \mathcal{O} / \partial \left(
{\varphi}_b + {\varphi}_a \right)$, there is no contribution for $\tau
< {\tau}^{\prime}$ by Eq.~(\ref{eq:dual_G_RA_eval}).  Therefore ${
\left( {\varphi}_b - {\varphi}_a \right) }_{\tau}$ is evaluated at its
stationary-point value, which by
Eq.~(\ref{eq:Crooks_vary_phi_dag_bulk}) and the initial
conditions~(\ref{eq:triv_vary_phi_dag_surf}) is zero.  In this way the
``anti-causality'' of what we have termed the nominal distribution in
Sec.~\ref{sec:dual_GF_SPsol} is extended to correlation functions of
arbitrary observables.  Variation of the parameters in the Liouville
operator ($\underline{x}$ in the examples) at times $\tau$ earlier
than the support ${\tau}^{\prime}$ of some observable $\mathcal{O}$
cannot be propagated to $\mathcal{O}$ because they are only carried on
the ${\varphi}^{\dagger}$ fields, which propagate them only to times
earlier than $\tau$.

\subsubsection{For free theories, short-cuts using the background
fields} 
\label{sec:BF_GF_shortcut}

For a free theory, where the stationary path is also the exact mean,
the result of a variation $\delta {\underline{x}}_{\tau}$ can be
computed directly by computing the dependence of
${\mathcal{O}}_{{\tau}^{\prime}}$ on the shifted stationary path.  In
the dual generating functional in variables $\left(
  {\varphi}^{\dagger} , \varphi \right)$, the result is
\begin{align}
  \delta
  \left<
    {\mathcal{O}}_{{\tau}^{\prime}}
  \right> 
& = 
  \frac{
    \partial  
    \left<
      {\mathcal{O}}_{{\tau}^{\prime}}
    \right> 
  }{
    \partial
    { 
    \left( {\bar{\varphi}}^{\dagger}_b - {\bar{\varphi}}^{\dagger}_a \right) 
    }_{{\tau}^{\prime}}
  }
  \delta 
  {
    \left( 
      {\bar{\varphi}}^{\dagger}_b - 
      {\bar{\varphi}}^{\dagger}_a 
    \right) 
  }_{{\tau}^{\prime}} + 
  \frac{
    \partial  
    \left<
      {\mathcal{O}}_{{\tau}^{\prime}}
    \right> 
  }{
    \partial 
    {
      \left( {\bar{\varphi}}_b + {\bar{\varphi}}_a \right) 
    }_{{\tau}^{\prime}}
  }
  \delta 
  {
    \left( {\bar{\varphi}}_b + {\bar{\varphi}}_a \right) 
  }_{{\tau}^{\prime}}
\nonumber \\
& = 
  \frac{
    \partial  
    \left<
      {\mathcal{O}}_{{\tau}^{\prime}}
    \right> 
  }{
    \partial
    { 
    \left( {\bar{\varphi}}^{\dagger}_b - {\bar{\varphi}}^{\dagger}_a \right) 
    }_{{\tau}^{\prime}}
  }
  {\Theta}_{\tau > {\tau}^{\prime}}
  e^{
    - \left( \tau - {\tau}^{\prime} \right) 
  } 
  {
    \left( 
      {\bar{\varphi}}^{\dagger}_b + 
      {\bar{\varphi}}^{\dagger}_a 
    \right) 
  }_{\tau} 
  \delta {\underline{x}}_{\tau} + 
  \frac{
    \partial  
    \left<
      {\mathcal{O}}_{{\tau}^{\prime}}
    \right> 
  }{
    \partial 
    {
      \left( {\bar{\varphi}}_b + {\bar{\varphi}}_a \right) 
    }_{{\tau}^{\prime}}
  }
  {\Theta}_{{\tau}^{\prime} > \tau}
  {
    \left( {\bar{\varphi}}_b - {\bar{\varphi}}_a \right) 
  }_{\tau}
  \delta {\underline{x}}_{\tau} 
\nonumber \\
& = 
  \frac{
    \partial  
    \left<
      {\mathcal{O}}_{{\tau}^{\prime}}
    \right> 
  }{
    \partial
    { 
    \left( {\bar{\varphi}}^{\dagger}_b - {\bar{\varphi}}^{\dagger}_a \right) 
    }_{{\tau}^{\prime}}
  }
  {\Theta}_{\tau > {\tau}^{\prime}}
  e^{
    - \left( \tau - {\tau}^{\prime} \right) 
  } 
  2 \delta {\underline{x}}_{\tau} . 
\label{eq:AA_stat_path_variation}
\end{align}
% \end{widetext}
The second and third lines use solutions for
${\bar{\varphi}}^{\dagger}$ and $\bar{\varphi}$ from
App.~\ref{app:Crooks_tilted_SP}, recovering the advanced and retarded
decay terms of Eq.~(\ref{eq:Ext_FDT_dual_solve}), including vanishing
of the second term for $\tau < {\tau}^{\prime}$.

Writing out the extended FDT directly in number fields clarifies the
way in which a shift from causality to anti-causality is accomplished
by a change in the way terms couple to boundary conditions.  The same
steps that lead to Eq.~(\ref{eq:AA_stat_path_variation}), starting
from the action~(\ref{eq:S_Crooks_AA_primes}), lead to
% \begin{widetext}
\begin{align}
  \frac{\delta}{\delta {\underline{x}}_{\tau}}
  \left<
    {\mathcal{O}}_{{\tau}^{\prime}}
  \right> 
& = 
  - 
  \left<
    {\mathcal{O}}_{{\tau}^{\prime}}
    \left[
      \left( 1 + e^{\tilde{\eta}} \right)
      \left( \frac{N}{2} - n \right) - 
      \left( 1 + e^{-\tilde{\eta}} \right)
      \left( \frac{N}{2} + n \right) 
    \right]
  \right> 
\nonumber \\
& \approx 
  2 
  \left[
    \begin{array}{cc}
      - \frac{N}{2} & 
      \left( 1 + \ch {\bar{\tilde{\eta}}}_{\tau} \right)
    \end{array}
  \right] 
  \left<
    {
      \left[
        \begin{array}{c}
          {\tilde{\eta}}^{\prime} \\
          n^{\prime}
        \end{array}
      \right]
    }_{\tau}
    \begin{array}{c}
      {
        \left[
          \begin{array}{cc}
            {\tilde{\eta}}^{\prime} & n^{\prime} 
          \end{array}
        \right]
      }_{{\tau}^{\prime}} \\
      \phantom{}
    \end{array}
  \right> 
  {
    \overline{
      \left[
        \begin{array}{c}
          \partial \mathcal{O} / 
          \partial \tilde{\eta} \\ 
          \partial \mathcal{O} / 
          \partial n
        \end{array}
      \right]
    }
  }_{{\tau}^{\prime}}
\nonumber \\
& =
  2 \left[
    \begin{array}{cc}
      - \frac{N}{2} & 
      \frac{2}{\left( 1 - {\bar{x}}_{\tau}^2 \right)}
    \end{array}
  \right] 
  \left<
    {
      \left[
        \begin{array}{c}
          {\tilde{\eta}}^{\prime} \\
          n^{\prime}
        \end{array}
      \right]
    }_{\tau}
    \begin{array}{c}
      {
        \left[
          \begin{array}{cc}
            {\tilde{\eta}}^{\prime} & n^{\prime} 
          \end{array}
        \right]
      }_{{\tau}^{\prime}} \\
      \phantom{}
    \end{array}
  \right> 
  {
    \overline{
      \left[
        \begin{array}{c}
          \partial \mathcal{O} / 
          \partial \tilde{\eta} \\ 
          \partial \mathcal{O} / 
          \partial n
        \end{array}
      \right]
    }
  }_{{\tau}^{\prime}}
\nonumber \\ 
& = 
  2 \left[
    \begin{array}{cc}
      - \frac{N}{2} & 
      \frac{2}{\left( 1 - {\bar{x}}_{\tau}^2 \right)}
    \end{array}
  \right] 
  \left\{ 
    {\Theta}_{\tau > {\tau}^{\prime}}
    e^{
      - \left( \tau - {\tau}^{\prime} \right) 
    }
    \frac{
      1 - {\bar{x}}^2_{\tau}
    }{
      1 - {\bar{x}}^2_{{\tau}^{\prime}}
    }
    \left[
      \begin{array}{cc}
        0 & 0 \\
        1 & {\tilde{M}}_{{\tau}^{\prime}}
      \end{array}
    \right] + 
    {\Theta}_{{\tau}^{\prime} > \tau}
    e^{
      - \left( {\tau}^{\prime} - \tau \right)
    }
    \frac{
      1 - {\bar{x}}^2_{{\tau}^{\prime}}
    }{
      1 - {\bar{x}}^2_{\tau}
    }
    \left[
      \begin{array}{cc}
        0 & 1 \\
        0 & {\tilde{M}}_{\tau}
      \end{array}
    \right]
  \right\} 
  {
    \overline{
      \left[
        \begin{array}{c}
          \partial \mathcal{O} / 
          \partial \tilde{\eta} \\ 
          \partial \mathcal{O} / 
          \partial n
        \end{array}
      \right]
    }
  }_{{\tau}^{\prime}}
\label{eq:delt_x_insert_AA}
\end{align}
\end{widetext}
In the first line, the stationary-path part of the term in square
brackets is zero by the equations of motion, and $\approx$ in the
second line indicates the leading expansion to second order in
fluctuations, the same order to which we have expanded Green's
functions in the action-angle variables.

The only nonzero term in the Green's function at ${\tau}^{\prime} >
\tau$ in the last line is  orthogonal, by Eq.~(\ref{eq:AA_M_both}), to
the vector  
$ \left[ 
  \begin{array}{cc} 
    - N / 2 & 
    2 / \left( 1 - {\bar{x}}_{\tau}^2 \right)
  \end{array} 
\right]$ in Eq.~(\ref{eq:delt_x_insert_AA}), ensuring that for $\tau <
{\tau}^{\prime}$ 
$\delta
  \left<
    {\mathcal{O}}_{{\tau}^{\prime}}
  \right> / \delta {\underline{x}}_{\tau} = 0$. 
In the particular case that $\mathcal{O} = \left[ \begin{array}{cc}
\tilde{\eta} & n \end{array} \right]$, Eq.~(\ref{eq:delt_x_insert_AA})
 recovers the variation of the stationary path for $\tau >
 {\tau}^{\prime}$,  
\begin{align}
  \frac{
    \delta {\bar{\tilde{\eta}}}_{{\tau}^{\prime}}
  }{
    \delta {\underline{x}}_{\tau}
  } 
& = 
  \frac{
    2 
  }{
    1 - {\bar{x}}_{{\tau}^{\prime}}^2
  }
  e^{- \left( \tau - {\tau}^{\prime} \right)} , 
\nonumber \\
  \frac{
    \delta {\bar{n}}_{{\tau}^{\prime}}
  }{
    \delta {\underline{x}}_{\tau}
  } 
& = 
  \frac{N}{2}
  e^{- \left( \tau - {\tau}^{\prime} \right)} .
\label{eq:AA_stat_path_vary_x}
\end{align}
The sensitivity of the response field in the first line includes the
measure term $ 1 - {\bar{x}}_{{\tau}^{\prime}}^2$ that we first saw in
the backward propagation of final-time tilting weights in
Eq.~(\ref{eq:plain_genfun_num_sol}).

The anti-causality of correlation functions demonstrated in this
section extends to more general non-linear rate laws, by using the
Wick expansion of higher-order perturbations about the stationary-path
background, wherever the perturbation series converges.

\section{Duality beyond time reversal}
\label{sec:duality_gen_CRN}

We now show how the constructions of the previous sections can be
applied at the level of the CRN framework of
Sec.~\ref{sec:2FFI_basics}.  The main new observation is the way
duality transformation for systems with non-linear rate laws
(corresponding to non-``free'' field theories) generalizes the
operation of transposing the kernel of the Liouville operator that was
exhibited in Eq.~(\ref{eq:S_Crooks_var}).

\subsection{Dualizing about steady states under the transition matrix}  
\label{sec:SS_dual_genform}

As we noted in Sec.~\ref{sec:2FFI_basics}, the adjacency matrix
${\mathbb{A}}_k$ on the complex network defines the relevant concept
of a graph Laplacian for stochastic processes on CRNs.  Acting on this
network with a suitable topology-preserving transformation exchanges
the roles of observable and response fields in the DP functional
integral.  

Since we will define duality transforms about steady states of the
generator~(\ref{eq:T_psi_from_A}) with strictly positive species
counts, we mention first the conditions under which such solutions are
ensured to exist.  The Feinberg deficiency-zero
criterion~\cite{Feinberg:notes:79,Feinberg:def_01:87} is a sufficient
condition as long as all rate constants in ${\mathbb{A}}_k$ are
nonzero.  Steady states may exist for a much wider range of CRNs than
these, but their existence can then depend quantitatively on the rate
constants.

Let $n^{\ast}$ denote the vector of steady-state numbers for $n$ under
the stochastic process with the topological adjacency matrix
${\mathbb{A}}^{\rm top}$ from Eq.~(\ref{eq:ref_A_genform}).  By this
choice of reference matrix, ${\psi}_{Yi} \! \left( n^{\ast} \right) =
\mbox{const.}$  is always a right eigenvector of ${\mathbb{A}}^{\rm
top}$, so we may choose $n^{\ast} \equiv 1 \Rightarrow {\psi}_{Yi} \! 
\left( n^{\ast} \right) = 1 ; \forall i$\footnote{This choice is WLOG
if the null complex never appears in the CRN as a source or sink; it
is forced if the null complex does appear, because for that complex
${\psi}_{\rm null} \equiv 1$.}.

Let $\underline{n}$ denote the steady state under the stochastic
process with matrix ${\mathbb{A}}_k$ from Eq.~(\ref{eq:A_genform}).
Whenever all ${\underline{n}}_p > 0$, the appropriate generalization
of Eq.~(\ref{eq:varphis_ulines}) to the more general CRN is
\begin{align}
  {\phi}_p
& \equiv 
  \left( 
    \frac{{\underline{n}}_p}{n^{\ast}_p}
  \right) 
  {\varphi}_p , 
\nonumber \\ 
  {\phi}^{\dagger}_p
& \equiv 
  \left( 
    \frac{n^{\ast}_p}{{\underline{n}}_p}
  \right) 
  {\varphi}^{\dagger}_p . 
\label{eq:CRN_duals_def_genform}
\end{align}
The evaluation of the Liouville operator~(\ref{eq:L_psi_from_A}) in
dual fields gives
\begin{widetext}
\begin{align}
& \phantom{=} 
  {\psi}_Y^T \! \left( {\phi}^{\dagger} \right) 
  {\mathbb{A}}_k \, 
  {\psi}_Y \! \left( \phi \right) 
\nonumber \\ 
& = 
  {\psi}_Y^T \! \left( {\phi}^{\dagger} \right) \, 
  \sum_{
    \left( i , j \right) 
  }
  \left( 
    w_j - w_i 
  \right) 
  k_{ji}
  w_i^T \, 
  {\psi}_Y \! \left( \phi \right) 
\nonumber \\ 
& = 
  {\psi}_Y^T \! \left( {\varphi}^{\dagger} \right) \, 
  \left[
    \diag 
    \left( 
      {\psi}_Y \! \left( n^{\ast} \right) / 
      {\psi}_Y \! \left( n \right) 
    \right)
  \right] 
  \sum_{
    \left( i , j \right) 
  }
  \left( 
    w_j - w_i 
  \right) 
  k_{ji}
  w_i^T
  \left[
    \diag 
    \left( 
      {\psi}_Y \! \left( n \right) / 
      {\psi}_Y \! \left( n^{\ast} \right) 
    \right)
  \right] \, 
  {\psi}_Y \! \left( \varphi \right) 
\nonumber \\ 
& \equiv  
  {\psi}_Y^T \! \left( {\varphi}^{\dagger} \right) \, 
  \sum_{
    \left( i , j \right) 
  }
  w_i 
  {\tilde{k}}_{ij}
  \left( 
    w_j^T - w_i^T 
  \right) \, 
  {\psi}_Y \! \left( \varphi \right) 
\nonumber \\
& \equiv 
  {\psi}_Y^T \! \left( {\varphi}^{\dagger} \right) 
  {\tilde{\mathbb{A}}}_k \, 
  {\psi}_Y \! \left( \varphi \right) .
\label{eq:tildeA_constr_genform}
\end{align}
\end{widetext}
Here $\left[ \diag \left( v \right)
\right]$, for a vector $v \equiv \left[ v_i \right]$, denotes the
diagonal matrix with $i$th entry $v_i$. In passing from the third to
the fourth line of Eq.~(\ref{eq:tildeA_constr_genform}), we have used
the fact that
\begin{equation}
  \sum_{\left( j \mid i \right)}
  {\tilde{k}}_{ji} = 
  \sum_{\left( j \mid i \right)}
  {\tilde{k}}_{ij}
\label{eq:CRN_duals_right_stoch}
\end{equation}
implied by ${\psi}_{Yi} \! \left( n^{\ast} \right) = 1 ; \forall i$.
${\tilde{\mathbb{A}}}_k$ is the rate matrix for the dual process,
defined from the fourth line in terms of dual rate constants
${\tilde{k}}_{ij}$.

The change coming from the kinetic term when the
action~(\ref{eq:CRN_L_genform}) is written in dual variables is
\begin{equation}
  - \left(
    d_{\tau} {\phi}^{\dagger}_p 
  \right) 
  {\phi}_p = 
  - \left(
    d_{\tau} {\varphi}^{\dagger}_p 
  \right) 
  {\varphi}_p + 
  d_{\tau} \log {\underline{n}}_p 
  {\varphi}^{\dagger}_p 
  {\varphi}_p , 
\label{eq:CRN_duals_deriv_term}
\end{equation}
and the term involving $d_{\tau} \log {\underline{n}}_p$ is the one
that must be subtracted by a path-weighting function to cause the
observable and response fields to exchange roles.  Note that this term
-- the path-integral version of the time derivative of the log-density
function of Hatano and Sasa~\cite{Hatano:NESS_Langevin:01} -- is
defined entirely from the overlap $\left( {\phi}_{\tau - \delta \tau}
\mid {\phi}_{\tau} \right)$ of the coherent states between insertions
of the Peliti representation of unity~(\ref{eq:field_int_ident}) at
adjacent times.  In the case of detailed balance as with the 2-state
example, it becomes the excess work of
Crooks~\cite{Crooks:path_ens_aves:00}.

The action for the dual generating functional, 
\begin{widetext}
\begin{align}
  S_{\rm Crooks} 
& = 
  \int d\tau 
  \left\{ 
    - \left( d_{\tau} {\phi}^{\dagger} \right)
    \phi + 
    {\psi}_Y^T \! \left( {\phi}^{\dagger} \right) 
    {\mathbb{A}}_k \, 
    {\psi}_Y \! \left( \phi \right) - 
    {\phi}^{\dagger}
    \left[ 
      \diag 
      \left( d_{\tau} \log \underline{n} \right) 
    \right]
    \phi 
  \right\} 
\nonumber \\ 
& = 
  \int d\tau 
  \left\{ 
    - \left( d_{\tau} {\varphi}^{\dagger} \right)
    \varphi + 
    {\psi}_Y^T \! \left( {\varphi}^{\dagger} \right) 
    {\tilde{\mathbb{A}}}_k \, 
    {\psi}_Y \! \left( \varphi \right) 
  \right\} , 
\label{eq:CRN_L_Crooksform}
\end{align}
\end{widetext}
is the general memory-erasing form for stochastic CRNs.  For the
two-state model, it recovers the
forms~(\ref{eq:S_Crooks_form},\ref{eq:S_Crooks_var}).

\subsection{Similarity transforms on the complex network and on
the state space}

\begin{figure}[ht]
  \begin{center} 
  \includegraphics[scale=0.3]{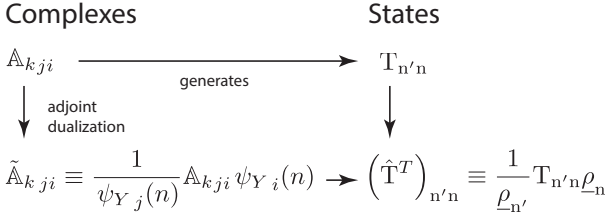} 
  \caption{
  The adjacency matrix ${\mathbb{A}}_k$ and the one-step transition
  matrix ${\rm T}$ under dualization.  The fixed matrix
  ${\mathbb{A}}_k$ generates a transition matrix ${\rm T}$ of rank
  equal to that of the state space.  Dualization to form the adjoint
  generating functional acts on both matrices by a rescaling
  similarity transform.  For the adjacency matrix, the scale factor
  is a function of average numbers, and for the transition matrix it
  is the vector of state probabilities.
  \label{fig:DP_comms_diagram_indices} 
  } 
  \end{center}
\end{figure}

Fig.~\ref{fig:DP_comms_diagram_indices} summarizes the parallel action
of the adjoint dualization transform on the adjacency matrix
${\mathbb{A}}_k$ and the transition rate matrix ${\rm T}$.  The fixed
set of reactions among complexes generate the full matrix of
transition rates among states.  As we note
in~\cite{Krishnamurthy:CRN_moments:17,Smith:CRN_moments:17}, the step
of interposing complexes between chemical species and reaction events,
thus placing stoichiometric constraints between species and complexes,
rather than between species and reactions directly, allows the complex
matrix to behave as an ordinary directed network for a random walk.
Therefore the vector of 1s on complexes $i$ annihilates
${\mathbb{A}}_k$ on the left, as the vector of 1s on states ${\rm n}$
annihilates ${\rm T}$ on the left in Eq.~(\ref{eq:hatT_properties}).

Under adjoint dualization, both matrices are similarity transformed by
rescaling, ${\mathbb{A}}_k$ with a vector of complex activities ${\psi}_Y
\! \left( \underline{n} \right)$, ${\rm T}$ with the vector of probabilities
$\underline{\rho}$ in which those activities would be observed.  The
adjoint adjacency matrix ${\tilde{\mathbb{A}}}_k$ generates the
adjoint transition rates among states.  In systems with detailed
balance, ${\tilde{\mathbb{A}}}_k = {\mathbb{A}}_k^T$, and thus both
${\mathbb{A}}_k$ and ${\rm T}$ are self-adjoint.  Details are provided
in App.~\ref{app:MR_sys_Aform}.  Self-duality of the Liouville
operator for networks with detailed-balance equilibria is similar to
Hermiticity of the Hamiltonian for systems with microscopic
reversibility.

We see both why nonlinear CRNs can transform as simply as linear
processes under dualization, and also that the adjoint construction
from Eq.~(\ref{eq:tildeA_constr_genform}) is much simpler than would
be expected from the general transformation~(\ref{eq:adjoint_update}).
The former is true because the adjacency matrix transforms as an
ordinary graph Laplacian for linear or non-linear processes; all
complexity is cordoned off in the estimates of $\underline{n}$.  Note,
however, that whereas the transition matrix is transformed by all
components of $\underline{\rho}$, Eq.~(\ref{eq:tildeA_constr_genform})
is defined from \emph{simple products} of the $P$ components of the
mean number vector $\underline{n}$.  Exact calculation of
$\underline{n}$ formally depends on calculation of $\underline{\rho}$,
because in general moment hierarchies do not
truncate~\cite{Krishnamurthy:CRN_moments:17,Smith:CRN_moments:17}.
However, approximate or parametric inversions based on mean values
are simpler and more robust than those that require explicit
estimation of all moments. 

The condition that assures many of the stronger simplifications among
fluctuation theorems -- principle among them the equivalence of the
backward and the adjoint versions of duality -- is detailed balance.
The simplification associated with the similarity transforms we have
just exhibited for CRNs -- that the low-dimensional information in the
scale factors ${\psi}_Y \! \left( \underline{n} \right)$ contain
\emph{all} the information in the densities $\underline{\rho}$ -- is
the weaker condition of complex balance.  For complex-balanced steady
states, the distributions are products of Poisson distributions on
individual ${\rn}_k$, or sections through such products, and ${\psi}_Y
\! \left( n \right) = {\left< {\Psi}_Y \! \left( \rn \right)
\right>}_{\underline{\rho}}$, a result known as the
Anderson-Craciun-Kurtz theorem~\cite{Anderson:product_dist:10}.

\subsection{Worked examples}
\label{sec:examples}

We close with a pair of worked examples to show how dual graphical
models characterizing the propagation of information in the response
fields are derived from the graphical models of the underlying
stochastic process.

\subsubsection{A simple cycle with no dynamical reversibility}
\label{sec:examp_3cyc}

The first example is a 3-cycle of purely irreversible events, with two
of the complexes involving pairs of particles.  The reaction schema
(with generally time-dependent rate constants) is
\begin{align}
  {\rm A}
& \overset{k_1}{\rightharpoonup}
  2 {\rm B} 
\nonumber \\ 
  2 {\rm B} 
& \overset{k_2}{\rightharpoonup}
  2 {\rm C} 
\nonumber \\ 
  2 {\rm C} 
& \overset{k_3}{\rightharpoonup}
  {\rm A} . 
\label{eq:cycle_system}
\end{align}
The rate matrix written explicitly is 
\begin{equation}
  {\mathbb{A}}_k = 
  \left[ 
    \begin{array}{rrr}
      -k_1 &  0\mbox{ } &  k_3 \\
       k_1 & -k_2 &    0\mbox{ } \\
         0\mbox{ } &  k_2 & -k_3
    \end{array}
  \right] , 
\label{eq:cycle_system_A_form}
\end{equation}
and the stoichiometric matrix is 
\begin{equation}
  Y = 
  \left[  
    \begin{array}{rrr}
      1 & 0 & 0 \\
      0 & 2 & 0 \\
      0 & 0 & 2 
    \end{array}
  \right] , 
\label{eq:cycle_Y_parms}
\end{equation}
in terms of which the stationary-path equation of motion for the field
$\phi$ becomes
\begin{align}
  0 
& \rightarrow 
  d_{\tau} \phi + 
  Y 
  {\mathbb{A}}_k
  {\psi}_Y \! \left( \phi \right)  
\nonumber \\ 
& = 
  d_{\tau} \phi + 
  Y 
  \left[  
    \begin{array}{rrr}
      -k_1 &  0\mbox{ } &  k_3 \\
       k_1 & -k_2 &    0\mbox{ } \\
         0\mbox{ } &  k_2 & -k_3
    \end{array}
  \right]
  \left[ 
    \begin{array}{l}
      {\psi}_A \! \left( \phi \right) \\
      {\psi}_{2B} \! \left( \phi \right) \\
      {\psi}_{2C} \! \left( \phi \right) 
    \end{array}
  \right] . 
\label{eq:cycle_system_EOM_class}
\end{align}

The vector of complex activities on the non-equilibrium steady-state
solution is given by 
\begin{equation}
  \left[ 
    \begin{array}{l}
      {\psi}_A \! \left( \underline{n} \right) \\
      {\psi}_{2B} \! \left( \underline{n} \right) \\
      {\psi}_{2C} \! \left( \underline{n} \right) 
    \end{array}
  \right] \equiv 
  \left[ 
    \begin{array}{l}
      {\underline{n}}_A \\
      {\underline{n}}_B^2 \\
      {\underline{n}}_C^2 
    \end{array}
  \right] \propto
  \left[ 
    \begin{array}{l}
      1 / k_1 \\ 
      1 / k_2 \\ 
      1 / k_3 
    \end{array}
  \right] . 
\label{eq:cycle_system_NESS}
\end{equation}
From this, the steady-state concentrations can be computed, and the
duality transform~(\ref{eq:CRN_duals_def_genform}) becomes 
\begin{align}
  {\phi}_A 
& \equiv 
  \frac{
    {\varphi}_A 
  }{
    k_1 
  } & 
  {\phi}_B 
& \equiv 
  \frac{
    {\varphi}_B 
  }{
    \sqrt{k_2} 
  } & 
  {\phi}_C 
& \equiv 
  \frac{
    {\varphi}_C 
  }{
    \sqrt{k_3} 
  } 
\nonumber \\
  {\phi}^{\dagger}_A 
& \equiv k_1 
  {\varphi}^{\dagger}_A & 
  {\phi}^{\dagger}_B 
& \equiv \sqrt{k_2} 
  {\varphi}^{\dagger}_B & 
  {\phi}^{\dagger}_C 
& \equiv \sqrt{k_3} 
  {\varphi}^{\dagger}_C . 
\label{eq:cycle_system_dual_def}
\end{align}
The adjacency matrix in the dual generating functional, from
Eq.~(\ref{eq:tildeA_constr_genform}), is then
\begin{equation}
  {\tilde{\mathbb{A}}}_k = 
  \left[ 
    \begin{array}{rrr}
      -k_1 &  0\mbox{ } &  k_1 \\
       k_2 & -k_2 &    0\mbox{ } \\
         0\mbox{ } &  k_3 & -k_3
    \end{array}
  \right] . 
\label{eq:cycle_system_tildeA}
\end{equation}

In the dual theory, the stationary-path background for the observable
field $\varphi$ is given by $\bar{\varphi} \equiv 1$.  We write the
stationary-path equation of motion for the response field in its
transpose form, for the sake of comparison to the un-weighted equation
of motion~(\ref{eq:cycle_system_EOM_class}), as
\begin{align}
  0 
& = 
  -d_{\tau} {{\varphi}^{\dagger}}^T + 
  \frac{
    \partial {\psi}_Y^T \! \left( \varphi \right) 
  }{
    \partial {\varphi}^T
  }
  {\tilde{\mathbb{A}}}_k^T
  {\psi}_Y \! \left( {\varphi}^{\dagger} \right) 
\nonumber \\ 
& \rightarrow
  -d_{\tau} {{\varphi}^{\dagger}}^T + 
  Y 
  \left[  
    \begin{array}{rrr}
      -k_1 &  k_2 & 0\mbox{ } \\
       0\mbox{ } & -k_2 & k_3 \\
       k_1 & 0\mbox{ } & -k_3
    \end{array}
  \right]
  \left[ 
    \begin{array}{l}
      {\psi}_A \! \left( {\varphi}^{\dagger} \right) \\
      {\psi}_{2B} \! \left( {\varphi}^{\dagger} \right) \\
      {\psi}_{2C} \! \left( {\varphi}^{\dagger} \right) 
    \end{array}
  \right]
\label{eq:cycle_system_dual_EOM}
\end{align}

\begin{figure}[ht]
  \begin{center} 
  \includegraphics[scale=0.5]{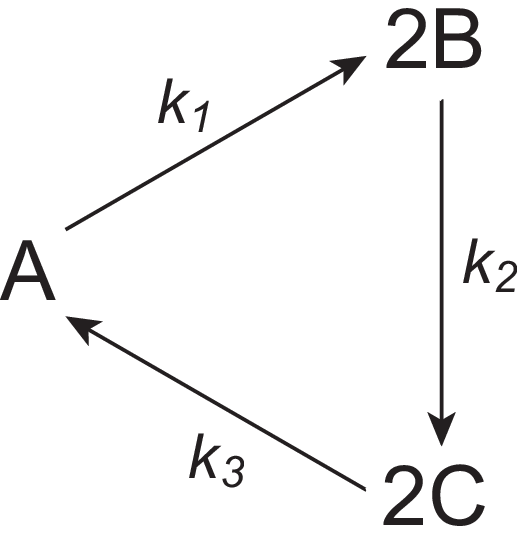} \qquad \qquad
  \includegraphics[scale=0.5]{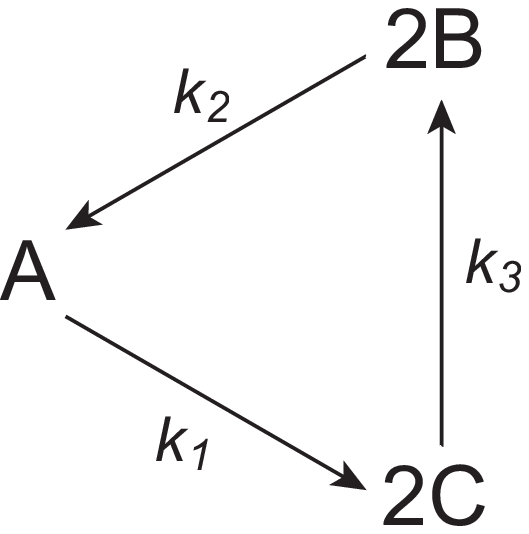} 
  \caption{
  Left graphic: original CRN with simple cycle.  Right graphic: dual
  CRN that propagates the inference field.  
  \label{fig:CRN_cycle} 
  } 
  \end{center}
\end{figure}

If the adjacency matrix had admitted a detailed-balance solution, then
by Eq.~(\ref{eq:DB_self_adjoint}),
Eq.~(\ref{eq:cycle_system_dual_EOM}) would simply have described the
process of Eq.~(\ref{eq:cycle_system_EOM_class}) with a mirror-image
trajectory for the rate constants.  In this fully-irreversible model,
the graphs for the original process, and for the dual process, shown
in Fig.~\ref{fig:CRN_cycle}, exhibit two further features.  In the
dual process, the only elementary moves are the opposites of those in
the underlying process.  The positions of the rate constants are also
moved to different links in the graph, as well as their time-courses'
becoming mirror images.

\subsubsection{An example in which a cycle co-occurs with symmetric
bi-directional links} 
\label{sec:examp_3cyc2bidir}

The second example combines the irreversible cycle of the previous
example with a conventional microscopically-reversible sub-network, to
show that duality is not always as simple as shuffling rate constants
on an existing complex network.

The reaction schema is 
\begin{align}
  {\rm A}
& \overset{k_1}{\rightharpoonup}
  2 {\rm B} 
\nonumber \\ 
  2 {\rm B} 
& \overset{k_2}{\rightharpoonup}
  2 {\rm C} 
\nonumber \\ 
  2 {\rm C} 
& \overset{k_3}{\rightharpoonup}
  {\rm A}
\nonumber \\
  2 {\rm B} 
& \overset{k_4}{\leftrightharpoons}
  {\rm D} 
\nonumber \\
  2 {\rm C} 
& \overset{k_4}{\leftrightharpoons}
  {\rm D} , 
\label{eq:hybrid_system}
\end{align}
and the rate matrix written explicitly is 
\begin{equation}
  {\mathbb{A}}_k = 
  \left[ 
    \begin{array}{rrrr}
      -k_1 & 0\mbox{ } & k_3 & 0\mbox{ } \\
       k_1 & - \left( k_2 + k_4 \right) & 0\mbox{ } & k_4 \\
         0\mbox{ } & k_2 & - \left( k_3 + k_4 \right) & k_4 \\
         0\mbox{ } & k_4 & k_4 & -2 k_4 
    \end{array}
  \right] . 
\label{eq:CRN_hybrid_Aform}
\end{equation}

The powers of species activities appearing in the complex-activity
vector $\psi \! \left( \underline{n} \right)$ on the non-equilibrium
steady state may be solved as 
\begin{equation}
  \left[ 
    \begin{array}{l}
      {\underline{n}}_A \\ 
      {\underline{n}}_B^2 \\ 
      {\underline{n}}_C^2 \\ 
      {\underline{n}}_D 
    \end{array}
  \right] \propto 
  \left[ 
    \begin{array}{l}
      \left( 
        2 k_2 + k_4 
      \right) / 2 k_1 k_2 \\
      \left( 
        2 k_3 + k_4 
      \right) / 2 k_2 k_3 \\
      \left( 
        2 k_2 + k_4 
      \right) / 2 k_2 k_3 \\
      \left( 
        k_2 + k_3 + k_4
      \right) / 2 k_2 k_3
    \end{array}
  \right]  
\label{eq:CRN_hybrid_NESS}
\end{equation}

\begin{figure}[ht]
  \begin{center} 
  \includegraphics[scale=0.5]{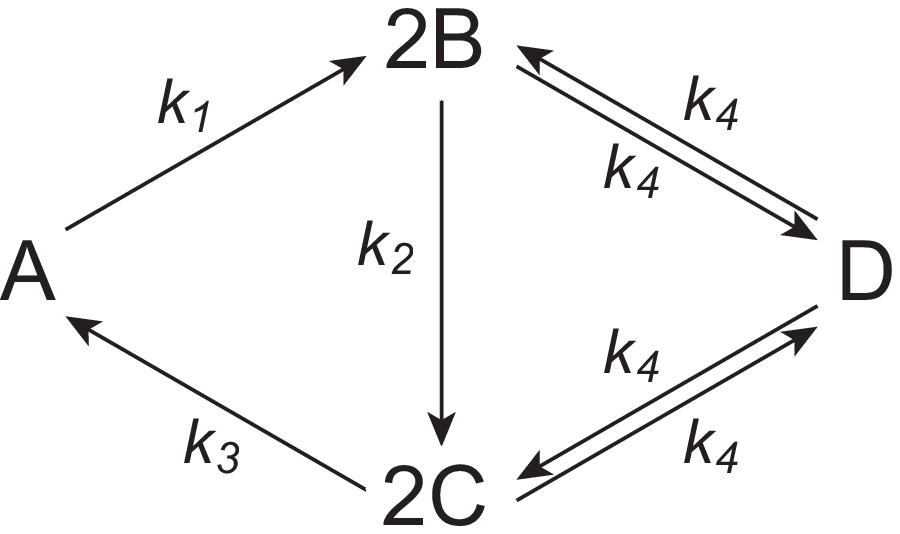} \\
  \bigskip
  \bigskip
  \includegraphics[scale=0.5]{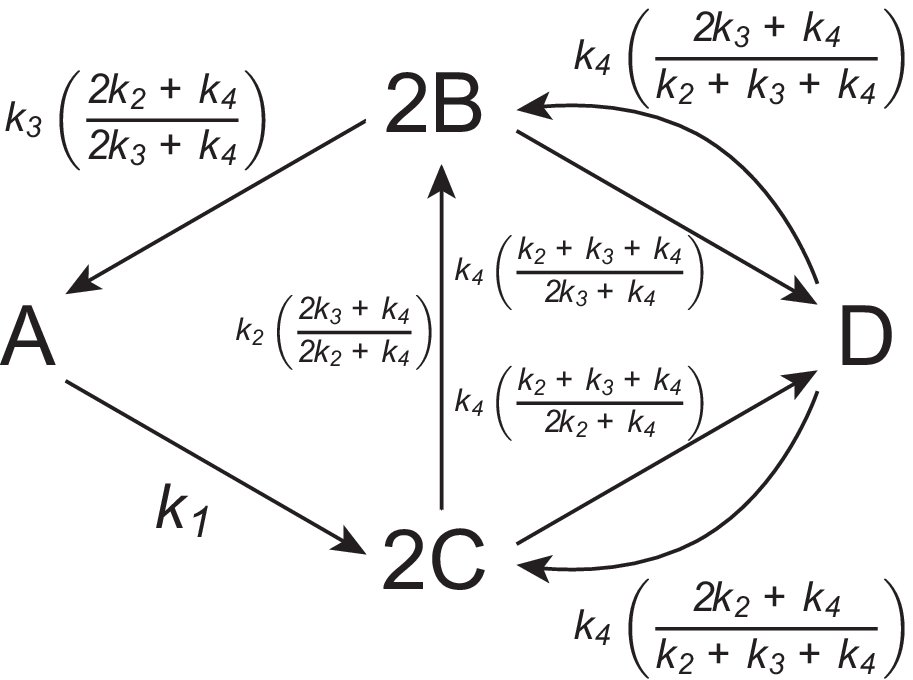} 
  \caption{
  Top graphic: original CRN with both a totally asymmetric loop and a
  totally-symmetric conversion.  Bottom graphic: dual CRN with respect
  to the canonical CRN on the same topology (see development in the
  text).  The dual graph now has structural asymmetry in the
  two-directional rates connected to complex $D$, emphasized by
  distinguishing their arcs.
  \label{fig:CRN_2_loop} 
  } 
  \end{center}
\end{figure}

The rate constants in the dual matrix ${\tilde{\mathbb{A}}}_k$ now
have a more complex form, shown on the complex-network graphs for the
original and dual processes in Fig.~\ref{fig:CRN_2_loop}.  The graphs
refer to writing the equation of motion for the response field
transposed, as in the previous example.  All but one of the dual
reaction rates are non-trivial functions of the equilibrium across the
network.

\section{Concluding remarks}
\label{sec:concl}

The substitution of measures on trajectories to produce an expression
of the form $E_P \left( P^{\ast} / P \right) \equiv 1$ is a tautology;
particular classes of such transformations rise to significance when
they are identified by an interpretation linking a class of
measurements.  The use of the underlying measure on reversed
trajectories was the first such group to be extensively developed in
fluctuation
theorems~\cite{Jarzynski:eq_FE_diffs:97,Jarzynski:neq_FE_diffs:97,%
Crooks:NE_work_relns:99,Crooks:path_ens_aves:00}; its interpretation
was time reversal and it linked work along paths with total entropy
production.  The entropy produced in the bath is the equilibrium
entropy by the local-equilibrium assumption.  The entropy in the
explicitly resolved stochastic process is still the usual Shannon
form.  It would become an equilibrium entropy if the system were held
in a fixed state by an external boundary condition, as it is
momentarily held in that state by the waiting time for transitions in
the stochastic process.

Hatano and Sasa~\cite{Hatano:NESS_Langevin:01} showed that the
underlying measure is not the only one with a fluctuation-theorem
interpretation, and even if it exists, it may not be a useful measure
for questions involving nonequilibrium systems.  However, they
provided an alternative in the adjoint measure.  Its interpretation is
the reversal of probability currents in the instantaneous steady-state
distribution, and it links the quantity identified by Oono and
Paniconi~\cite{Oono:NESS_thermo:98} as housekeeping heat, with the
change in Shannon entropy of these steady states.  Housekeeping heat
is still defined in terms of a near-equilibrium approximation for the
bath, making this interpretation of the adjoint fluctuation theorem
somewhat contingent on the system~\cite{Seifert:stoch_thermo_rev:12}.
We have therefore developed the adjoint fluctuation theorems in terms
that are independent of any external thermodynamic interpretations of
the rate constants, emphasizing instead generating functionals, memory
erasure, and inference, and we have developed some of the consequences
for correlation functions at the level of generality of stochastic
CRNs.

We would like to regard the two measures studied so far as two early
cases within a more general duality concept captured by fluctuation
theorems.  One can imagine designing adjoint transformations to yield
measures of other aspects of non-equilibrium structure, as the
Hatano-Sasa construction measures changes in system Shannon entropy
under a Langevin equation, using excess heat.  We hope that a
generating-functional framing is helpful in developing that
generalization systematically.

\subsection*{Acknowledgments}

DES thanks the Physics Department at Stockholm University for support
during visits in 2014, 2016, 2017 and 2018 when the bulk of this work
was carried out.  The authors are grateful to Massimiliano Esposito,
David Lacoste, Sreekanth Manikandan, Luca Peliti and Gatien Verley for
very helpful discussions and references.

\appendix 

\section{Definitions, notation, and standard constructions for the
  Doi-Peliti generating functional}
\label{app:DP_details}

Here we briefly review the operator-algebra construction of
Doi~\cite{Doi:SecQuant:76,Doi:RDQFT:76} for moment-generating
functions, and the Peliti coherent-state
expansion~\cite{Peliti:PIBD:85,Peliti:AAZero:86} that creates a
functional-integral representation of extended-time generating
functions and functionals.  More didactic reviews may be found
in~\cite{Mattis:RDQFT:98,Cardy:FTNEqSM:99,Smith:LDP_SEA:11}.

\subsection{The Doi operator algebra construction}
\label{sec:Doi_algebra_rev}

Following the notations~(\ref{eq:a_adag_defs}) to represent complex
arguments and their partial derivatives as raising and lowering
operators, with the operator commutator~(\ref{eq:comm_relns}), the Doi
algebra~\cite{Doi:SecQuant:76,Doi:RDQFT:76} defines a Hilbert space of
generating functions and an inner product that corresponds to
projection.  These are written as right and left ``ground states'',
with the correspondences
\begin{align}
  1 
& \rightarrow 
  \left| 0 \right)
% \nonumber \\
& 
  \int d^P \! z \, 
  {\delta}^P \! \left( z \right) 
& \rightarrow 
  \left( 0 \right| . 
\label{eq:null_states}
\end{align}
A moment-generating function is a polynomial in the components of $z$
multiplying the number 1, which in the Doi algebra is the action of a
polynomial of the raising operators acting on the left ground state.
Each monomial is a basis vector for this space, 
\begin{align}
  \prod_{p = 1}^P
  z_p^{{\rn}_p} \times 
  1 
& \rightarrow 
  \prod_{p = 1}^P
  {
    a_p^{\dagger} 
  }^{{\rn}_p} 
  \left| 0 \right) \equiv 
  \left| \rn \right) ; 
\label{eq:number_states}
\end{align}
the basis vectors are termed \textit{number states}.  

Under the mapping~(\ref{eq:number_states}) the analytic function $\Phi
\! \left( z \right)$ becomes a state vector $\left| \Phi \right)$,
given by 
\begin{align}
  \Phi \! \left( z \right) = 
  \sum_{\rn}
  {\rho}_{\rn}
  \prod_{p = 1}^P
  z_p^{{\rn}_p} \times 1 
& \rightarrow 
  \sum_{\rn}
  {\rho}_{\rn}
  \left| \rn \right) \equiv 
  \left| \Phi \right) . 
\label{eq:genfun_to_state}
\end{align}
All number states have unit normalization in an inner product known as
the \textit{Glauber norm}, given by 
\begin{align}
  \left( 0 \right|
  e^{\sum_p a_p}
  \left| \rn \right) = 1 , \quad
  \forall \rn .
\label{eq:Glauber_inn_prod}
\end{align}
As a consequence, the Glauber norm of each generating function is
simply the trace of the underlying probability density:
\begin{align}
  \left( 0 \right|
  e^{\sum_p a_p}
  \left| \Phi \right) = 
  \sum_{\rn}
  {\rho}_{\rn} = 1 . 
\label{eq:Glauber_is_trace}
\end{align}

The Doi algebra may be seen simply as a way to use integration and
$\delta$-functions to change the variable argument of a generating
function from a complex argument $z$ to a formal argument
$a^{\dagger}$.  The compact notation it provides for an inner product
becomes very convenient when an evolving generating function must be
projected onto a basis at each of a large sequence of time intervals,
as is done to integrate the equations of motion.  The inverse
transform, from variables $a^{\dagger}$ back to analytic argument $z$,
is given by the inner product Eq.~(\ref{eq:Glauber_to_FG}) in the
text, of which the Glauber norm~(\ref{eq:Glauber_is_trace}) is a
special case.

\subsection{The Peliti coherent-state expansion and path integral}
\label{sec:Peliti_integral_rev}

In the Doi Hilbert-space representation, time
evolution~(\ref{eq:Liouville_eq_aadag}) is formally reduced to
quadrature by exponentiation in the Liouville operator:
\begin{equation}
  \left| {\Phi}_T \right) = 
  \mathbb{T}
  e^{
    - \int_0^T d\tau
    {\mathcal{L}}_{\tau}
  }
  \left| {\Phi}_0 \right) .
\label{eq:GF_quadrature}
\end{equation}
Here the subscript ${\mathcal{L}}_{\tau}$ indicates that the
parameters in $\mathcal{L}$ may be explicit functions of time, and
$\mathbb{T}$ stands for time-ordering of the exponential, written as a
product of terms over successive small time slices.  

The 2FFI representation introduced by
Peliti~\cite{Peliti:PIBD:85,Peliti:AAZero:86} provides a way to
evaluate the quadrature~(\ref{eq:GF_quadrature}) by expanding the
arbitrary evolving generating function in a basis of \textit{coherent
  states}, which were introduced in Eq.~(\ref{eq:coh_state_def}).
Coherent states are eigenstates of all coefficients in the lowering
operator:
\begin{equation}
  a_p 
  \left| \phi \right) = 
  {\phi}_p
  \left| \phi \right) . 
\label{eq:phi_coherent_eigs}
\end{equation}

A sequence of insertions of the representation of
unity~(\ref{eq:field_int_ident}), formed from the outer product of
left and right coherent states (projection operators and their
conjugate generating functions), defines a skeletonized functional
integral, in which the values of $\phi$ and ${\phi}^{\dagger}$ at each
time become field-variables of integration.  In the continuum limit
defined by letting the spacing between insertions $\delta \tau
\rightarrow 0$, the functional measure of integration is denoted and
defined as
\begin{equation}
  \int_0^T
  \mathcal{D} {\phi}^{\dagger}
  \mathcal{D} \phi \equiv 
  \lim_{\delta \tau \rightarrow 0}
  \prod_{k = 0}^{T/\delta \tau}
  \int 
  \frac{
    d^p \! {\phi}^{\dagger}_{k \delta \tau}
    d^p \! {\phi}_{k \delta \tau}
  }{
    {\pi}^p
  } . 
\label{eq:ctm_measure}
\end{equation}
From the quadrature~(\ref{eq:GF_quadrature}), the insertion of this
skeletonized measure, with Eq.~(\ref{eq:Glauber_to_FG}) used to change
variable back from $a^{\dagger}$ to a complex surface argument $z$,
produces the functional integral representation~(\ref{eq:PI_genform})
for the generating function of the time-dependent density $\rho$
evolved from time $\tau = 0$ to time $\tau = T$.

\section{The Feinberg decomposition for stochastic CRNs}
\label{app:CRN_bkgnd}

The main feature in our use of Doi-Peliti representations for
stochastic Chemical Reaction Networks is a decomposition due to
Feinberg~\cite{Feinberg:notes:79,Feinberg:def_01:87}, which recognizes
\textit{complexes} as formal objects distinct from species, and
factors the representation of reactions into independent events that
occur on the complex network, and stoichiometric relations that
connect removal or addition of a complex to removal or addition of a
set of species.  The complex network can be represented by an ordinary
directed graph, and the stochastic process on this network is
equivalent to a simple random walk on this graph.  The stoichiometric
relations are represented by links of a distinct type, giving a way to
represent the multi-hypergraph of the CRN with a bipartite ordinary
graph, with two kinds of nodes (species and complexes) and two kinds
of links (reactions and stoichiometric connections).  The work of
Feinberg typically represents only the complex network explicitly, and
we follow this convention in Sec.~\ref{sec:examples}.
Elsewhere~\cite{Smith:geo13:16,Krishnamurthy:CRN_moments:17,%
Smith:CRN_moments:17} we have used the bipartite graph representation
to more fully represent the multi-hypergraph.

To obtain a representation for the transition matrix of a stochastic
CRN, we begin with a matrix representation for the
\textit{adjacency/rate} structure on the complex network:
\begin{equation}
  {\mathbb{A}}_k \equiv 
  \sum_{
    \left( i , j \right) 
  }
  \left( 
    w_j - w_i 
  \right) 
  k_{ji}
  w_i^T . 
\label{eq:A_genform}
\end{equation}
Here $i$ and $j$ index complexes, and the ordered pair $\left( i , j
\right)$ indexes a reaction from complex $i$ to complex $j$.  We will
later assign a probability of formation to each complex in state
$\rn$, and imagine these probabilities written as a column vector
indexed by $i$.  $w_i$ is then a vector which is an indicator function
with value 1 at index $i$, and $w_i^T$ is a transpose which selects
the formation probability at complex $i$.  The parameter $k_{ji}$ is
the \textit{rate constant} for complex $i$, if formed, to be converted
to complex $j$. 

An important theorem for
CRNs~\cite{Feinberg:notes:79,Feinberg:def_01:87} is that the existence
and uniqueness of steady states of the time
evolution~(\ref{eq:ME_genform}) is for some cases determined by the
topology of the adjacency matrix, independent of its rate structure as
long as the rates are nonzero.  We will represent the topology by
referring to an adjacency matrix with all rates set (arbitrarily) to
unity, denoted by
\begin{equation}
  {\mathbb{A}}^{\rm top} \equiv 
  \sum_{
    \left( i , j \right) 
  }
  \left( 
    w_j - w_i 
  \right) 
  w_i^T . 
\label{eq:ref_A_genform}
\end{equation}

For the simple case where complexes are formed from species by
sampling without replacement in a well-mixed reactor (the only case we
will develop here; some more general cases are considered
in~\cite{Anderson:product_dist:10}), the probability to form complex
$i$ is given by a truncated factorial of the species numbers $\left\{
  {\rn}_p \right\}$, which we denote by the function
\begin{equation}
  {{\Psi}_Y}_i \! 
  \left( \rn \right) \equiv 
  \prod_p
  \frac{
    {\rn}_p ! 
  }{
    \left( {\rn}_p - y_p^i \right) !
  } . 
\label{eq:interpretation_act_prod}
\end{equation}
Here the $y_p^i$ are non-negative integer-valued
\textit{stoichiometric coefficients}, giving the number of members
from species $p$ that form complex $i$. 

In the Doi algebra, the operator that will produce the combinatorial
factor~(\ref{eq:interpretation_act_prod}) for complex $i$ when acting
on a number state $\left| \rn \right)$, and the conjugate operator
that will create the particles in complex $i$, are both given by the
same function taking respectively $a$ and $a^{\dagger}$ as arguments.
We denote these by
\begin{align}
  {{\psi}_Y}_i \! 
  \left( a \right) 
& \equiv 
  \prod_p
  {a}_p^{y_p^i} , & 
  {\psi}_Y^i \! 
  \left( a^{\dagger} \right) 
& \equiv 
  \prod_p 
  {a^{\dagger}}_p^{y^i_p} . 
\label{eq:interpretations_a}
\end{align}
In the transition matrix, when complex $i$ is consumed and complex $j$
is produced, the generator $\rm T$ adds probability to the state
$\left| \rn \right)$ from the state $\left| \rn + y^i - y^j \right)$,
for which each ${\rn}_p$ is offset to ${\rn}_p + y^i_p - y^j_p$.  The
shift operator that produces the index offset ${\rn}_p \rightarrow
{\rn}_p + y^i$ is also given by
\begin{equation}
  {{\psi}_Y}_i \! 
  \left( 
    e^{\partial / \partial \rn}
  \right) \equiv 
  \prod_p 
  e^{y^i_p \partial / \partial {\rn}_p} = 
  e^{
    {y^i}^T \partial / \partial \rn 
  } .  
\label{eq:psi_of_shift_def}
\end{equation}
Here $y^i \equiv \left[ y^i_p \right]$ is a column vector with
components $y^i_p$, and ${y^i}^T$ is its transpose, forming an inner
product with the vector $\partial / \partial \rn \equiv \left[
  \partial / \partial {\rn}_p \right]$.  The complementary shift
operator for ${\rn}_p \rightarrow {\rn}_p - y^j$ is ${{\psi}_Y}_j \!
\left( e^{- \partial / \partial \rn} \right)$, and the two
operators commute.

From these definitions, the expression~(\ref{eq:T_psi_from_A}) for the
transition matrix and~(\ref{eq:L_psi_from_A}) for the Liouville
operator follow.

\section{Importance Sampling and relations to saddle-point methods}
\label{sec:IS_SP}

Here we review basic concepts and terms associated with Importance
Sampling, which are helpful in interpreting the roles and meanings of
the observable and response fields in DP functional integrals.
Although there is no process identical to sample estimation in the
2FFI representation, many of the same criteria and interpretations
apply to stationary-point expansions, which are a type of saddle-point
approximation.  In the interest of brevity, we simplify from the
general notation of the text for binomial distributions, and assign
reduced names for distribution parameters.

\subsection{Altering the performance of sample estimators by
  Importance Sampling}

Consider the problem of estimation by sampling a variable $\rn \in 0 ,
\ldots , N$ from a binomial distribution with parameter $p$:
\begin{equation}
  {\rho}_{\rn} = 
  p^{\rn}
  {\left( 1 - p \right)}^{N - \rn}
  \left(
    \begin{array}{c}
      N \\ \rn 
    \end{array}
  \right) . 
\label{eq:binomial_min_form}
\end{equation}
Expectations in $\rho$ are defined as in Eq.~(\ref{eq:EA_min_form}).

Sample estimation may perform poorly if the support of the observable
$\mathcal{O}$ falls outside the range where $\rho$ produces many
samples; or if the mean depends on an interaction between the shape of
$\mathcal{O}$ and the shape of $\rho$, as is often the case if $\rho$
must be sampled in a ``tail''.

It may be possible to mitigate both problems by drawing samples, not
from the original distribution $\rho$, but rather from a tilted
distribution $\tilde{\rho}$.  In order to obtain an unbiased estimator
for the mean, the observable $\mathcal{O}$ must be multiplied by a
weight function that compensates for the tilt of the sample
distribution.  The key observation behind the protocol known as
\textit{Importance Sampling} (IS)~\cite{Owen:mcbook:13} is that
the tilt may be chosen so that sample values of the weighted
observable cluster closer to their mean than did those of the
unweighted observable.  If the binomial parameter $p$ is tilted to a
new parameter $q$, the expectation~(\ref{eq:EA_min_form}) becomes
\begin{align}
  \left< \mathcal{O} \right> 
& = 
  \sum_{\rn = 0}^N
  {
    \left( \frac{p}{q} \right)
  }^{\rn}
  {
    \left( \frac{1-p}{1-q} \right)
  }^{N - \rn}
  {\mathcal{O}}_{\rn}
  q^{\rn}
  {\left( 1 - q \right)}^{N - \rn}
  \left(
    \begin{array}{c}
      N \\ \rn 
    \end{array}
  \right)  
\nonumber \\ 
& \equiv 
  \sum_{\rn = 0}^N
  {\tilde{\mathcal{O}}}_{\rn}
  q^{\rn}
  {\left( 1 - q \right)}^{N - \rn}
  \left(
    \begin{array}{c}
      N \\ \rn 
    \end{array}
  \right)  
\nonumber \\ 
& \equiv 
  \sum_{\rn = 0}^N
  {\tilde{\mathcal{O}}}_{\rn}
  {\tilde{\rho}}_{\rn} . 
\label{eq:EA_split_rho}
\end{align}
The tilted density $\tilde{\rho}$ is termed the \textit{importance
distribution}, and the residual correction factor that converts
$\mathcal{O}$ into $\tilde{\mathcal{O}}$ is termed the
\textit{likelihood ratio}.  The original density $\rho$ is termed the
\textit{nominal distribution}.  The likelihood ratio under an
exponential tilting is also known as the \textit{Radon-Nikodym}
derivative of the map from the nominal probability measure $\rho$ to
the importance measure $\tilde{\rho}$.

Suppose, for instance, that ${\mathcal{O}}_{\rn} = {\varphi}^{\rn}$
for some constant $\varphi$.  Then if we choose 
\begin{equation}
  q = 
  \frac{
    \varphi p 
  }{
    1 - p + \varphi p 
  } , 
\label{eq:shift_choice_binom}
\end{equation}
the likelihood-weighted observable 
\begin{align}
  {\tilde{\mathcal{O}}}_{\rn} 
& = 
  {
    \left( \frac{1-p}{1-q} \right)
  }^N 
  {
    \left(
      \frac{
        p \left( 1 - q \right) \varphi
      }{
        q \left( 1 - p \right)
      }
    \right)
  }^{\rn} = 
  {
    \left( \frac{1-p}{1-q} \right)
  }^N \times
  1^{\rn}
\nonumber \\ 
& = 
  {
    \left( 1 - p + \varphi p \right)
  }^N = 
  \left< \mathcal{O} \right> , 
\label{eq:shifted_A_eval}
\end{align}
\emph{at all $\rn$.}  Whereas the values of the original observable
$\mathcal{O}_{\rn}$ are dispersed in ${\rn}$, and the shape of
${\rho}_{\rn}$ is needed to compute the mean from them, the values of
${\tilde{\mathcal{O}}}_{\rn}$ are tightly compressed around the mean
(here, they are perfectly collapsed onto it), and the shape of
${\tilde{\rho}}_{\rn}$ does not matter.  

\subsection{Use of tilting in saddle-point methods}

The criterion of minimizing variance in sample-estimation is related
to approximation procedures such as saddle-point expansions, for which
exponential tilting also is commonly
employed~\cite{Goutis:saddlepts:95}.  For the same distribution $\rho$
and exponential observable ${\mathcal{O}}_{\rn} = {\varphi}^{\rn}$,
the leading exponential dependence in the sum~(\ref{eq:EA_min_form})
defining $\left< \mathcal{O} \right>$ is given by
\begin{equation}
  - \frac{1}{N}
  \log \left( 
    {\mathcal{O}}_{\rn} {\rho}_{\rn}
  \right)
\sim 
  \frac{\rn}{N}
  \log \left(
    \frac{\rn}{N p \varphi}
  \right) + 
  \frac{N-\rn}{N}
  \log \left(
    \frac{N-\rn}{N \left(1 - p \right)}
  \right) , 
\label{eq:EA_log_expand}
\end{equation}
(where $\sim$ indicates the omission of higher-order corrections in
the Stirling formula for factorials).

The saddle point of the argument in Eq.~(\ref{eq:EA_min_form}) is the
value $\bar{n}$ where $d \log \left( {\mathcal{O}}_{\rn} {\rho}_{\rn}
\right) / d{\rn} = 0$, given by
\begin{equation}
  \frac{\bar{n}}{N - \bar{n}} = 
  \frac{p \varphi}{1 - p} .
\label{eq:EA_saddle_place}
\end{equation}
Thus $\bar{n} / N = q$ from Eq.~(\ref{eq:shift_choice_binom}), the
same saddle-point value that would be obtained by approximating the
tilted distribution $\tilde{\rho}$ alone.  In the saddle-point
approximation $\sum_{\rn} {\mathcal{O}}_{\rn} {\rho}_{\rn}$ is
replaced by the leading-exponential term
\begin{align}
  \sum_{\rn} {\mathcal{O}}_{\rn} {\rho}_{\rn} \sim 
  {\mathcal{O}}_{\bar{n}} {\rho}_{\bar{n}} \approx 
  e^{
    - N \log \left(
      \frac{N - \bar{n}}{N \left( 1 - p \right)}
    \right)
  } 
& = 
  {
    \left(
      \frac{1 - \bar{n} / N}{1-p}
    \right)
  }^N 
\nonumber \\
& = 
  \left< \mathcal{O} \right> . 
\label{eq:saddle_EA_eval}
\end{align}
For this case, the optimal IS-tilt is the one that, in
stationary-point approximation, maximally insulates the resulting
observable $\tilde{\mathcal{O}}$ from sample fluctuations, and in
complementary fashion brings the identification of the saddle point to
depend only on the importance distribution.  

The duality transformation~(\ref{eq:varphis_ulines}), which can be
used to absorb path weighting terms in the class of the NEWRs, has
exactly the effect of exchanging a Radon-Nikodym derivative between an
observable field that behaves as the mean of an importance
distribution, and a response field that carries a complementary
likelihood ratio.  When the path weighting is done in such a way that
the stationary points of either the observable or the response field
become fixed points of the forward or backward equations
(respectively, in the dual or the original variables), it has the
effect of maximally separating the values of one field from the
distribution dynamics of the other, as seen in the above example.

\section{Supporting algebra for elementary stationary-path solutions}
\label{app:stat_pt_alg}

This appendix provides supporting algebra for the stationary-point
evaluations of the four free field theories introduced in
Sec.~\ref{sec:duality_SPs}.  We keep surface arguments $z_{aT}$ and
$z_{bT}$ and their variations explicit whether they are $\sim
\mathcal{O} \! \left( {\delta \tau}^0 \right)$ or $1 + \mathcal{O} \!
\left( \delta \tau \right)$.  The four cases are contrasted according
to whether the absence or presence of dynamics (respectively)
preserves or dissipates the terminal values of stationary paths, and
according to whether (by duality transformation) the response fields
or observable fields are fixed points of the respective backward and
forward equations.

\subsection{Identity path-integral at a single time, followed by
  discrete tilting}
\label{app:null_SP}

The first variational derivatives in the action~(\ref{eq:S_null_form})
lead to stationary-path conditions for ${\phi}^{\dagger}$ of 
\begin{align}
  d_{\tau}
  {\bar{\phi}}_a^{\dagger} 
& = 
  0 , 
& 
  d_{\tau}
  {\bar{\phi}}_b^{\dagger} 
& = 
  0 .
\label{eq:null_vary_phi_bulk}
\end{align}
The surface terms at $\tau = T$ from variation of $\phi$ set the
boundary values of ${\bar{\phi}}^{\dagger}$ equal to the arguments
\begin{align}
  {({\bar{\phi}}_a^{\dagger})}_T
& = 
  z_a , 
& 
  {({\bar{\phi}}_b^{\dagger})}_T
& = 
  z_b .
\label{eq:null_vary_phi_surf}
\end{align}

The stationary-path conditions for $\phi$ are likewise 
\begin{align}
  d_{\tau}
  {\bar{\phi}}_a
& = 
  0 , 
& 
  d_{\tau}
  {\bar{\phi}}_b
& = 
  0 .
\label{eq:null_vary_phi_dag_bulk}
\end{align}
Their boundary values are set by the variations of
${\phi}^{\dagger}$ at $\tau = 0$, to give 
\begin{align}
  {(\bar{\phi}_a)}_0
& = 
  \frac{
    \partial \log {\psi}_0
  }{
    \partial {({\bar{\phi}}_a^{\dagger})}_0
  } = 
  \frac{
    N 
    \left( 1 - {\bar{x}}_0 \right)
  }{
    \left( 1 - {\bar{x}}_0 \right)
    {({\bar{\phi}}_a^{\dagger})}_0 + 
    \left( 1 + {\bar{x}}_0 \right)
    {({\bar{\phi}}_b^{\dagger})}_0 
  } , 
\nonumber \\
  {(\bar{\phi}_b)}_0
& = 
  \frac{
    \partial \log {\psi}_0
  }{
    \partial {({\bar{\phi}}_b^{\dagger})}_0
  } = 
  \frac{
    N 
    \left( 1 + {\bar{x}}_0 \right)
  }{
    \left( 1 - {\bar{x}}_0 \right)
    {({\bar{\phi}}_a^{\dagger})}_0 + 
    \left( 1 + {\bar{x}}_0 \right)
    {({\bar{\phi}}_b^{\dagger})}_0 
  } .
\label{eq:null_vary_phi_dag_surf}
\end{align}
Because we have chosen to vary $z_a$, $z_b$ within the
contour~(\ref{eq:z_shift_vals_assn}), the denominators in
Eq.~(\ref{eq:null_vary_phi_dag_surf}) equal 2 at all values of $z_b -
z_a$.  The solutions $\bar{\phi}$, ${\bar{\phi}}^{\dagger}$, and
$\bar{n}$ are then constant at the values in
Eq.~(\ref{eq:SPs_null_evals}).

\subsection{Continuous tilting at a single time using a 2-field
functional integral} 
\label{app:tilt_only_SP}

Because the duality transform~(\ref{eq:varphis_ulines}) converts the
action~(\ref{eq:S_triv_form}) into the stationary form
~(\ref{eq:S_triv_var}), the variations are the same as in the previous
case except for their boundary values.  First variational derivatives
give 
\begin{align}
  d_{\tau}
  {\bar{\varphi}}_a^{\dagger} 
& = 
  0 , 
& 
  d_{\tau}
  {\bar{\varphi}}_b^{\dagger} 
& = 
  0 ; 
\label{eq:triv_vary_phi_bulk} \\
  d_{\tau}
  {\bar{\varphi}}_a
& = 
  0 , 
& 
  d_{\tau}
  {\bar{\varphi}}_b
& = 
  0 .
\label{eq:triv_vary_phi_dag_bulk}
\end{align}

The surface terms at $\tau = T$ from variation of the $\varphi$ fields
set the response fields equal to the arguments
\begin{align}
  {({\bar{\varphi}}_a^{\dagger})}_T
& = 
  z_{aT} 
  \left( 1 - {\underline{x}}_{T - \delta \tau} \right) 
& 
  {({\bar{\varphi}}_b^{\dagger})}_T
& = 
  z_{bT} 
  \left( 1 + {\underline{x}}_{T - \delta \tau} \right) 
\nonumber \\
& = 
  \left( 1 - {\underline{x}}_T \right) ,
& 
& = 
  \left( 1 + {\underline{x}}_T \right) . 
\label{eq:triv_vary_phi_surf}
\end{align}
The $\tau = 0$ boundary values for $\bar{\varphi}$ take a simple form
due to the duality transform because we have chosen ${\underline{x}}_0
= {\bar{x}}_0$:
\begin{align}
  {(\bar{\varphi}_a)}_0
& = 
  \frac{
    \partial \log {\psi}_0
  }{
    \partial {({\bar{\varphi}}_a^{\dagger})}_0
  } = 
  \frac{
    N   
  }{
    {({\bar{\varphi}}_a^{\dagger})}_0 + 
    {({\bar{\varphi}}_b^{\dagger})}_0
  } , 
\nonumber \\
  {(\bar{\varphi}_b)}_0
& = 
  \frac{
    \partial \log {\psi}_0
  }{
    \partial {({\bar{\varphi}}_b^{\dagger})}_0
  } = 
  \frac{
    N   
  }{
    {({\bar{\varphi}}_a^{\dagger})}_0 + 
    {({\bar{\varphi}}_b^{\dagger})}_0
  } . 
\label{eq:triv_vary_phi_dag_surf}
\end{align}

Thus both $\bar{\varphi}$ and ${\bar{\varphi}}^{\dagger}$ are constant
at their boundary values.  In particular, $\bar{n}$, which is an
invariant under the duality transform, is given by 
\begin{align}
  \bar{n}_a 
& = 
  {\bar{\phi}}_a^{\dagger}
  {\bar{\phi}}_a = 
  {\bar{\varphi}}_a^{\dagger}
  {\bar{\varphi}}_a = 
  \frac{N}{2}
  \left( 1 - {\underline{x}}_T \right) , 
\nonumber \\
  \bar{n}_b 
& = 
  {\bar{\phi}}_b^{\dagger}
  {\bar{\phi}}_b = 
  {\bar{\varphi}}_b^{\dagger}
  {\bar{\varphi}}_b = 
  \frac{N}{2}
  \left( 1 + {\underline{x}}_T \right) .
\label{eq:triv_exp_num}
\end{align}
Inverting the duality transform gives the $\tau$-dependent values for
$\bar{\phi}$ and ${\bar{\phi}}^{\dagger}$ of
Eq.~(\ref{eq:SPs_tilt_evals}).  

Note in particular, with reference to
Eq.~(\ref{eq:psi_prod_part_evolve}) that the stationary values for the
response fields correspond to
\begin{align}
  {({\bar{\phi}}_a^{\dagger})}_{\tau}
& = 
  \frac{
    1 - {\underline{x}}_T
  }{
    1 - {\underline{x}}_{\tau}
  } = 
  \prod_{{\tau}^{\prime} = \tau + d\tau}^T 
  \left( 
    \frac{
      {{\underline{\nu}}_a}_{{\tau}^{\prime}}
    }{
      {{\underline{\nu}}_a}_{{\tau}^{\prime} - d\tau}
    } 
  \right) ,
\nonumber \\ 
  {({\bar{\phi}}_b^{\dagger})}_{\tau}
& = 
  \frac{
    1 + {\underline{x}}_T
  }{
    1 + {\underline{x}}_{\tau}
  } = 
  \prod_{{\tau}^{\prime} = \tau + d\tau}^T 
  \left( 
    \frac{
      {{\underline{\nu}}_b}_{{\tau}^{\prime}}
    }{
      {{\underline{\nu}}_b}_{{\tau}^{\prime} - d\tau}
    } 
  \right) ,
\label{eq:triv_evol_phi_daggers}
\end{align}
the values of the likelihood ratio from the IS decomposition of the
tilted generating function.

\subsection{Retarded time evolution under the un-tilted stochastic
process} 
\label{app:untilted_SP}

The introduction of dynamics along the contour $\tau$ changes the
character of stationary paths by dissipating the boundary data at
$\tau = T$ (the weight variables $z_a$, $z_b$) and $\tau = 0$ (the
distribution parameter ${\bar{x}}_0$), and gradually substituting the
influence of the rate parameter ${\underline{x}}_{\tau}$ in the
transition matrix~(\ref{eq:T_two_state_genform}).  The observable
fields are evolved with retarded dynamics, and the response fields
with advanced dynamics.  Unlike either of the single-time weighting
protocols, the nominal distribution -- the distribution reported by
$\bar{n}$ -- becomes a dynamical function of $\tau$, sensitive to both
initial and final data as well as the history of rates
$\underline{x}$.

Variation of the observable fields $\phi$ in the action $S_{\rm dyn}$
from Eq.~(\ref{eq:S_plain_form}) gives advanced equations of motion
for the response fields with source $\underline{x}$,
\begin{align}
  \left( 
    - d_{\tau} + 1 
  \right)
  \left( 
    {\bar{\phi}}_b^{\dagger} - 
    {\bar{\phi}}_a^{\dagger} 
  \right)
& = 
  0 , 
\nonumber \\
  - d_{\tau} 
  \left( 
    {\bar{\phi}}_b^{\dagger} + 
    {\bar{\phi}}_a^{\dagger} 
  \right)
& = 
  \underline{x}
  \left( 
    {\bar{\phi}}_b^{\dagger} - 
    {\bar{\phi}}_a^{\dagger} 
  \right) .
\label{eq:plain_vary_phi_bulk}
\end{align}
The surface variation at $\tau = T$ again produces the
assignments~(\ref{eq:null_vary_phi_surf}).  The solutions for these in
terms of the arguments $z_a$, $z_b$ of the generating function at time
$T$ are
\begin{align}
  {
    \left( 
      {\bar{\phi}}_b^{\dagger} - 
      {\bar{\phi}}_a^{\dagger} 
    \right) 
  }_{\tau} 
& = 
  \left( z_b - z_a \right)
  e^{
    - \left( T - \tau \right) 
  } , 
\nonumber \\
  {
    \left( 
      {\bar{\phi}}_b^{\dagger} + 
      {\bar{\phi}}_a^{\dagger} 
    \right) 
  }_{\tau} 
& = 
  \left( z_b + z_a \right) + 
  \left( z_b - z_a \right)
  \int_{\tau}^T
  d {\tau}^{\prime}
  e^{
    - \left( T - {\tau}^{\prime} \right) 
  }
  {\underline{x}}_{{\tau}^{\prime}} . 
\label{eq:plain_phi_dag_sol}
\end{align}
At $z_a = z_b = 1$ we have the
usual~\cite{Mattis:RDQFT:98,Cardy:FTNEqSM:99} fixed-point solutions
${\bar{\phi}}_b^{\dagger} = {\bar{\phi}}_a^{\dagger} \equiv 1$.

Variation of the response fields ${\bar{\phi}}^{\dagger}$ fields gives
the retarded equations of motion for the observable fields 
\begin{align}
  d_{\tau}
  \left( 
    {\bar{\phi}}_b + 
    {\bar{\phi}}_a
  \right)
& = 
  0 , 
\nonumber \\
  \left( 
    d_{\tau} + 1 
  \right) 
  \left( 
    {\bar{\phi}}_b - 
    {\bar{\phi}}_a
  \right)
& = 
  \underline{x}
  \left( 
    {\bar{\phi}}_b + 
    {\bar{\phi}}_a
  \right) ,
\label{eq:plain_vary_phi_dag_bulk}
\end{align}
and the variation at $\tau = 0$ again gives the surface
conditions~(\ref{eq:null_vary_phi_dag_surf}). 

The contour for variation of $\left( z_a, z_b \right)$ that preserves
a normalized generating function is no longer determined only from the
starting and ending values ${\bar{x}}_0$, ${\underline{x}}_T$, because
the fields ${\bar{\phi}}^{\dagger}$ have an extended dependence on
$\underline{x}$.  However, at whatever the value the integral in
Eq.~(\ref{eq:plain_phi_dag_sol}) gives at $\tau = 0$, we can vary
$\left( z_a, z_b \right)$ along the contour passing through $\left( 1
  , 1 \right)$ and holding the denominator in
Eq.~(\ref{eq:null_vary_phi_dag_surf}) fixed at 2.  Then the solutions
for the observable fields at all times will be
\begin{align}
  {\bar{\phi}}_b + 
  {\bar{\phi}}_a
& \equiv 
  N , 
\nonumber \\
  {\bar{\phi}}_b - 
  {\bar{\phi}}_a
& = 
  N \bar{x} , 
\label{eq:piain_phi_solve}
\end{align}
for ${\bar{x}}_{\tau}$ solving Eq.~(\ref{eq:binoms_MFT_evol}) and
${\bar{x}}_0 = {\underline{x}}_0$. 

It can be shown that ${\bar{n}}_b + {\bar{n}}_a \equiv N$ at all
times, while if we write $z_b - z_a$ in terms of the displacement
${\left( {\bar{n}}_b - {\bar{n}}_a \right)}_T - N {\bar{x}}_T$, then
${\left( {\bar{n}}_b - {\bar{n}}_a \right)}_{\tau}$ is given by
Eq.~(\ref{eq:plain_genfun_num_sol}).  If $z_a = z_b = 1$, the number
fields as well as $\bar{\phi}$ obey
\begin{equation}
  d_{\tau}
  \left(
    {\bar{n}}_a -
    {\bar{n}}_b
  \right) = 
  - \left[ 
    \left(
      {\bar{n}}_a -
      {\bar{n}}_b
    \right) - 
    N \underline{x} 
  \right] . 
\label{eq:plain_n_EOM}
\end{equation}

\subsection{Advanced time evolution under the Crooks-tilted generating
functional} 
\label{app:Crooks_tilted_SP}

When cumulative tilting is matched to the rate parameters in the
transition matrix~(\ref{eq:T_two_state_genform}), we arrive at the
same final distribution as in the absence of dynamics, and even the
same sequence of stationary values for the observable fields.  The
nominal distribution remains dynamical as in the previous case, but
instead of being a sum of terms from the evolving distribution and a
final weight factor, it evolves with advanced dynamics from the final
distribution.

As for the single-time Green's function with cumulative tilting, the
stationary-path solutions are most easily solved in the dual
variables.  Starting from the action $S_{\rm Crooks}$ in 
Eq.~(\ref{eq:S_Crooks_var}), the equations of motion for the response
field are 
\begin{align}
  - d_{\tau}
  \left( 
    {\bar{\varphi}}_b^{\dagger} + 
    {\bar{\varphi}}_a^{\dagger} 
  \right)
& = 
  0 , 
\nonumber \\
  \left( 
    - d_{\tau} + 1 
  \right)
  \left( 
    {\bar{\varphi}}_b^{\dagger} - 
    {\bar{\varphi}}_a^{\dagger} 
  \right)
& = 
  \underline{x}
  \left( 
    {\bar{\varphi}}_b^{\dagger} + 
    {\bar{\varphi}}_a^{\dagger} 
  \right) , 
\label{eq:Crooks_vary_phi_bulk}
\end{align}
and the surface condition at $\tau = T$ is again given by
Eq.~(\ref{eq:triv_vary_phi_surf}).  $\left(
  {\bar{\varphi}}_b^{\dagger} + {\bar{\varphi}}_a^{\dagger} \right)$
is constant at its the terminal value 2, while $\left(
  {\bar{\varphi}}_b^{\dagger} - {\bar{\varphi}}_a^{\dagger} \right)$
is solved by
\begin{widetext}
\begin{align}
  {
    \left( 
      {\bar{\varphi}}_b^{\dagger} - 
      {\bar{\varphi}}_a^{\dagger} 
    \right)
  }_{\tau} 
& = 
  {
    \left( 
      {\bar{\varphi}}_b^{\dagger} - 
      {\bar{\varphi}}_a^{\dagger} 
    \right)
  }_T 
  e^{
    - \left( T - \tau \right) 
  } +  
  \left( 
    {\bar{\varphi}}_b^{\dagger} + 
    {\bar{\varphi}}_a^{\dagger} 
  \right)
  \int_{\tau}^T
  d {\tau}^{\prime}
  e^{
    - \left( {\tau}^{\prime} - \tau \right)
  }
  {\underline{x}}_{{\tau}^{\prime}} 
\nonumber \\
& = 
  \left( z_{bT} - z_{aT} \right)
  \left( 1 - {\underline{x}}_T^2 \right)
  e^{
    - \left( T - \tau \right)
  } + 
  \left( 
    {\bar{\varphi}}_b^{\dagger} + 
    {\bar{\varphi}}_a^{\dagger} 
  \right)
  \left[ 
    {\underline{x}}_{\tau} + 
    \int_{\tau}^T
    d {\tau}^{\prime}
    e^{
      - \left( {\tau}^{\prime} - \tau \right)
    }
    {\partial}_{{\tau}^{\prime}}
    {\underline{x}}_{{\tau}^{\prime}} 
  \right] . 
\label{eq:Crooks_varphi_dag_diff_sol}
\end{align}
\end{widetext}
When $z_{bT} = z_{aT}$, the first term in the second line of
Eq.~(\ref{eq:Crooks_varphi_dag_diff_sol}) vanishes, $\left(
{\bar{\varphi}}_b^{\dagger} + {\bar{\varphi}}_a^{\dagger}
\right) = z_{bT} + z_{aT}$, and $\left( {\bar{\varphi}}_b^{\dagger} -
{\bar{\varphi}}_a^{\dagger} \right)$ evolves under an \emph{advanced}
response to $\underline{x}$, dual in both time and field conjugation
to the retarded response that ${\bar{\phi}}_b - {\bar{\phi}}_a$ shows
in Eq.~(\ref{eq:plain_vary_phi_dag_bulk}).

The variation of response fields ${\varphi}^{\dagger}$ gives the
equations of motion for the observable fields $\bar{\varphi}$
\begin{align}
  \left( 
    d_{\tau} + 1 
  \right) 
  \left(
    {\bar{\varphi}}_b - 
    {\bar{\varphi}}_a
  \right)
& = 
  0 , 
\nonumber \\
  d_{\tau}
  \left(
    {\bar{\varphi}}_b + 
    {\bar{\varphi}}_a
  \right)
& = 
  \underline{x} 
  \left(
    {\bar{\varphi}}_b - 
    {\bar{\varphi}}_a
  \right) , 
\label{eq:Crooks_vary_phi_dag_bulk}
\end{align}
and again recovers the surface
conditions~(\ref{eq:triv_vary_phi_dag_surf}).  Thus $\bar{\varphi}_b =
\bar{\varphi}_a = N / 2$, independent of the trajectory
$\underline{x}$.  The \emph{observable} field in the dual variables
has become a fixed point of its forward equation -- the role played by
the \emph{response} function (with respect to the backward equation)
in an unweighted time evolution of a plain distribution as noted
following Eq.~(\ref{eq:plain_phi_dag_sol}).  

The components of the number field again satisfy ${\bar{n}}_b +
{\bar{n}}_a = N$ at all times, but the difference value $\left(
  {\bar{n}}_a - {\bar{n}}_b \right)$ satisfies
Eq.~(\ref{eq:SP_n_crooks}), with $\bar{x}$ obeying the advanced
equation of motion~(\ref{eq:binoms_MFT_adv_evol}) and ${\bar{x}}_T =
{\underline{x}}_T$.

To compare the field interpretations to the non-dynamical case with
cumulative tilting, we convert back using the duality
transform~(\ref{eq:varphis_ulines}), to obtain
\begin{align}
  \frac{1}{2}
  {
    \left( 
      {\bar{\phi}}_b^{\dagger} + 
      {\bar{\phi}}_a^{\dagger} 
    \right)
  }_{\tau} 
& = 
  1 - 
  \frac{
    {\underline{x}}_{\tau}
  }{
    \left( 1 - {\underline{x}}^2_{\tau} \right)
  }
  \int_{\tau}^T
  d {\tau}^{\prime}
  e^{
    - \left( {\tau}^{\prime} - \tau \right)
  }
  {\partial}_{{\tau}^{\prime}}
  {\underline{x}}_{{\tau}^{\prime}} , 
\nonumber \\
  \frac{1}{2}
  {
    \left( 
      {\bar{\phi}}_b^{\dagger} - 
      {\bar{\phi}}_a^{\dagger} 
    \right)
  }_{\tau} 
& = 
  \frac{
    1 
  }{
    \left( 1 - {\underline{x}}^2_{\tau} \right)
  }
  \int_{\tau}^T
  d {\tau}^{\prime}
  e^{
    - \left( {\tau}^{\prime} - \tau \right)
  }
  {\partial}_{{\tau}^{\prime}}
  {\underline{x}}_{{\tau}^{\prime}} , 
\nonumber \\
  {
    \left( 
      \bar{\phi}_b + 
      \bar{\phi}_a 
    \right)
  }_{\tau}
& = 
  N , 
\nonumber \\
  {
    \left( 
      \bar{\phi}_b - 
      \bar{\phi}_a 
    \right)
  }_{\tau}
& = 
  N {\underline{x}}_{\tau} . 
\label{eq:Crooks_non_var_allt}
\end{align}
By construction the observable fields are the same.  For the response
fields, the quantity ${\underline{x}}_T - {\underline{x}}_{\tau}$ in
Eq.~(\ref{eq:triv_evol_phi_daggers}) is replaced with the integral
$\int_{\tau}^T d {\tau}^{\prime} e^{ - \left( {\tau}^{\prime} - \tau
  \right) } {\partial}_{{\tau}^{\prime}}
{\underline{x}}_{{\tau}^{\prime}}$ in
Eq.~(\ref{eq:Crooks_non_var_allt}).

\section{Supporting algebra for Green's function computation in
action-angle variables}
\label{app:AA_Gs_comps}

The retarded and advanced Green's functions for number fields are
solutions to the inhomogeneous differential equations involving
$n^{\prime}$ in the second and fourth lines of
Eq.~(\ref{eq:etas_vary_AA_twoforms}), corresponding to
Eq.~(\ref{eq:G_RA_solve}) in coherent-state fields.

However, whereas the diffusion kernel $D_{\tau}$ in
Eq.~(\ref{eq:G_RA_solve}) leads to the characteristic decay rate 1
(the lower-right entry in Eq.~(\ref{eq:plain_D_form})), the two decay
rates appearing in the second and fourth lines of
Eq.~(\ref{eq:etas_vary_AA_twoforms}) are non-obvious functions of
$\underline{x}$ and $\bar{\eta}$ or ${\bar{\tilde{\eta}}}$, which
evaluate respectively to 
\begin{align}
  \ch \bar{\eta} + \underline{x} \sh \bar{\eta} 
& = 
  1 , 
\nonumber \\
  \ch \bar{\tilde{\eta}} - \underline{x} \sh \bar{\tilde{\eta}} 
& = 
  \frac{
    \left( 1 + {\bar{x}}^2 \right) - 
    2 \bar{x} \underline{x}
  }{
    \left( 1 - {\bar{x}}^2 \right)
  }
\nonumber \\
& = 
  1 - 
  d_{\tau}
  \log 
  \left( 1 - {\bar{x}}^2 \right) . 
\label{eq:AA_time_consts}
\end{align}

The argument of the decaying exponential, which is $\tau -
{\tau}^{\prime}$ in Eq.~(\ref{eq:plain_G_RA_eval}) for coherent-state
fields, remains $\tau - {\tau}^{\prime}$ for number fields in the
un-tilted stochastic process, but is replaced with the integral
\begin{equation}
  \tilde{Y}_{{\tau}^{\prime \prime}}^{\tau} \equiv 
  \int_{{\tau}^{\prime}}^{\tau} 
  d {\tau}^{\prime \prime}
  \left( 
    \ch \bar{\tilde{\eta}} - \underline{x} \sh \bar{\tilde{\eta}}
  \right) = 
  \left( \tau - {\tau}^{\prime} \right) + 
  \log 
  \left( 
    \frac{
      1 - {\bar{x}}^2_{{\tau}^{\prime}}
    }{
      1 - {\bar{x}}^2_{\tau}
    }
  \right)  
\label{eq:AA_dual_diss_rate}
\end{equation}
in the tilted generating functional.  

Using these in the equations~(\ref{eq:AA_GK_M_plain_dual}) for the
Keldysh kernel, one can check that both are solved by the functions
\begin{align}
  M 
& = 
  \frac{N}{4}
  \left( 1 - {\bar{x}}^2 \right) , 
& 
  \tilde{M} 
& = 
  \frac{N}{4}
  \left( 1 - {\bar{x}}^2 \right) , 
\label{eq:AA_M_both}
\end{align}
with the appropriate retarded solution~(\ref{eq:binoms_MFT_evol}) for
${\bar{x}}_{\tau}$ in the case of $M$ and the advanced
solution~(\ref{eq:binoms_MFT_adv_evol}) for ${\bar{x}}_{\tau}$ in the
case of $\tilde{M}$.  The quadrature form of
Eq.~(\ref{eq:AA_GK_M_plain_dual}) for the kernel $\tilde{M}$ is 
\begin{equation}
  {\tilde{M}}_{\tau} = 
  e^{
    - 2 \tilde{Y}_{{\tau}^{\prime}}^{\tau}
  }
  {\tilde{M}}_{{\tau}^{\prime}} + 
  \int_{{\tau}^{\prime}}^{\tau}
  d {\tau}^{\prime \prime}
  e^{
    - 2 \tilde{Y}_{{\tau}^{\prime \prime}}^{\tau}
  }
  \frac{N}{2}
  {
    \left( 1 - \bar{x} \underline{x} \right)
  }_{{\tau}^{\prime \prime}} . 
\label{eq:AA_tildeM_genform}
\end{equation}
Collecting these evaluations together gives the
expression~(\ref{eq:AA_dual_Gs_eval}) for the Green's functions in the
text.

\section{Rate matrix and duality for microscopically reversible systems}
\label{app:MR_sys_Aform}

Microscopic reversibility requires that the rate matrix have a form
\begin{equation}
  {\mathbb{A}}_k \equiv 
  \sum_{
    \left< i , j \right> 
  }
  \left( 
    w_j - w_i 
  \right) 
  \left( 
    k_{ji}
    w_i^T - 
    k_{ij}
    w_j^T
  \right) , 
\label{eq:A_DB_form}
\end{equation}
in which the rate constants can be expressed in terms of
single-complex and transition-state chemical potentials, as 
\begin{align}
  k_{ji}
& = 
  e^{
    - \beta 
    \left[ 
      {\underline{\mu}}_{\left< ij \right>} - 
      {\underline{\mu}}_i
    \right]
  } = 
  e^{
    - \beta 
    \left[ 
      {\underline{\mu}}_{\left< ij \right>} - 
      \left( {\underline{\mu}}_i + {\underline{\mu}}_j \right) / 2 
    \right]
  }
  e^{
    \beta \left( {\underline{\mu}}_i - {\underline{\mu}}_j \right) / 2 
  } , 
\nonumber \\ 
  k_{ij}
& = 
  e^{
    - \beta 
    \left[ 
      {\underline{\mu}}_{\left< ij \right>} - 
      {\underline{\mu}}_j
    \right]
  } = 
  e^{
    - \beta 
    \left[ 
      {\underline{\mu}}_{\left< ij \right>} - 
      \left( {\underline{\mu}}_i + {\underline{\mu}}_j \right) / 2 
    \right]
  }
  e^{
    \beta \left( {\underline{\mu}}_j - {\underline{\mu}}_i \right) / 2 
  } . 
\label{eq:DB_ks_form}
\end{align}
Here the single-complex chemical potentials ${\underline{\mu}}_i$ are
expressed in terms of the one-particle potentials
${\underline{\mu}}_p$ for each species, and the stoichiometric
coefficients, as
\begin{equation}
  {\underline{\mu}}_i \equiv 
  \sum_p 
  y_p^i 
  {\underline{\mu}}_p .
\label{eq:DB_complex_cp}
\end{equation}
$\left< ij \right>$ denotes the \emph{un-ordered} pair of $i$ and $j$,
and ${\underline{\mu}}_{\left< ij \right>}$ denotes the single-complex
transition-state free energy (equal to a chemical potential in energy
units) for the reactions $\left( ij \right)$ and $\left( ji \right)$,
the same in both directions under the assumption that these are
elementary reactions.  Eq.~(\ref{eq:A_DB_form}) then becomes 
\begin{widetext}
\begin{equation}
  {\mathbb{A}}_k \equiv 
  \sum_{
    \left< i , j \right> 
  }
  \left( 
    w_j - w_i 
  \right) 
  e^{
    - \beta 
    \left[ 
      {\underline{\mu}}_{\left< ij \right>} - 
      \left( {\underline{\mu}}_i + {\underline{\mu}}_j \right) / 2 
    \right]
  }
  \left( 
    e^{
      \beta \left( {\underline{\mu}}_i - {\underline{\mu}}_j \right) / 2 
    }
    w_i^T - 
    e^{
      \beta \left( {\underline{\mu}}_j - {\underline{\mu}}_i \right) / 2 
    }
    w_j^T
  \right) . 
\label{eq:A_DB_from_mus}
\end{equation}
\end{widetext}

The detailed-balance equilibrium satisfies 
\begin{equation}
  {\underline{n}}_p \propto 
  e^{
    - \beta {\underline{\mu}}_p
  } , 
\label{eq:DB_species_c}
\end{equation}
from which (choosing the normalization for convenience), the dual
variables~(\ref{eq:CRN_duals_def_genform}) become 
\begin{align}
  {\phi}_p 
& \equiv 
  e^{
    - \beta {\underline{\mu}}_p
  } 
  {\varphi}_p , 
\nonumber \\ 
  {\phi}^{\dagger}_p 
& \equiv 
  e^{
    \beta {\underline{\mu}}_p
  } 
  {\varphi}^{\dagger}_p . 
\label{eq:DB_duals_def}
\end{align}
The complex activities in Eq.~(\ref{eq:tildeA_constr_genform}) then
transform as
\begin{align}
  {\psi}_i \! \left( \phi \right) 
& =  
  e^{
    - \beta {\underline{\mu}}_i
  } 
  {\psi}_i \! \left( \varphi \right) , 
\nonumber \\ 
  {\psi}_i \! \left( {\phi}^{\dagger} \right) ,
& =  
  e^{
    \beta {\underline{\mu}}_i
  } 
  {\psi}^{\dagger}_i \! \left( {\varphi}^{\dagger} \right) .
\label{eq:DB_complex_c}
\end{align}
For the form~(\ref{eq:A_DB_form}) of ${\mathbb{A}}_k$, the rate matrix
for the dual complex network becomes 
\begin{equation}
  {\tilde{\mathbb{A}}}_k = 
  {\mathbb{A}}_k^T .
\label{eq:DB_self_adjoint}
\end{equation}

\vfill 
\eject 

% \bibliographystyle{unsrt} 
% \bibliography{DES}

\begin{thebibliography}{10}

\bibitem{Evans:shear_SS:93}
Denis~J. Evans, E.~G.~D. Cohen, and G.~P. Morriss.
\newblock Probability of second law violations in shearing steady states.
\newblock {\em Phys.~Rev.~Lett.}, 71:2401--2404, 1993.

\bibitem{Gallavotti:dyn_ens_NESM:95}
G.~Gallavotti and E.~D.~G. Cohen.
\newblock Dynamical ensembles in non-equilibrium statistical mechanics.
\newblock {\em Phys.~Rev.~Lett.}, 74:2694--2697, 1995.

\bibitem{Gallavotti:dyn_ens_SS:95}
G.~Gallavotti and E.~D.~G. Cohen.
\newblock Dynamical ensembles in stationary states.
\newblock {\em J.~Stat.~Phys.}, 80:931--970, 1995.

\bibitem{Cohen:NESM_2_thms:99}
E.~D.~G. Cohen and G.~Gallavotti.
\newblock Note on two theorems in nonequilibrium statistical mechanics.
\newblock {\em J.~Stat.~Phys.}, 96:1343--1349, 1999.

\bibitem{Jarzynski:eq_FE_diffs:97}
C.~Jarzynski.
\newblock Equilibrium free-energy differences from nonequilibrium measurements:
  A master-equation approach.
\newblock {\em Phys.~Rev.~E}, 56:5018--5035, 1997.

\bibitem{Jarzynski:neq_FE_diffs:97}
C.~Jarzynski.
\newblock Nonequilibrium equality for free energy differences.
\newblock {\em Phys.~Rev.~Lett.}, 78:2690--2693, 1997.

\bibitem{Kurchan:fluct_thms:98}
Jorge Kurchan.
\newblock Fluctuation theorem for stochastic dynamics.
\newblock {\em J.~Phys.~A}, 31:3719, 1998.

\bibitem{Searles:fluct_thm:99}
Denis~J. Evans and Debra~J. Searles.
\newblock Fluctuation theorem for stochastic systems.
\newblock {\em Phys.~Rev.~E}, 60:159--164, 1999.

\bibitem{Crooks:NE_work_relns:99}
Gavin~E. Crooks.
\newblock Entropy production fluctuation theorem and the nonequilibrium work
  relation for free energy differences.
\newblock {\em Phys.~Rev.~E}, 6:2721--2726, 1999.

\bibitem{Crooks:path_ens_aves:00}
Gavin~E. Crooks.
\newblock Path-ensemble averages in systems driven far from equilibrium.
\newblock {\em Phys.~Rev.~E}, 61:2361--2366, 2000.

\bibitem{Kurchan:NEWRs:07}
Jorge Kurchan.
\newblock Non-equilibrium work relations.
\newblock {\em J.~Stat.~Mech.}, 2007:P07005, 2007.

\bibitem{Jarzynski:fluctuations:08}
C.~Jarzynski.
\newblock Nonequilibrium work relations: foundations and applications.
\newblock {\em Eur.~Phys.~J.~B}, 64:331--340, 2008.

\bibitem{Chetrite:fluct_diff:08}
Rapha{\"{e}}l Chetrite and Krzysztof Gawedzki.
\newblock Fluctuation relations for diffusion processes.
\newblock {\em Commun.~Math.~Phys.}, 282:469--518, 2008.

\bibitem{Esposito:fluct_theorems:10}
Massimiliano Esposito and Christian Van~den Broeck.
\newblock Three detailed fluctuation theorems.
\newblock {\em Phys.~Rev.~Lett.}, 104:090601, 2010.

\bibitem{Hatano:NESS_Langevin:01}
Takahiro Hatano and Shin-ichi Sasa.
\newblock {Steady state thermodynamics of Langevin systems}.
\newblock {\em Phys.~Rev.~Lett.}, 86:3463--3466, 2001.

\bibitem{Chernyak:PI_fluct_thms:06}
V.~Chernyak, M.~Chertkov, and C.~Jarzynski.
\newblock {Path-integral analysis of fluctuation theorems for general Langevin
  processes}.
\newblock {\em J.~Stat.~Mech.}, page P08001, 2006.
\newblock doi:10.1088/1742-5468/2006/08/P08001.

\bibitem{Harris:fluct_thms:07}
R.~J. Harris and G.~M. Sch{\"{u}}tz.
\newblock Fluctuation theorems for stochasticdynamics.
\newblock {\em J.~Stat.~Mech.}, page P07020, 2007.
\newblock doi:10.1088/1742-5468/2007/07/P07020.

\bibitem{Oono:NESS_thermo:98}
Yoshitsugu Oono and Marco Paniconi.
\newblock Steady state thermodynamics.
\newblock {\em Prog.~Theor.~Phys.~Supplement}, 130:29--44, 1998.

\bibitem{Chetrite:fluct_nonstat_bath:09}
Rapha{\"{e}}l Chetrite.
\newblock Fluctuation relations for diffusion that is thermally driven by a
  nonstationary bath.
\newblock {\em Phys.~Rev.~E}, 80:051107, 2009.

\bibitem{Seifert:FDT:10}
Udo Seifert and Thomas Speck.
\newblock Fluctuation-dissipation theorem in nonequilibrium steady states.
\newblock {\em Europhysics Lett.}, 89:10007, 2010.

\bibitem{Chetrite:refreshing_fluct:11}
Rapha{\"{e}}l Chetrite and Shamik Gupta.
\newblock Two refreshing views of fluctuation theorems through kinematics
  elements and exponential martingale.
\newblock {\em J.~Stat.~Phys.}, 143:543--584, 2011.

\bibitem{Verley:FDT_motors:11}
G.~Verley, K.~Mallick, and D.~Lacoste.
\newblock Modified fluctuation-dissipation theorem for non-equilibrium steady
  states and applications to molecular motors.
\newblock {\em Europhys.~Lett.}, 93:10002, 2011.

\bibitem{Verley:FDT_ising:11}
Gatien Verley, Rapha{\"{e}}l Ch{\'{e}}trite, and David Lacoste.
\newblock {Modified fluctuation-dissipation theorem for general non-stationary
  states and applications to the Glauber-Ising chain}.
\newblock {\em J.~Stat.~Mech.}, page P10025, 2011.
\newblock arXiv:1108.1135v2.

\bibitem{Verley:HS_FDT:12}
Gatien Verley and David Lacoste.
\newblock Fluctuations and response from a hatano and sasa approach.
\newblock {\em Physica Scripta}, 86:058505, 2012.

\bibitem{Evans:fluct_thm:02}
Denis~J. Evans and Debra~J. Searles.
\newblock The fluctuation theorem.
\newblock {\em Adv.~Phys.}, 51:1529--1585, 2002.

\bibitem{Seifert:stoch_thermo_rev:12}
Udo Seifert.
\newblock Stochastic thermodynamics, fluctuation theorems, and molecular
  machines.
\newblock {\em Rep.~Prog.~Phys.}, 75:126001, 2012.
\newblock arXiv:1205.4176v1.

\bibitem{Espigares:inf_2nd_law_ineq:12}
Carlos P{\'{e}}rez-Espigares, Alejandro~B. Kolton, and Jorge Kurchan.
\newblock Infinite family of second-law-like inequalities.
\newblock {\em Phys.~Rev.~E}, 85:031135, 2012.

\bibitem{Perunov:adaptation:15}
Nikolai Perunov, Robert Marsland, and Jeremy England.
\newblock Statistical physics of adaptation.
\newblock 2015.
\newblock arXiv:1412.1875v1 [physics.bio-ph].

\bibitem{Horowitz:fine_tuning_NESM:17}
Jordan~M. Horowitz and Jeremy~L. England.
\newblock Spontaneous fine-tuning to environment in many-species chemical
  reaction networks.
\newblock {\em Proc.~Nat.~Acad.~Sci.~USA}, 114:7565--7570, 2017.

\bibitem{Onsager:RRIP1:31}
Lars Onsager.
\newblock {\protect Reciprocal Relations in Irreversible Processes. I.}
\newblock {\em Phys.~Rev.}, 37:405--426, 1931.

\bibitem{Onsager:RRIP2:31}
Lars Onsager.
\newblock {\protect Reciprocal Relations in Irreversible Processes. II.}
\newblock {\em Phys.~Rev.}, 38:2265--2279, 1931.

\bibitem{Nicolis:fluct_NEQ:71}
G.~Nicolis and I.~Prigogine.
\newblock Fluctuations in nonequlibrium systems.
\newblock {\em Proc.~Nat.~Acad.~Sci.~USA}, 68:2102--2107, 1971.

\bibitem{Glansdorff:structure:71}
P.~Glansdorff and I.~Prigogine.
\newblock {\em Thermodynamic Theory of Structure, Stability, and Fluctuations}.
\newblock Wiley, New York, 1971.

\bibitem{DeGroot:NET:84}
S.~R. de~Groot and P.~Mazur.
\newblock {\em Non-equilibrium Thermodynamics}.
\newblock Dover, New York, 1984.

\bibitem{Touchette:large_dev:09}
Hugo Touchette.
\newblock The large deviation approach to statistical mechanics.
\newblock {\em Phys.~Rep.}, 478:1--69, 2009.
\newblock arxiv:0804.0327.

\bibitem{Touchette:LDP_NEQ_sys:13}
Hugo Touchette and Rosemary~J. Harris.
\newblock Large deviation approach to nonequilibrium systems.
\newblock In Rainer Klages, Wolfram Just, and Christopher Jarzynski, editors,
  {\em Nonequilibrium Statistical Physics of Small Systems: Fluctuation
  Relations and Beyond}, pages 335--360. Wiley-VCH, Weinham, 2013.

\bibitem{Seifert:entP_traj:05}
Udo Seifert.
\newblock Entropy production along a stochastic trajectory and an integral
  fluctuation theorem.
\newblock {\em Phys.~Rev.~Lett.}, 95:040602, 2005.

\bibitem{Horn:mass_action:72}
F.~Horn and R.~Jackson.
\newblock General mass action kinetics.
\newblock {\em Arch.~Rat.~Mech.~Anal}, 47:81--116, 1972.

\bibitem{Feinberg:def_01:87}
Martin Feinberg.
\newblock {Chemical reaction network structure and the stability of complex
  isothermal reactors -- I. The deficiency zero and deficiency one theorems}.
\newblock {\em Chem.~Enc.~Sci.}, 42:2229--2268, 1987.

\bibitem{Weinberg:phenom_Lagr:79}
Steven Weinberg.
\newblock {Phenomenological Lagrangians}.
\newblock {\em Physica A}, 96:327--340, 1979.

\bibitem{Gardiner:stoch_meth:96}
Crispin Gardiner.
\newblock {\em Stochastic Methods: A Handbook for the Natural and Social
  Sciences}.
\newblock Springer, Heidelberg, 1996.

\bibitem{Doi:SecQuant:76}
M.~Doi.
\newblock Second quantization representation for classical many-particle
  system.
\newblock {\em J.~Phys.~A}, 9:1465--1478, 1976.

\bibitem{Doi:RDQFT:76}
M.~Doi.
\newblock Stochastic theory of diffusion-controlled reaction.
\newblock {\em J.~Phys.~A}, 9:1479--, 1976.

\bibitem{Peliti:PIBD:85}
L.~Peliti.
\newblock Path-integral approach to birth-death processes on a lattice.
\newblock {\em J.~Physique}, 46:1469, 1985.

\bibitem{Peliti:AAZero:86}
L.~Peliti.
\newblock Renormalization of fluctuation effects in $a + a \rightarrow a$
  reaction.
\newblock {\em J.~Phys.~A}, 19:L365, 1986.

\bibitem{Kamenev:DP:02}
Alex Kamenev.
\newblock Keldysh and doi-peliti techniques for out-of-equilibrium systems.
\newblock In I.~V. Lerner, B.~L. Althsuler, V.~I. Fal${}^{\prime}$ko, and
  T.~Giamarchi, editors, {\em Strongly Correlated Fermions and Bosons in
  Low-Dimensional Disordered Systems}, pages 313--340, Heidelberg, 2002.
  Springer-Verlag.

\bibitem{Martin:MSR:73}
P.~C. Martin, E.~D. Siggia, and H.~A. Rose.
\newblock Statistical dynamics of classical systems.
\newblock {\em Phys.~Rev.~A}, 8:423--437, 1973.

\bibitem{Schwinger:MBQO:61}
J.~Schwinger.
\newblock Brownian motion of a quantum oscillator.
\newblock {\em J.~Math.~Phys.}, 2:407--32, 1961.

\bibitem{Keldysh:noneq_diag:65}
L.~V. Keldysh.
\newblock Diagram technique for nonequilibrium processes.
\newblock {\em Sov.~Phys.~JETP}, 20:1018, 1965.

\bibitem{Freidlin:RPDS:98}
M.~I. Freidlin and A.~D. Wentzell.
\newblock {\em Random Perturbations in Dynamical Systems}.
\newblock Springer, New York, second edition, 1998.

\bibitem{Smith:LDP_SEA:11}
Eric Smith.
\newblock Large-deviation principles, stochastic effective actions, path
  entropies, and the structure and meaning of thermodynamic descriptions.
\newblock {\em Rep.~Prog.~Phys.}, 74:046601, 2011.
\newblock http://arxiv.org/submit/199903.

\bibitem{Smith:evo_games:15}
Eric Smith and Supriya Krishnamurthy.
\newblock {\em Symmetry and Collective Fluctuations in Evolutionary Games}.
\newblock IOP Press, Bristol, 2015.

\bibitem{Goutis:saddlepts:95}
Constantinos Goutis and George Casella.
\newblock Explaining the saddlepoint approximation.
\newblock {\em Am.~Statistician}, 53:216--224, 1999.

\bibitem{Owen:mcbook:13}
Art~B. Owen.
\newblock {\em Monte Carlo theory, methods and examples}.
\newblock http://statweb.stanford.edu/~owen/mc/, 2013.

\bibitem{Krishnamurthy:CRN_moments:17}
Supriya Krishnamurthy and Eric Smith.
\newblock Solving moment hierarchies for chemical reaction networks.
\newblock {\em J.~Phys.~A: Math.~Theor.}, 50:425002, 2017.

\bibitem{Smith:CRN_moments:17}
Eric Smith and Supriya Krishnamurthy.
\newblock Flows, scaling, and the control of moment hierarchies for stochastic
  chemical reaction networks.
\newblock {\em Phys.~Rev.~E}, 96:062102, 2017.

\bibitem{Baish:DP_duality:15}
Andrew~James Baish.
\newblock Deriving the jarzynski relation from doi-peliti field theory.
\newblock Bucknell University Honors Thesis, 2015.

\bibitem{Polettini:trans_fluct:14}
Matteo Polettini and Massimiliano Esposito.
\newblock Transient fluctuation theorems for the currents and initial
  equilibrium ensembles.
\newblock {\em J.~Stat.~Mech.}, page P10033, 2014.

\bibitem{Andersen:comp_rules:13}
Jakob~L. Andersen, Christoph Flamm, Daniel Merkle, and Peter~F. Stadler.
\newblock Inferring chemical reaction patterns using rule composition in graph
  grammars.
\newblock {\em J.~Sys.~Chem.}, 4:4:1--14, 2013.

\bibitem{Andersen:generic_strat:14}
Jakob~L. Andersen, Christoph Flamm, Daniel Merkle, and Peter~F. Stadler.
\newblock Generic strategies for chemical space exploration.
\newblock {\em Int.~J.~Comput.~Biol.~Drug Des.}, 7:225--258, 2014.

\bibitem{Andersen:NP_autocat:12}
Jakob~L. Andersen, Christoph Flamm, Daniel Merkle, and Peter~F. Stadler.
\newblock {Maximizing output and recognizing autocatalysis in chemical reaction
  networks is NP-complete}.
\newblock {\em J.~Sys.~Chem.}, 3:1, 2012.

\bibitem{Feinberg:notes:79}
Martin Feinberg.
\newblock Lectures on chemical reaction networks.
\newblock lecture notes, 1979.
\newblock https://crnt.osu.edu/LecturesOnReactionNetworks.

\bibitem{Mattis:RDQFT:98}
Daniel~C. Mattis and M.~Lawrence Glasser.
\newblock The uses of quantum field theory in diffusion-limited reactions.
\newblock {\em Rev.~Mod.~Phys}, 70:979--1001, 1998.

\bibitem{Cardy:FTNEqSM:99}
J.~Cardy.
\newblock Field theory and non-equilibrium statistical mechanics.
\newblock 1999.
\newblock http://www-thphys.physics.ox.ac.uk/users/JohnCardy/home.html.

\bibitem{Goldstein:ClassMech:01}
Herbert Goldstein, Charles~P. Poole, and John~L. Safko.
\newblock {\em Classical Mechanics}.
\newblock Addison Wesley, New York, third edition, 2001.

\bibitem{Weinberg:QTF_I:95}
Steven Weinberg.
\newblock {\em {The Quantum Theory of Fields, Vol.~I: Foundations}}.
\newblock Cambridge, New York, 1995.

\bibitem{Anderson:product_dist:10}
David~F. Anderson, George Craciun, and Thomas~G. Kurtz.
\newblock Product-form stationary distributions for deficiency zero chemical
  reaction networks.
\newblock {\em Bull.~Math.~Bio.}, 72:1947--1970, 2010.

\bibitem{Smith:geo13:16}
Eric Smith and Harold~J. Morowitz.
\newblock {\em The origin and nature of life on Earth: the emergence of the
  fourth geosphere}.
\newblock Cambridge U.~Press, London, 2016.

\bibitem{Gardiner:Poisson_rep:77}
C.~W. Gardiner and S.~Chaturvedi.
\newblock The poisson representation. i. a new technique for chemical master
  equations.
\newblock {\em J.~Stat.~Phys.}, 17:429--468, 1977.

\bibitem{Chaturvedi:Poisson_rep:78}
S.~Chaturvedi and C.~W. Gardiner.
\newblock The poisson representation. ii two-time correlation functions.
\newblock {\em J.~Stat.~Phys.}, 18:501--522, 1978.

\bibitem{Petrosyan:NEQ_Lyap:14}
K.~G. Petrosyan and Chin-Kun Hu.
\newblock {Nonequilibrium Lyapunov function and a fluctuation relation for
  stochastic systems: Poisson-representation approach}.
\newblock {\em Phys.~Rev.~E}, 89:042132, 2014.

\bibitem{Flajolet:anal_comb:09}
Philippe Flajolet and Robert Sedgewick.
\newblock {\em Analytic Combinatorics}.
\newblock Cambridge U.~Press, London, 2009.

\bibitem{Wilf:gen_fun:06}
Herbert~S. Wilf.
\newblock {\em Generatingfunctionology}.
\newblock A K Peters, Wellesley, MA, third edition, 2006.

\bibitem{Li:CRN_path_integrals:16}
Tiejun Li and Feng Lin.
\newblock Two-scale large deviations for chemical reaction kinetics through
  second quantization path integral.
\newblock {\em J.~Phys.~A}, 49:135204, 2016.

\end{thebibliography}

\end{document}